\documentclass[english,11pt,oneside]{article}

\pdfoutput=1
\usepackage[T1]{fontenc}
\usepackage[latin9]{inputenc}
\usepackage[usenames,dvipsnames]{xcolor}
\usepackage{color}
\usepackage{babel}
\usepackage{setspace}
\usepackage{amsmath}
\usepackage{amstext}
\usepackage{amssymb}
\PassOptionsToPackage{normalem}{ulem}
\usepackage{ulem}
\usepackage[unicode=true,pdfusetitle,
 bookmarks=true,bookmarksnumbered=false,bookmarksopen=false,citecolor=Turquoise,
 breaklinks=false,pdfborder={0 0 1},backref=false,colorlinks=true,pdfpagemode=FullScreen,linktoc=page]
 {hyperref}
\usepackage{graphicx}
\usepackage{mathtools}
\usepackage[small]{caption}
\usepackage{changepage}
\usepackage{slashed}
\definecolor{note_fontcolor}{rgb}{0.80078125, 0.80078125, 0.80078125}
\usepackage[top=3 cm, bottom=3 cm, left=3.1416 cm, right=3.1416 cm]{geometry}

\def\beq{\begin{equation}}
\def\eeq{\end{equation}}
\def\bea{\begin{eqnarray}}
\def\eea{\end{eqnarray}}

\begin{document} 

\baselineskip=17pt


\thispagestyle{empty}
\vspace{20pt}
\font\cmss=cmss10 \font\cmsss=cmss10 at 7pt

\begin{flushright}
%
UMD-PP-017-018 \\
\end{flushright}

\hfill

\begin{center}
{\Large \textbf
{LHC Signals for Singlet Neutrinos from a Natural Warped Seesaw (II)}}
\end{center}

\vspace{15pt}

\begin{center}
{\large Kaustubh Agashe$\, ^{a}$, Peizhi Du$\, ^{a}$, Sungwoo Hong$\, ^{a}$ \\
\vspace{15pt}
$^{a}$\textit{Maryland Center for Fundamental Physics,
     Department of Physics,
     University of Maryland,
     College Park, MD 20742, U.~S.~A.}
      \\ 
      
      \vspace{0.3cm}
      
{\it email addresses}: kagashe@umd.edu; pdu@umd.edu; sungwoo83hong@gmail.com}

\end{center}

\vspace{5pt}

\begin{center}
\textbf{Abstract}
\end{center}
\vspace{5pt} {\small \noindent

A \emph{natural} seesaw mechanism for obtaining the observed size of SM neutrino masses can arise in a warped extra dimensional/composite Higgs framework. In a previous paper, we initiated the study of signals at the LHC for the associated $\sim$ TeV mass SM singlet neutrinos, within a canonical model of $SU(2)_L \times SU(2)_R \times U(1)_{ B - L }$ (LR) symmetry in the composite sector, as motivated by consistency with the EW precision tests.
Here, we investigate LHC signals in a different region of parameter space for the \emph{same} model, where
production of singlet neutrinos can occur from particles \emph{beyond} those in usual LR models. 
Specifically, we assume that composite $(B - L)$ gauge boson is lighter than all the others in the EW sector. 
We show that the composite $(B - L)$ gauge boson can acquire a significant coupling to light quarks 
simply via mixing with elementary hypercharge gauge boson.
Thus, the singlet neutrino can be pair-produced via decays of $(B - L)$ gauge boson, \emph{without} a charged current counterpart. 
Furthermore, there is \emph{no} decay for $(B - L)$ gauge boson directly into dibosons, unlike for the usual case of $W_R^{ \pm }$ and $Z^{ \prime }$. Independently of the above extension of the EW sector, we analyze production of singlet neutrinos in decays of composite partners of $SU(2)_L$ \emph{doublet} leptons, which are absent in the usual LR models. In turn, these doublet leptons can be produced in composite $W_L$ decays. We show that $4 - 5 \sigma$ signal can be achieved for both cases described above for the following spectrum with 3000 fb$^{-1}$ luminosity: $2 - 2.5$ TeV composite gauge bosons, $1$ TeV composite doublet lepton (for the second case) and $500 - 750$ GeV singlet neutrino.


}
 
\vfill\eject
\noindent


\section{Introduction}

Numerous versions of seesaw mechanism for explaining the extreme smallness of Standard Model (SM) neutrino masses have been proposed over the last few decades.
In a previous paper \cite{Agashe:2015izu}, we discussed how a fully {\em natural} avatar arises with the SM fields propagating in a warped extra dimension. By the AdS/CFT correspondence, this framework is dual to the
SM Higgs boson being composite of some new strong dynamics, while rest of the SM fields are {\em partially} composite, i.e., admixtures of composites and elementary particles external to the strong dynamics.
In fact, this realization of the seesaw paradigm combines two specific models therein. Namely, it is basically an {\em inverse} seesaw \cite{inverse}, i.e., SM neutrino mass is generated by exchange of weak scale pseudo-Dirac singlet neutrinos, but with the tiny Majorana mass term for the relevant component of the singlet being the result of a {\em type I} high-scale seesaw \cite{original}.
Finally, the effective seesaw scale being several orders of magnitude below Planck scale can also be accommodated naturally in this scenario: for more explanation and more references, see reference \cite{Agashe:2015izu}.

All in all, a study of the signals at the Large Hadron Collider (LHC) from production of these TeV-mass singlets, which play crucial role in generating SM neutrino masses, is then highly motivated.
We would like to emphasize here that earlier analyses \cite{Huber:2003sf} of the same framework did not use the {\em mass} basis for these singlet modes, which resulted in a suggestion that it is a (purely) {\em high}-scale seesaw instead. With this {\em in}correct impression, even though a KK tower of singlet particles starting at $\sim$ TeV is still present, the focus would be instead on a super-heavy singlet mode, i.e., seemingly beyond {\em direct} reach of current/future experiments.
In this sense, as a consequence of the realization in reference \cite{Agashe:2015izu} that it is physically an inverse seesaw, the status of the TeV-mass singlets changed from being mere ``vestiges'' of SM neutrino mass generation to central players therein.

In a very recent paper \cite{Agashe:2016ttz}, we took the first step in this direction. There, we focussed on a specific five-dimensional (5D) model with a $SU(2)_L \times SU(2)_R \times U(1)_X$ {\em bulk} electroweak (EW) gauge symmetry, where $SU(2)_R \times U(1)_X$ is broken down to $U(1)_Y$ on the Planck brane \cite{Agashe:2003zs}.
This choice is dual to the composite sector respecting a $SU(2)_L \times SU(2)_R \times U(1)_X$ {\em global}
symmetry, whereas only its $SU(2)_L \times U(1)_Y$ subgroup is gauged by the elementary sector.
The purpose of this extension of the EW symmetry to a left-right symmetric (LR) structure was to ameliorate constraints from EW precision tests.
We analyzed the production of the composite singlet neutrinos via decays of {\em on}-shell composite gauge bosons, namely, the $W^{ \pm }_R$ and $Z^{ \prime }$. 
We assumed there that these neutrinos, along with $SU(2)_L$ \emph{singlet} composite charged leptons, are doublets of $SU(2)_R$, denoted by $\left( N^{ (1) }, \tilde{\ell}^{ (1) } \right)$. \footnote{ We use this notation for these composites since they correspond to KK modes of the 5D model: the qualifier ``tilde'' on $l$ will be explained later.}

In addition, in the above-mentioned work, we made two assumptions mostly for simplicity. The first one was that 
\begin{itemize}

\item
{\em all} the EW spin-1 composites are approximately degenerate, with the composite sector taken in isolation, i.e., neglecting the small mixing with the elementary sector,
{\em and} considering EW symmetry breaking (EWSB) effects also as a perturbation.

\end{itemize}
As we will explain below, the role of the above choice was crucial in inducing a significant coupling of 
light quarks to $W_R^{ \pm }$ and $Z^{ \prime }$, as required for their production at the LHC.
The net result is that for 2 TeV composite gauge boson mass, 750 GeV singlet neutrinos and with 300 fb$^{ -1 }$ luminosity, we can have discovery of singlet neutrinos via decay of $W_R^{ \pm }$ in this case \cite{Agashe:2016ttz}.

The second choice we implicitly made was that 
\begin{itemize}

\item
$SU(2)_L$ {\em doublet} composite leptons, which are mandatory in this framework, are heavier than one-half of the composite gauge boson, i.e, only singlets are in the game.

\end{itemize}
In summary, we see that the above signals are roughly similar to usual LR models\footnote{For a review, see \cite{Mohapatra:2016twe}.}, even though 
quantitative details are different, that too significantly. For example, note that $W_R^{ \pm }$ decay produces {\em composite} singlet charged lepton (in association with singlet neutrino), which, in turn, decays into SM charged lepton and Higgs (including longitudinal $W/Z$). That is, we get an {\em extra} identifiable final state particle as compared to usual LR model, where $W_R^{ \pm }$ {\em directly} decays into SM charged lepton (and singlet neutrino).

In the {\em present} paper, which is to be considered as the {\em second} installment of this series, 
we study LHC signals of singlet neutrinos in the {\em same} set-up as above, but now moving on to different region of its parameter space.
As we will see, even though this step looks simply like a ``quantitative'' change, we show that it will 
lead to variations in the {\em qualitative} features of the signals, in particular, involving particles
for which there is {\em no} counterpart in the usual LR models. In this sense,  the search channels will be 
even {\em more} different from the usual LR models than those discussed in the earlier paper.

In the first part of this paper, 
\begin{itemize}

\item
we will relax the earlier assumption of near-degeneracy of the composite spin-1 states. In particular, we keep some at 2 TeV so that we get sufficient production rate at the LHC, but raise the others above this value.

\end{itemize}
Indeed, the spirit here is simply to explore other options as compared to the previous paper. 
However, at first sight, this seems like a drastic step to take, namely, this could reduce the coupling of light quarks to $W^{ \pm }_R$, thus seems to render {\em negligible} the $W^{ \pm }_R$ signal (similarly for the $Z^{ \prime }$).
Hence, {\em naively} we {\em might} then have to look for alternate avenues for production of singlet neutrino in this case, recalling that the singlet neutrino couples only to $W^{ \pm }_R$ and $Z^{ \prime }$ among the spin-1 composites.\footnote{On the other hand, the composite charged lepton $\tilde{\ell}^{ (1) }$, couples also to composite hypercharge so that only external-composite mixing suffices for its production via light quark initial state, i.e., the EWSB induced mixing, which is suppressed for this case of {\em non}-degeneracy is {\em not} needed here. However, $\tilde{\ell}^{ (1) }$ is not {\em directly} related to the mechanism of generation of SM neutrino mass.}
Indeed, this is what actually happens for $W_R^{ \pm }$. 
\begin{itemize}

\item
Remarkably, we discover that in the process of making composite $( B - L )$ boson lighter\footnote{A factor of $\sim 1.5$ is enough here} than composite $W_R^3$ (or vice versa), a ``new'' {\em neutral} current channel can emerge. Namely, the lighter of these two composite gauge boson potentially still has significant production rate, with{\em out} requiring EWSB. 

\end{itemize}
%
This ``twist'' arises in a subtle manner as follows.
For this purpose, it is worthwhile recapping what are the various couplings of composite spin-1 particles to fermions.
First of all, the matter particles in a given sector, either elementary or composite, only couple to the corresponding gauge bosons. 
However, there is mass mixing between these two sectors, both for fermions and gauge bosons, resulting in modifications of these couplings as well. The light SM quarks are mostly elementary so that the coupling of these quarks to the heavy gauge bosons induced by the elementary-composite fermionic mixing is negligible.
However, a sizable (even if mildly suppressed as compared to SM) such coupling can result from mixing of elementary and composite {\em gauge bosons}\footnote{The amount of  {\em elementariness} of the {\em heavy} gauge bosons, even if small, is much larger than the degree of compositeness of light quarks.}: straightforwardly, we have elementary $W_L$ (charged {\em and} neutral) mixing with their composite counterparts.
Obviously, this opens the door for production of these composite particles at the LHC. Needless to say, the composite singlet neutrino does not couple to these composite $W_L$'s; on the other hand, it does have a coupling to composite $W_R^{ \pm }$. 
However, there is no mixing effect analogous to $W_L$ in the $W_R^{ \pm }$ sector, where there is {\em only} the composite side, thus making $W_R^{ \pm }$ (and, in turn, the singlet neutrino) {\em in}accessible to the LHC at this level.
Finally, there remain the {\em neutral} gauge bosons $W^3_R$ and $(B-L)$, where the situation is rather subtle as follows.
The  elementary hypercharge (denoted by $B$ henceforth) gauge boson mixes with a specific {\em combination} of the composite $W^3_R$ and $(B-L)$. The point is that, in the approximation of composite $W_R^3$ and $(B-L)$ being degenerate, this ``superposition'' {\em is} a mass eigenstate, dubbed ``composite'' hypercharge. Thus, the composite $B$ does couple to light quarks via this mixing, just like for the case of $W_L$'s discussed above.
Whereas, the {\em orthogonal} combination of composite $W_R^3$ and $(B-L)$, 
usually denoted as $Z^{ \prime }$, 
does not mix with elementary hypercharge, thus being decoupled from light quarks at this order.
Recall that the composite SM singlet neutrino couples to {\em both} the composite $W^3_R$ 
and $(B-L)$ gauge bosons, in such a manner that in the degenerate case, there is then only a coupling of singlet neutrino to $Z^{ \prime }$, but not to composite hypercharge. 
%
%
The upshot here then {\em seems} to be that, including both neutral and charged channels, there is no coupling of singlet neutrino to light quarks for its production at the LHC via spin-1 intermediaries at this stage. 

Turning on the Higgs VEV mixes the various composites amongst themselves, in both charged and neutral sectors. 
{\em Combined} with elementary-composite mixing, this effect of EWSB then {\em does} induce a coupling of light quarks to $Z^{ \prime }$: explicitly, this proceeds via $Z^{ \prime }$ mixing with composite $B$ and $W^3_L$, followed by elementary-composite $B$, $W^3_L$ mixing.
A similar argument applies to $W^{ \pm }_R$, now involving mixing with various $W^{ \pm }_L$'s. Hence, singlet neutrino production at the LHC can then take place in {\em both} neutral and charged channels.
However, naively this coupling still seems to be suppressed, since the Higgs VEV is somewhat smaller than the compositeness scale.
Remarkably, the {\em same} composite {\em degeneracy} comes to the rescue here, since it can result in {\em large} mixing {\em angle} between $Z^{ \prime }$ and composite $B$ and $W^3_L$, even with a smaller mass mixing {\em term}.
Once again, a parallel consideration holds for $W_R^{ \pm }-W_L^{ \pm }$ mixing. Thus, we finally get a {\em non}-negligible coupling of $Z^{ \prime}$ and $W_R^{ \pm }$ to light quarks inside the proton.
Furthermore, the above argument suggests that a significant {\em non}-degeneracy of spin-1 composites might reduce these signals. 

However, when the spin-1 composites are not degenerate, we {\em have to} go back to their  ``original'' identities, chosen as per the symmetries of the strong dynamics in order to be in the physical, mass basis. 
Namely, we have a degenerate triplet of composite $W_L$'s, another one for $W_R$ and finally a neutral composite $(B-L)$, with three different masses in general. Of course, compared to the above one, there is {\em no} change in the basis for composite $W_L$'s and $W^{ \pm }_R$ here, but there is a crucial difference for composite $W_R^3$,  $(B-L)$ as follows.
As before, the external $B$ mixes with {\em both} composite $W^3_R$ and $(B-L)$, with $N^{ (1) }$ coupling to both of them. However, the composite $W^3_R$ and $(B-L)$ now have different masses
so that this combination of composites is {\em not} even close to a mass eigenstate. This situation is to be contrasted with the degenerate case, where this admixture {\em is} a mass eigenstate, i.e., the composite $B$, and in fact the singlet neutrino decouples from it, due to a ``cancellation'' between couplings to the constituent $W^3_R$ and $(B-L)$.
We can then contemplate {\em two} cases, i.e., composite $(B-L)$ is lighter or heavier than composite $W_R^3$, say, $\sim 2$ vs.~$\sim 3$ TeV.

So, as anticipated earlier, we can have non-negligible {\em production} of composite $(B-L)$ (assuming that is lighter) {\em simply} via its mixing with external hypercharge gauge boson which, in turn, 
couples to light quarks inside proton.
In this way, there is {\em no} need to involve EWSB for generating this coupling, cf.~for composite $Z^{ \prime }$ in the degenerate case.
Clearly, production of the heavier composite $W_R^3$ (or composite $(B-L)$ in the other case) via a similar mechanism can then be neglected in comparison: coupling of this heavier state to light quarks is similar 
to that of the lighter one so that suppression in the rate is simply due to the masses.
Furthermore, composite $(B-L)$ (or $W^3_R$ in the other case) will decay into pair of $N^{ (1) }$ with a sizable branching ratio. 

Even though the final state looks similar to production of singlet neutrinos from decay of $Z^{ \prime }$ in usual LR models, the composite $(B-L)$ (or $W^3_R$) production involves different couplings/branching ratios etc.~and thus the two should be distinguishable. 
In fact, in this regard
\begin{itemize}

\item
the composite $(B-L)$ production is extremely interesting, since it does {\em not} decay into Higgs/$W/Z_{ \rm long}$: only channels are SM fermions and their composite partners \footnote{In this case, it is dominated by SM top simply because other SM fermions are mostly elementary and hence couple weakly to composite $(B-L)$.} and the singlet neutrino, composite RH charged lepton (assuming other composite fermions are heavier than one-half the mass of composite $(B-L)$).

\end{itemize}
So, the singlet production could serve not only as a test of the seesaw mechanism, but also as a {\em first} signal for this EW composite boson. Recall that usually this role is played by the dibosons, i.e., Higgs/$W/Z_{ \rm long}$ instead (including in the degenerate case studied earlier).

In addition, suppose we {\em also} make composite $W_L$'s heavier than composite $W_R$'s or vice versa. 
In this case, production of $W_R^{ \pm }$ from light quarks might become negligible, since this coupling requires EWSB mixing, whose effect is damped by the non-degeneracy. 
Note that elementary $W_L^{ \pm }$ mixes {\em only} with its composite counterpart, cf.~neutral case, where elementary $B$ mixes with {\em both} composite $W_R^3$ and $(B-L)$.\footnote{We lose $Z b \bar{b}$ custodial symmetry in this case, since that requires invariance under $L \leftrightarrow R$ exchange, i.e., 
$(t,b)_L$ is bi-doublet of $SU(2)_L \times SU(2)_R$ {\em and} degenerate composite $W_L$ and $W_R$ \cite{Agashe:2006at}. However, we can still ensure that shift in $Zb \bar{b}$ is small as follows. In the case of composite $(B-L)$ being light, say, $\sim 2$ TeV, it suffices to assume that {\em both} $W_{ L, R }^3$ are heavy, say, $\sim 5$ TeV, {\em ir}respective of representations of $b_L$ under $SU(2)_R$, since these are the only states which couple to Higgs VEV and thus can cause $Z b \bar{b}$ shift.
Whereas in the case of $\sim 2$ TeV composite $W_R^3$ (with composite $(B-L)$ at $\gtrsim 3$ TeV),
we might have to revert to canonical representations instead, i.e., $(t,b)_L$ is {\em singlet} of $SU(2)_R$
so that only composite $W^3_L$ couples to {\em both} Higgs and $b_L$ causing shift in $Z b \bar{b}$, but
which can be heavy enough (again, $\sim 5$ TeV) in order to make this shift small.
In other words, light (but still $\gtrsim$ 2 TeV) composite $W_R^3$ or $(B-L)$ can be consistent with
$Zb \bar{b}$ shift, since these particles can {\em de}couple {\em either} from Higgs or $b_L$.}
In other words, there is no analog of above neutral channel effect in the charged case.
So,
\begin{itemize}

\item
another striking feature of this case is singlet neutrino production via neutral gauge boson -- either composite $(B-L)$ or $W_R^3$ -- with{\em out} being accompanied by similar contribution from the {\em charged}, $W_R^{ \pm }$, channel, cf.~usual LR models, where $W_R^{ \pm }$ signal is typically larger than $Z^{ \prime }$ due to former's smaller mass.

\end{itemize}

For simplicity and for clearly illustrating the above two distinctive signatures, we will focus our analysis on the case of composite $(B-L)$ gauge boson being light (say, $\sim 2$ TeV), with {\em both} the $W_R$ and $W_L$ (charged and neutral) being heavier ($\gtrsim 4$ TeV) so that latter's production at the LHC is
negligible. Hence, we will only observe a {\em neutral} spin-1 heavy particle in the EW sector, that too {\em not} decaying into dibosons, but with sizable singlet neutrino production from it.

Independent of above non-degeneracy of spin-1 composites, 
\begin{itemize}

\item we consider the possibility of composite singlet neutrino signal from production and decay of 
composite lepton $SU(2)_L$ {\em doublet}, denoted by $L_L^{ (1) }$.

\end{itemize}
Note that $L_L^{ (1) }$ couples both to composite $W_L$, via a gauge coupling, and to $N^{ (1) }, \tilde{\ell}^{ (1) }$ and the Higgs doublet, via the coupling {\em same} as involved in the Dirac part of the SM neutrino mass seesaw formula.
For example, we could choose a spectrum where $L_L^{ (1) }$ is at $\sim 1$ TeV so that it can be pair-produced at the LHC in decays of {\em on}-shell $\sim 2.5$ TeV composite $W_L$ (either charged or neutral), which couple to light quarks using elementary-composite $W_L$ mixing. 
And, we take $N^{ (1) }, \tilde{\ell}^{ (1) }$ mass to be a bit smaller than $L_L^{ (1) }$, say, 500 GeV 
 so that the dominant decays of $L_L^{ (1) }$ will be 
to $N^{ (1) }, \tilde{\ell}^{ (1) }$ and Higgs/$W/Z_{ \rm long}$. 
It may be worthwhile to emphasize that 
\begin{itemize}

\item
composite lepton doublet is absent in usual LR models, whereas their presence is required in the composite (or 5D) seesaw being studied here.

\end{itemize}
Also, we would like to reiterate that have not at all changed the model compared to the one used in our previous paper. The composite lepton doublet was still present even earlier, but was simply assumed to be heavy.

Finally, we emphasize the following model-independence of the above signal. Suppose the production of singlet neutrino via {\em direct} decays of spin-1 composites, e.g. $W_R^{ 3, \; \pm }$ or $(B-L)$, is suppressed, for example, due to latter being heavy, say, $\gtrsim 3$ TeV. 
An even more extreme case 
is the ``absence'' of the singlet neutrino gauge couplings altogether, either because the RH neutrino is singlet even under extended EW symmetry of the strong dynamics or we do not even have such an extension in the first place.
Even in these cases, we will still have available the above-mentioned avenue of singlet neutrino production via decay of doublet composite lepton, in turn, originating from decays of composite $W_L$. 
The point is that this {\em entire} reaction, production of $W^{ \pm }_L$ followed by decay into composite lepton doublet {\em and} decay of composite lepton doublet into singlet neutrino, proceeds via couplings (gauge and Yukawa, respectively) which are {\em always} present {\em and} sizable. This feature is to be contrasted with the singlet neutrino signal via $W_R^{ 3, \; \pm }$ or $(B-L)$. In the latter case, the size of couplings in production of spin-1 states depends on the amount of degeneracy when EWSB effects are important, whereas their decay into singlet neutrino is dictated by the choice of the representation of the singlet neutrino under the extended gauge symmetry.
Of course if composite $W_L$ is also heavy, then we can resort to {\em SM} $W_L$ exchange for production of composite lepton {\em doublet}, followed by its decay into singlet neutrino as above, although in this case cross-section will be smaller, since it will be {\em non}-resonant.\footnote{An even more model-independent
production of singlet neutrino, involving neither heavy spin-1 {\em nor} other spin-1/2 states, is via its coupling to {\em SM} $W$ and LH electron. This coupling arises from mixing of {\em SM} doublet and singlet neutrinos induced by the Higgs VEV, i.e. using the {\em same} Yukawa coupling which gives SM neutrino Dirac mass term in the seesaw \cite{Das:2015toa}.}

Here is the outline of the rest of this paper.
We begin in Sec.~\ref{sec:5D} with a brief review of the basic seesaw model in the warped extra dimensional framework, emphasizing the region of parameter space which is {\em new} compared to the one studied in our earlier paper.
In Sec.~\ref{sec:twosite}, we present details of the ``simplified'', two-site approach \cite{Contino:2006nn} to studying the 5D model. We will employ this two-site model in our actual analysis of LHC signals. 
We then discuss our main results, starting with production mechanism and decay widths of various heavy particles in Sec.~\ref{sec:Overview}, followed by analyses of SM backgrounds and thus the discovery potential for the new particles in Sec.~\ref{sec:analysis_results}.
We conclude and present some directions for future work in Sec.~\ref{conclude}. Appendices contain the more technical details of the mixing between elementary hypercharge and composite $W^3_R, (B - L)$ gauge bosons.
%


\section{Review of 5D Model}
\label{sec:5D}

We study a model implementing the 
seesaw mechanism for SM neutrino mass in the context of SM fields propagating in a warped
extra dimension.
As mentioned in the introduction, the present paper is part of a series on this topic.
We will continue using the same basic model with left-right symmetry structure in the EW sector as our previous paper 
on LHC signals \cite{Agashe:2016ttz}. So in this section, we just give a brief review, simply referring to \cite{Agashe:2016ttz} for further details and more references.
The main purpose is to discuss new parts of parameter space or features which were {\em not} elaborated upon in the earlier paper.
Also, the reader is just referred to \cite{Huber:2003sf, Agashe:2015izu} for the basic warped seesaw model with{\em out} extended EW gauge symmetry.

The
bulk EW gauge symmetry is $SU(2)_L \times SU(2)_R \times U(1)_X$. 
The 
$SU(2)_L \times SU(2)_R$ subgroup is broken by the Higgs VEV on the IR brane down to $SU(2)_V$ custodial symmetry,
while 
$U(1)_X$ is unbroken here.
On the UV brane, $SU(2)_R \times U(1)_X$ broken down by boundary conditions to $U(1)_Y$, where
$Y = T_{ 3 R } + X$.
%
%
The bulk fermions are taken to be similar to before. In particular, 
the SM $SU(2)_L$ doublet lepton is a singlet of $SU(2)_R$, whereas the SM 
right-handed [$SU(2)_L$ singlet] charged lepton is embedded in a doublet of $SU(2)_R$.
%
%
%
%
%

In this framework, the new states beyond SM are the Kaluza-Klein (KK) excitations of the SM particles, as well as 
singlet neutrino modes, all 
with mass at the $\sim$ TeV scale.
In this and the previous paper, we study the 
LHC signals
from production and decay of these {\em singlet} neutrinos arising from the decay of other heavier KK particles, in particular, belonging to the extended EW sector alluded to above.
In the current paper, we actually perform two signal analyses.
In both of cases, as for the most general possibility,
we assume presence of non-negligible 
brane-localized kinetic terms (BKTs) for bulk {\em gauge} fields, which
change their masses and couplings \cite{Carena:2002dz} .\footnote{Another possibility
for such modifications of properties of gauge KK is to assume that the various gauge fields propagate in
different bulk regions \cite{Agashe:2016rle}.}
%
%
%
In particular, this will result in essentially {\em three} independent KK masses for the $U(1)_X$, $SU(2)_L$ and $SU(2)_R$
gauge bosons.
%
%
In this work, all we need is these masses differing by $ O(1)$ factors for which $ O(1)$ BKT's suffices
(i.e., no larger hierarchy is called for here).
This is to be contrasted with our earlier paper where BKTs for gauge fields were implicitly neglected (or 
assumed same for all) so that these KK fields would instead be approximately degenerate up to boundary conditions on UV brane and EWSB effects.
Note that in the earlier paper, BKT's {\em were} invoked for the {\em singlet} neutrino\footnote{For a discussion of fermion BKT's in general, see reference \cite{Carena:2004zn}} in order to make it lighter than
the EW KK modes such that former can be produced in decays of the latter: we continue to do so in this paper also.

We will 
work out details in two-site model in the next section, but here we would like just to summarize the impact of
the above non-degeneracy. Firstly, the KK $W^{ \pm }_R$-KK $W^{ \pm }_L$
mixing due to Higgs VEV is now suppressed as $\propto v^2 / ( \hbox{KK mass splitting} )^2$, with KK mass splitting of order KK mass itself. 
%
A
more subtle effect happens in the {\em extended} neutral gauge sector which couples to singlet neutrino.
%
%
Namely, without degeneracy in their masses, re-organizing
KK modes of $W^3_R$ and $X$ 
into KK hypercharge and $Z^{ \prime }$ is no longer valid. 
Instead, KK modes of $W^3_R$ and KK $X$ are approximately mass eigenstates separately. Although they mix on UV brane
, where $U(1)_R \times U(1)_X$
is broken down to $U(1)_Y$, this UV-brane localized mixing effect can be treated as a {\em perturbation} given that the KK mode profiles are peaked
near the IR brane. 
%
Note that the light quarks are effectively localized on the UV brane.
In this way, both approximated mass eigenstates, mostly made of KK $X$ and KK $W_R^3$,  couple to light quarks ($\propto$ hypercharge coupling), thus can be produced at the LHC,  and decay into
pair of singlet neutrinos along with other channels.
This scenario is the 
focus of the first part of this paper, where we assume the canonical choice $X =\frac{1}{2} ( B - L ) $.  
Remarkably, 
KK $X$ does {\em not} decay to di-bosons, cf.~typical EW KK gauge bosons.

In the second study of LHC signals in this paper, 
%
%
the main new feature 
compared to before is to assume {\em KK} excitations of SM $SU(2)_L$ {\em doublet} are light due to BKT's.
%
In particular, the mass of KK doublet lepton is taken to be 
$\lesssim \frac{1}{2}$ of KK $W_L$ so that these can be pair-produced in KK $W_L$ decays. 
These $SU(2)_L$ KK doublet leptons can then decay into the SM singlet neutrinos, assuming it is kinematically allowed. 
%
%
Note that these states always exist in 5D models, but are simply assumed to be heavy in previous paper.

Having pointed out new features of the 5D models we consider in this paper, 
we now switch to two-site model for an actual analysis.
%
%

\section{Two-Site Approach to 5D Model}
\label{sec:twosite}

The two-site (or sector) model is an economical, effective description of the above 5D model, motivated by deconstruction and
AdS/CFT correspondence. It is roughly equivalent to keeping only first KK and zero modes of the 5D model.
%
We start with a brief review of the two-site model used in our previous paper \cite{Agashe:2016ttz}. 
%
The composite sector has $SU(2)_L \times SU(2)_R \times U(1)_X$ {\em global} symmetry
so that there are massive spin-1 composites in the adjoint representation of this group.
Only the SM subgroup of this symmetry, $SU(2)_L \times U(1)_Y$, is gauged.
%
%
Namely, in elementary sector, $W_L^{ 3, \; \pm }$ and hypercharge $B$ gauge bosons
mix with the appropriate spin-1 composites.
Before EWSB effects, which we discuss in the next subsection, the SM EW gauge fields are the resulting massless eigenstates, which have a small admixture of the composites.
Similarly, the heavy spin-1 fields are mostly the composites, but also have a bit elementary components.

%
%
Moving on to the fermionic sector, we have massive Dirac composites in suitable representations of the global symmetry, having couplings to the relevant composite gauge bosons.
%
%
Elementary {\em chiral} fermions 
mix with these composites, producing the SM fermions (before EWSB) as the massless eigenstates after diagonalization, plus the heavy fermions, which are mostly composite,
but have a small admixture of the elementary fermions.

Next, we will elaborate further on the sub-sector whose LHC signals will be analyzed, again highlighting the difference from before. 
As mentioned above, we work out two different cases in this regard so that it is convenient to
spell out the Lagrangians separately.

\vspace{0.1in}

\noindent {\bf (a) Composite $(B-L)$ gauge boson }

Here, we assume the canonical LR structure in the lepton sector as in \cite{Agashe:2016ttz}, i.e., identify $X$ with $\frac{1}{2} ( B - L )$,
taking the two sets of composite leptons to be $(2,1)_{-1/2}$ and $(1,2)_{-1/2}$ under $SU(2)_L \times SU(2)_R \times U(1)_X$, respectively.
The modification from \cite{Agashe:2016ttz} is allowing for {\em non}-degeneracy of EW spin-1 composites, $SU(2)_L$, $SU(2)_R$ and $U(1)_X$. Such non-degeneracy is ``inspired'' by the effect of BKT's in the 5D model mentioned above.
%
%
%
%

One 
consequence of this departure from our earlier work takes place {\em after} EWSB which we will briefly consider in the next section.
Here, we focus on an interesting effect which takes place even before EWSB in the spin-1 sector coupled to
the SM singlet neutrino, i.e., $W^3_R$ and $(B-L)$ gauge bosons.
The details of this process are given in the Appendix \ref{non-deg_mix}, from which we extract the bottomlines as follows.

The  
elementary hypercharge mixes with a combination of composite $(B-L)$ and $W^3_R$ as before\footnote{This matches mixing between these KK's on UV brane in 5D model, which is is not an EWSB effect.}, but the crucial point is that these two composites are not degenerate anymore.
The two heavy mass eigenstates resulting from this procedure thus contain mostly composite $(B-L)$ and
$W^3_R$, respectively, with small admixture of elementary hypercharge, followed
by even smaller bit of the {\em other} composite. 
Such small admixture of elementary hypercharge, however, turns out to be large enough to secure significant production of heavy gauge bosons at the LHC.
These are then dubbed simply (with slight abuse of notation) as ``heavy $(B-L)$'' and ``heavy $W^3_R$''.
Note that light quarks essentially couple only to the elementary hypercharge.
%
%
As a result, 
{\em both} the heavy spin-1 fields acquire a coupling to light quarks, roughly at similar level as in the $W_L$ sector, 
i.e., $\sim g^2 / g_{ \star }$, where $g$ and $g_{ \star }$ are the appropriate SM and composite gauge couplings, respectively.
And, 
both heavy $(B-L)$ and $W^3_R$ can decay into
singlet neutrinos.

The same analysis in the degenerate limit shows that one heavy mass eigenstate decouples from light quarks, but not from the singlet neutrino, and vice versa for the other. 
This is in agreement with \cite{Agashe:2016ttz}. Former would be combination of composite $W_R^3$ and $(B-L)$
which does not mix with elementary hypercharge (called $Z^{ \prime }$), whereas latter is simply the ``heavy hypercharge boson''
(corresponding to KK hypercharge of the 5D model).

For simplicity and because signal is then more dramatic, we assume composite $W^3_R$ (and $W^{ \pm }_R$) is heavier than composite $(B-L)$,
thus keeping only the heavy $(B - L)$. 
Similarly, 
composite $W_L$'s are also taken to be heavier than composite $(B - L)$ and therefore neglected.
Finally, composite $SU(2)_L$ doublet is made heavier than one half mass of heavy $(B - L)$, whereas
composite SM singlet neutrino lighter so that it can be pair-produced in 
decays of heavy $(B - L)$.

Hence, {\em before} EWSB, the Lagrangian in the mass eigenstate basis for the fermion-gauge sector relevant for our first analysis is given by 
\bea
{\cal L}_{ \rm gauge-fermion } & = & - \frac{1}{4} \Big[ F^{ (0) \; 2}_{ \mu \nu } + \rho_{ B - L }^{ \mu \nu \; 2 } \Big] + \frac{1}{2}m_{ \star }^2 \rho_{ B - L \; \mu }^2 
\nonumber \\
& & + \overline{ \psi^{ (0) } } i \slashed{D} \psi^{ (0) } + \overline{ \tilde{L}_R^{ (1) } } \left( i \slashed{D} - m_R \right) \tilde{L}^{ (1) }_R 
\nonumber \\
& & - Q_Y \frac{  g^{2}_Y }{ g_{ \star } }  \; \overline{  \psi^{ (0) }_{ \rm light } } \rho^{ \mu }_{ B - L } \gamma_{ \mu } \psi^{ (0) }_{ \rm light }
\nonumber \\
& & + \frac{1}{6} \left( -\frac{ g^{2}_Y }{ g_{ \star } }  \; \cos^2 \phi_{Q^3_L } + g_{ \star } \sin ^2 \phi_{Q^3_L}  \right) \overline{  Q^{ (0) \; 3 }_L } \rho^{ \mu }_{ B - L } \gamma_{ \mu } Q^{ (0) \; 3 }_L + \frac{1}{6} g_{ \star } \overline{ t_R^{ (0) } }
\rho^{ \mu }_{ B - L } \gamma_{ \mu } t_R^{ (0) } \nonumber \\
& & -\frac{1}{2} g_{ \star } \overline{ \tilde{L}_R^{ (1) } } \rho^{ \mu }_{ B - L } \gamma_{ \mu } \tilde{L}_R^{ (1) }.
\label{L_gauge-fermion_1}
\eea
An explanation of the notation is in order here. 
The massless (before EWSB) SM fields, including gauge bosons and fermions, are denoted by superscript ``$(0)$'', since they correspond to zero-modes of the 5D model. 
%
In particular, 
$\psi^{ (0) }_{ \rm light }$ stands for $\left( \psi^{ (0) } - \left\{ Q_L^{ (0) \; 3 }, t^{(0)}_R \right\} \right)$, which are 
assumed to be mostly elementary.
We have assumed $t^{ (0) }_R$ is fully composite, 
whereas $\phi_{ Q^3_L }$ is the elementary-composite mixing angle for $Q^{ (0) \; 3 }_L$.
The only heavy fermion is 
$\tilde{L}_R^{ (1) }$, the vector-like/heavy $SU(2)_R$ doublet fermion which contains the SM singlet neutrino relevant for the SM neutrino mass seesaw (along with a charged lepton-like partner). The superscript ``$(1)$'' again is a reminder that this would be the KK mode of the corresponding 5D field,
while the ``tilde'' notation will be explained a bit later.
Moving onto spin-1 sector, 
$\rho_{ B - L }$ stands for the heavy $(B - L)$ gauge boson (roughly the KK $(B-L)$ of the 5D model), with $g_{ \star }$ being the associated composite gauge coupling. The elementary hypercharge-composite $(B - L)$ mixing angle is $\approx g_Y / g_{ \star }$ (see Appendix \ref{non-deg_mix} for details), where $g_Y$ denotes SM hypercharge coupling. Since $g_Y / g_{ \star }\ll 1$, we simply set cosine of this mixing angle to be 1 for a good approximation.
Finally, 
$D_{ \mu }$ stands for the covariant derivative with respect to the SM gauge group, i.e., $D_{ \mu } = \partial_{ \mu } - 
i g A^{ (0) }_{ \mu }$, where $A^{(0)}_\mu$ is in matrix form with appropriate generators.

\vspace{0.1in}

\noindent {\bf (b) Composite $SU(2)_L$ doublet lepton }

In the second study, we focus on the composite, heavy partner of $SU(2)_L$ {\em doublet} lepton, which is 
coupled to composite $W_L$ gauge boson. We assume mass of former is less than one-half of latter, but larger than 
that of the composite singlet neutrino.
Also, this heavy $SU(2)_L$ {\em doublet} lepton couples to the 
Higgs field and singlet neutrino. This coupling is related to that entering neutrino mass seesaw, i.e., with {\em SM} replacing composite doublet lepton.
In this manner, 
the singlet neutrino production  can occur via the decay of composite doublet lepton using Yukawa coupling, with the latter  produced via
decay of heavy $W_L$ in addition to the exchange of {\em SM} $W_L$.
Note that 
the couplings in {\em both} production (related to SM gauge) 
and decay (related to Yukawa) of composite doublet lepton are {\em in}dependent of representation of SM singlet neutrino under
$SU(2)_R \times U(1)_X$.
So, for {\em simplicity} and in order to re-iterate the above model-independence of this signal, we just 
drop the extended ($SU(2)_R \times U(1)_X$) structure, while  
consider only a $SU(2)_L \times U(1)_Y$ composite sector model with composite singlet neutrino for {\em this study of composite $SU(2)_L$ doublet} .
We also assume composite hypercharge is heavy.
Thus, Lagrangian in fermion-gauge sector for this signal is 
\bea
{\cal L}_{ \rm gauge-fermion } & = & 
- \frac{1}{4} F^{ (0) \; 2 }_{ \mu \nu } + \frac{1}{2} \left( D_{ \mu } \rho_{ W_L \; \nu } - D_{ \nu } \rho_{ W_L \; \mu } \right)^2 + 
%
%
m_{ \star }^2 \rho_{ W_L \; \mu }^2 
\nonumber \\
& & + \overline{ \psi^{ (0) } } i \slashed{D} \psi^{ (0) } + \overline{ L_L^{ (1) } } \left( i \slashed{D} - m_L \right) L^{ (1) }_L
+ \overline{ N^{ (1) } } \left( i \slashed{\partial} - m_N \right) N^{ (1) }
\nonumber \\
& & -\frac{ g^2_W }{ g_{ \star } }  \; \overline{  \psi^{ (0) }_{ \rm light \; L } } \rho^{ \mu }_{ W_L } \gamma_{ \mu } \psi^{ (0) }_{ \rm light \; L }
\nonumber \\
& & + \left(  -\frac{ g^2_W }{ g_{ \star } } \; \cos^2_{ \phi_{Q^3_L }} + g_{ \star } \sin^2_{ \phi_{Q^3_L} } \right) \overline{  Q^{ (0) \; 3 }_L } \rho^{ \mu }_{ W_L } \gamma_{ \mu } Q^{ (0) \; 3 }_L \nonumber \\
& & + g_{ \star } \overline{ L_L^{ (1) } } \rho^{ \mu }_{ W_L } \gamma_{ \mu } L_L^{ (1) }
\label{eq:gauge_fermion_SU2L}
\eea
where 
$\rho_{ W_L }$ denotes the heavy $W_L$ gauge boson with coupling $g_{ \star }$, which corresponds to 
KK $W_L$ of the 5D model.  Note that in the above Lagrangian, $\rho_{W_L}$ is the matrix of gauge bosons with appropriate generators. One should take the trace of the pure gauge part (the first row) of the above Lagrangian to get the final answer. Terms with more than two heavy spin-1 fields are dropped, simply because they are much smaller than the dominate interactions we consider.
$L_L^{ (1) }$ is the vector-like/heavy $SU(2)_L$ doublet  lepton and 
the singlet neutrino is denoted by $N^{ (1) }$ which is also vector-like, but has no charged lepton partner (cf.~composite $(B-L)$ model or previous paper).
Also, the elementary-composite $W_L$ mixing angle is given approximately by $\frac{g_W}{ g_{ \star }}$, where $g_W$ is SM $W$ coupling. Similarly to the case in composite $(B-L)$ model, we simply treat cosine of this angle to be 1, provided this mixing angle is small.
Note that $N^{ (1) }$ in this model does not have {\em any} gauge interactions (before EWSB).
Finally, although it is not shown explicitly, the isospin charges are to be understood in the gauge couplings.

We would like to emphasize that the above channel exists even in the original model, i.e., with SM singlet neutrino being doublet of $SU(2)_R$. Qualitatively the signal will be similar to what we discuss here, although details at the $O(1)$ factors will be different.
In fact, we choose a few TeV mass scale for heavy $W_L$ in order to get enough signal for this model. With such a mass, EW precision tests actually {\em require} the extended EW structure.
In particular, even though $T$ parameter constraint can be satisfied with a suitably heavy hypercharge, suppressing the shift in $Z \to b \bar{b}$ coupling with a custodial
symmetry mandates composite $SU(2)_R$ gauge bosons which are degenerate with $SU(2)_L$. This will 
result in modifications of the signal.
In this sense, our study 
could be taken as a toy or simplified version of the fully realistic case, but is clearly sufficient for the purpose of illustrating the basics of this signal.

\subsection{Higgs sector}
\label{subsec:Higgssector}

The SM 
Higgs doublet is taken to be purely in the composite sector, with {\em no} direct coupling to the elementary
gauge bosons or fermions. 
In the cases that we study here, it has the following effects. The Higgs VEV gives mass to the 
SM EW gauge fields and SM fermions.
The gauge and Yukawa couplings of the Higgs field also 
result in decay channels for heavy spin-1 and fermion 
into physical Higgs, as well as into longitudinal $W/Z$. In the unitary gauge, this actually arises 
via mass mixing (induced by the Higgs VEV) between various fields. The relevant cases here are heavy-heavy 
mixing for fermions and heavy-massless mixing for fermions and spin-1 states.
In principle, the Higgs VEV induces {\em heavy-heavy} spin-1 mixing. Indeed, in \cite{Agashe:2016ttz}, this effect was enhanced by degeneracy of spin-1
composites and played a crucial role in the LHC signals, in particular,
resulting in significant coupling of composite $W_R^{ \pm }$ 
to light quarks.
However, this particular Higgs induced mixing is rendered negligible for the purposes of LHC signals in this paper by our assumption of significant {\em non}-degeneracy among composite spin-1 states. 

For a detailed explanation of 
%
%
the above features, it is better to treat the two LHC signal models separately.

\vspace{0.1in} 

\noindent {\bf (a) Composite $(B - L)$ gauge boson }

The bosonic part of Higgs Lagrangian is 
\bea
{\cal L}_{ \rm gauge-Higgs } & = & | D_{ \mu } H |^2 - V ( H ) 
\eea
We can show (see Appendix \ref{non-deg_mix}) that 
$\rho_{ B - L }$ coupling to Higgs 
is {\em doubly}-suppressed in elementary-composite mixing angles, so that it is neglected here.
Thus, the only effect of Higgs here is to give masses to SM EW gauge fields.

Next, we consider the Yukawa couplings. while this is similar to \cite{Agashe:2016ttz}, for the sake of completeness, we repeat it here.
As already mentioned above, in this case study, we assume that the composite $SU(2)_L$ doublet $ L_L^{ (1) }
$ is too heavy to be relevant for LHC signal, keeping only singlet neutrino (and its charged lepton-like partner).
Thus, the 
relevant part of 
Lagrangian is 
\bea\label{eq:LYukawa-B-L}
{\cal L}_{ \rm Yukawa } & = & y_{ 0 0 } \overline{ L^{ (0) }_L }  \ell_R^{ (0) } H
+ y_{ 01 } \overline{ L^{ (0) }_L } \tilde{L}_R^{(1)} H 
\eea
where $y_{00}(y_{01})$ denotes Yukawa coupling between two massless (one massless and one heavy) modes. Here we drop flavor indices for simplicity, but there should be one copy of the above form for each generation of leptons.
%
Also, 
$L^{ (0) }_L$ and $\tilde{L}_R^{ (1) }$ are given 
in doublet of $SU(2)_L$ and $SU(2)_R$ respectively. Their components are
\bea
L^{ (0) }_L & = & \left( \nu^{ (0) }_L, \ell_L^{ (0) } \right) \nonumber \\
\tilde{L}_R^{ (1) }  & = & \left(  N^{ (1) }, \; \tilde{\ell}^{ (1) } \right) 
\eea
where first line is chiral $SU(2)_L$ doublet lepton, while the
second line is heavy and vector-like.
Note that we are invoking 
``split'' multiplets in the $SU(2)_R$ doublet lepton sector. That is, we introduce separate multiplets for
SM right-handed charged lepton and for the singlet neutrino involved in seesaw, so that $\tilde{\ell}^{(1)}$ in second line above is not the heavy version (or KK excitation) of 
%
%
$\ell^{ (0) }_R$ (see \cite{Agashe:2016ttz} for more details).

Clearly, the above couplings 
lead directly to decay of heavy fermions into SM fermions and the physical Higgs boson.
In addition, after Higgs VEV, we get heavy-massless fermion mass mixing as follows .
We 
begin with the neutrino sector, where 
these mass
terms (along with vector-like masses) can be written as 
\bea
{\cal L}_{ \rm mass } & = & \frac{ y_{ 01 } v }{ \sqrt{2} } \overline{ \nu^{ (0) }_L } 
N_R^{ (1) } + m_N \overline{ N_L^{ (1) } } N_R^{ (1) }. 
\eea

\noindent The resulting 
mass eigenstates then (approximately) are
\bea\label{eq:Nvmixing_B-L}
N_R & \approx & N_R^{ (1) } 
\nonumber \\
N_L & \approx & N_L^{ (1) } + V_{ \ell N } \nu^{ (0) }_L 
\nonumber \\
\nu_L & \approx & \nu^{ (0) }_L - V_{ \ell N } N_L^{ (1) }
\eea
where $N$ is the heavy, mostly SM singlet, mass eigenstate, while $\nu_L$ is the SM/massless one.
The mixing angle is given by $V_{ \ell N } \approx y_{ 01 } v /(\sqrt{2} m_N)$. Here, we are treating the
Higgs VEV effect as a perturbation, i.e., assuming the mass mixing terms above to be smaller than
the vector-like mass.

For the 
charged lepton sector, 
the 
relevant mass terms, including Higgs VEV-induced mixing, but dropping the negligible SM charged lepton Yukawa, i.e. the first term in Eq.~(\ref{eq:LYukawa-B-L}), are given by 
\bea
\frac{ y_{ 01 } v }{ \sqrt{2} } \overline{ \ell_L^{ (0) } } \tilde{\ell}_R^{ (1) } 
+ m_N \overline{ \tilde{\ell}_L^{ (1)} } \tilde{\ell}_R^{ (1) } 
\eea
Thus, the
mass eigenstates are
\bea\label{eq:eemixing_B-L}
\tilde{\ell}_L & \approx & \tilde{\ell}_L^{ (1) } + V_{ \ell N } \ell_L^{ (0) } \nonumber \\
\tilde{\ell}_R & \approx & \tilde{\ell}_R^{ (1) } \nonumber \\
\ell_L & \approx & \ell_L^{ (0) } - V_{ \ell N } \tilde{\ell}^{ (1) }_L 
\eea
Again, $\ell_L$ here is the SM field, whereas $\tilde{\ell}$ is the heavy, mostly $SU(2)_L$ singlet, mass eigenstate.

Recall that the fields with superscripts ``$(0)$'' and ``$(1)$'' on the RHS of the mass eigenstate equations above 
are the mass eigenstates {\em before} EWSB. Equivalently, they are the weak/gauge eigenstates, with gauge couplings given in Eq.~(\ref{L_gauge-fermion_1}).
So, we 
need to re-express the gauge couplings in Eq.~(\ref{L_gauge-fermion_1}) in terms of the {\em final} mass eigenstate via the 
inverse of the above mass eigenstate equations.
%
%
In particular, the EWSB-induced $SU(2)_L$ doublet-singlet mass mixing will then result in the $N$ (similarly for $\tilde{\ell}$)
coupling to (and thus decaying into) 
the SM lepton $SU(2)_L$ doublet and $W/Z$. Of course,
this unitary gauge effect (i.e., via gauge coupling) is equivalent to thinking of longitudinal $W/Z$ as
unphysical Higgs so that this decay proceeds simply via the {\em Yukawa} coupling instead.

We will return to all these couplings in the next section.

\vspace{0.1in}

\noindent {\bf (b) Composite $SU(2)_L$ doublet lepton }

As a reminder, for simplicity, we neglect here the $SU(2)_R \times U(1)_X$ structure so that 
the bosonic part of Higgs Lagrangian is 
\bea
{\cal L}_{ \rm gauge-Higgs } & = & | D_{ \mu } H + i g_{\star } \rho_{ W_L } H |^2 - V( H ).
\eea
We thus get decay of $\rho_{ W_L }$ into physical Higgs boson and $W/Z$ simply from the cross-term in 
the above covariant derivative.
In addition, the Higgs VEV leads to 
mass mixing between $W_L^{ (0) }$ and $\rho_{ W_L }$, resulting in
decay of the heavier mass eigenstate into {\em pair} of $W/Z$ 
via the SM tri-linear 
%
%
coupling of the $W_L^{ (0) }$ component
of the heavy state/combination\footnote{Equivalently, as done above for fermions, we can obtain the same effect by
thinking of longitudinal $W/Z$ as unphysical
Higgs.}: this is the same phenomenon as in \cite{Agashe:2016ttz}  and earlier studies,
so will just refer to it for details.
However, the {\em mass} of the final heavy spin-1 state (and its couplings to {\em fermions}) are negligibly shifted compared to those of 
$\rho_{ W_L }$. Thus  for simplicity of notation, we will continue to denote this particle by ``$\rho_{ W_L }$'' .


Moving onto Yukawa couplings, 
the 
heavy $SU(2)_L$ doublet lepton in weak/gauge basis is denoted in component form as
\bea
L^{ (1) }_L & = & \left( \nu^{ (1) }, \ell^{ (1) } \right) 
\eea
which is vector-like.
Once again, the heavy/vector-like SM singlet neutrino is labelled as $N^{ (1) }$.
The relevant Lagrangian becomes
\bea\label{eq:LYukawa_SU2L}
{\cal L}_{ \rm Yukawa } & = & 
y_{ 00 } \overline{ L_L^{ (0) } } \ell_R^{ (0) } H + y_{ 01 } \overline{ L_L^{ (0) } } N_R^{ (1) } H + y_{ 1 0 } \overline{ L_L^{ (1) } } \ell_R^{ (0) } 
H 
+ y_{ 11 } \overline{ L_L^{ (1) } } N_R^{ (1) }  H
\eea
where $y_{10}(y_{11})$ denotes Yukawa coupling between one massless and one heavy (two heavy) modes. $y_{11}$ is naturally greater than $y_{10}$ or $y_{01}$ due to the later being suppressed by one power of elementary-composite mixing. 
Including mass mixing terms (along with the dominant vector-like masses) in the neutrino sector, one can obtain:
\bea
\frac{ y_{ 01 } v }{ \sqrt{2} } \overline{ \nu_L^{ (0) } } N_R^{ (1) } + \frac{ y_{ 11 } v }{ \sqrt{2} } \overline{ \nu_L^{ (1) } } N_R^{ (1) } 
+ m_N \overline{ N_L^{ (1) } } N_R^{ (1) } + m_L \overline{ \nu_L^{ (1) } } \nu^{ (1) }_R 
\label{L_Yukawa_2}
\eea
Assuming $y_{11} \gg y_{01}$ and $m_L > m_N > v$, the 
resulting mass eigenstates can be approximated as
\bea
\nu_L & \approx & \nu^{ (0) }_L  - V_{ \ell N } N_L^{ (1) } 
 \nonumber \\
 N_L & \approx & N_L^{ (1) }  - V_{ LN } \frac{ m_N }{ m_L }  \nu^{ (1) }_L +V_{ \ell N } \nu^{ (0) }_L
\nonumber \\
N_R & \approx & N_R^{ (1) } -\nu_R^{ (1) } V_{ LN } \nonumber \\
\nu^h_L & \approx & \nu^{ (1) }_L + V_{ LN } \frac{ m_N }{ m_L }  N^{ (1) }_L \nonumber \\
\nu^h_R &  \approx & \nu^{ (1) }_R + V_{ LN } N_R^{ (1) }
\label{eq:EWSB_SU2L}
\eea
%
%
%
where $V_{ \ell N }$ is the same as before and $V_{ LN } \approx y_{ 11 } v m_L/\left(\sqrt{2} (m^2_L-m_N^2)\right)$. $\nu^h$ denotes heavy, approximately $SU(2)_L$ doublet, mass eigenstate (rest of the notation is as before).

Similarly, in the charged lepton sector, we have the relevant mass terms 
\bea
\frac{ y_{ 10 } v }{ \sqrt{2} } \overline{ \ell_L^{ (1) } } \ell_R^{ (0) } + m_L \overline{ \ell_L^{ (1) } } \ell_R^{ (1) } 
\eea
giving
\bea
\ell^h_R & \approx & \ell_R^{ (1) } + V_{ \ell L } \ell_R^{ (0) } 
\nonumber \\
\ell_R & \approx & \ell_R^{ (0) } - V_{ \ell L } \ell_R^{ (1) } 
\nonumber \\
\ell_L & \approx & \ell_L^{ (0) } \nonumber \\
\ell^h_L & \approx & \ell_L^{ (1) } 
\eea
where $\ell^h$ denotes the heavy, approximately $SU(2)_L$ doublet, charged lepton mass eigenstate and
$V_{ \ell L } \approx  y_{ 10 } v /(\sqrt{2}m_L)$. 

Similarly to the above case, the decay of the heavy $SU(2)_L$ doublet states (i.e., $\nu^h$ and $\ell^h$) into 
 singlet neutrino state ($N$) and physical Higgs boson follows directly from
Eq.~(\ref{L_Yukawa_2}), whereas decay into longitudinal $W/Z$ (plus singlet neutrino) is the result of  
Higgs VEV induced mass mixing in the unitary gauge.


\section{Overview of LHC signals}
\label{sec:Overview}

In this section, we describe our signal channels for each of the two models presented above. They both involve production of the singlet neutrino via decays of heavier states. 
We first explicitly show the interactions which are directly relevant for our signals. Then, we introduce parameter choices for each process which are safe from current bounds. Analytic expressions for the decay widths for the new, heavy 
%
%
particles are listed. Since we work on two different models (depending on the immediate parent particle for
the singlet neutrino), we will discuss each model in a separate subsection. Composite $(B-L)$ model is 
%
%
analyzed 
in Sec.~\ref{subsec:compositeB-L}, while Sec.~\ref{subsec:compositeSU2L} is devoted to the composite $SU(2)_L$  doublet lepton model.

\subsection{Composite $(B-L)$}
\label{subsec:compositeB-L}
In this section, we consider the scenario where the composite sector only has 
%
%
$(B-L)$ gauge field, i.e., we assume that all $SU(2)_L \times SU(2)_R $ gauge bosons are heavy and hence irrelevant for our signal study. 
%
%
For producing this new particle at the LHC, we need a significant coupling to light quarks inside the proton. 
However, as discussed in detail in Sec.~\ref{sec:twosite},
the light quarks being elementary do not directly couple to the composite gauge bosons including $(B - L)$.
Nonetheless, the composite $(B - L)$ gauge boson mixes with the elementary hypercharge 
gauge boson. The heavy mass eigenstate (denoted by $\rho_{B-L}$) can then couple to light quarks via this small admixture of elementary hypercharge gauge boson.
%
%
In short, our signal channel is then the produced $\rho_{B-L}$ decaying to a pair of $N$'s. 
Subsequently, each $N$ decaying to $\ell$ and $W$ or $\nu$ and 
$Z/H$. 
%
%


\subsubsection{Relevant Couplings}
\label{subsec:relevant_couplings_X}

Here, we summarize the relevant couplings for our signal. For notational details,
the reader is referred to Sec.~\ref{sec:twosite}. There are three types of couplings: (1) couplings between $\rho_{B-L}$ and SM fermions (2) couplings of $\rho_{B-L}$ to a  pair of $N$ or a pair of $\tilde \ell$, and (3) couplings among $N$($\tilde \ell$), $H/W/Z$ and SM $\ell$($\nu$) via Yukawa coupling.  

(1) The first type of coupling can be obtained from Eq.~(\ref{L_gauge-fermion_1}) :
\bea
\delta \mathcal L_{(1)}&=& -Q_Y \frac{g_Y^2}{g_{\star } } \rho_{B-L}^{\mu}\bar\psi_{\rm light} \gamma_\mu \psi_{\rm light} + \frac{ 1}{6} g_{\star } \sin^2 \phi_{Q^3_L} \rho_{B-L}^{\mu}\bar Q^3_L \gamma_\mu Q^3_L + \frac{1}{6} g_{\star}  \rho_{B-L}^{\mu}\bar t_R \gamma_\mu t_R \nonumber \\
~
\label{eq:Xproduction}
\eea
where $g_Y$ is SM hypercharge coupling and $g_{\star}$ is composite $(B-L)$ coupling. $Q_Y$ is SM hypercharge of corresponding light fermion $\psi_{\rm light}$ and factor of $\frac{1}{6}$ in the last two terms arises from the $(B-L)$ charge of $Q^3_L$ and $t_R$.  
%
%
In the 2nd coupling, we have assumed that $ g_\star \sin^2 \phi_{Q^3_L}$,
is large enough
that the component arising from spin-1 mixing, i.e. $\propto \frac{g^2_Y}{g_{\star }} \cos^2 \phi_{Q^3_L}$ 
, can be neglected here.
All fermions in all Lagrangians shown in this section are mass eigenstates after EWSB. Since EWSB has negligible effects for all $\psi^{(0)}$s, so we simply drop superscript ``(0)'' to show their mass eigenstates. 

%
%
%
%
These couplings are responsible for the production of $\rho_{B-L}$ via light quarks inside proton and also constitute decays channels
for the $\rho_{B-L}$.

(2) The second type of coupling can be understood from Eq.~(\ref{L_gauge-fermion_1}) : 
\bea\label{eq:Xdecay}
\delta \mathcal L_{(2)}= -\frac{1}{2} g_{\star } \rho_{B-L}^{\mu}\bar N \gamma_\mu N -\frac{1}{2} g_{\star } \rho_{B-L}^{\mu}\bar {\tilde \ell} \gamma_\mu \tilde \ell
\eea
where $-\frac{1}{2}$ is the $(B-L)$ charge of $N$ and $\tilde \ell$. These couplings lead to the dominant decays of $\rho_{B-L}$ to pair of $N$ and pair of $\tilde \ell$.

(3) The third type of couplings are similarly obtained from Eq.~(\ref{L_gauge-fermion_1}) and (\ref{eq:LYukawa-B-L}) and mixing induced by EWSB Eq.~(\ref{eq:Nvmixing_B-L}) and (\ref{eq:eemixing_B-L}):
\bea \label{eq:Ndecay}
\delta \mathcal L_{(3)}&=& \frac{g_{ W}}{\sqrt{2}} V_{\ell N} W^+_{\mu}\bar N_L \gamma^\mu \ell_L+\{ N \leftrightarrow \nu ; \ell \leftrightarrow \tilde\ell\}  \nonumber \\
&+ &\frac{g_{ Z}}{2} V_{\ell N} Z_{\mu}\bar N_L\gamma^\mu \nu_L + \frac{y_{01}}{\sqrt{2}} H \bar N_R \nu_L +\{ N \leftrightarrow \tilde \ell ; \nu \leftrightarrow \ell\} + \textrm {h.c.}
\eea
where $g_{W/Z}$ is the SM $W/Z$ gauge coupling and $y_{01}$ is the Yukawa coupling defined in Eq.~(\ref{eq:LYukawa-B-L}). These couplings lead to the decays of $N$ and $\tilde{\ell}$ to $H/W/Z$ and $\ell/\nu$: note that this part is the same as in our previous paper on LHC signals. 
%
%

\subsubsection{Parameter Choice}
\label{subsec:Parameter_Choice_B-L}

For composite $(B-L)$ model, the only relevant composite gauge boson is $\rho_{B-L}$, thus the only relevant parameters in spin-1 composite sector being $g_{\star }$ and $m_\star$. 
%
%
In general, there is a lower bound on these $g_{\star }$'s from the requirement that the Landau poles for the SM gauge couplings 
are below the GUT/Planck scale, which
turns out to be $\approx 3$ \cite{Agashe:2005vg} for hypercharge in 
{\em in $SU(5)$ normalization} (this was studied in the context of gauge coupling unification, hence this choice of normalization).
Converting this to {\em SM} normalization for hypercharge used here, we get 
$g_{\star Y} \geq 3\sqrt{\frac{3}{5}}$.
%
%
In our model,  the composite sector has $SU(2)_{\rm L} \times SU(2)_{\rm R}\times U(1)_{B-L}$ symmetry, with composite $U(1)_{Y}$ obtained from a linear combination of $U(1)_R$(the $U(1)$ part of $SU(2)_R$) and $U(1)_{B-L}$. This structure leads to the relation among couplings as $g_{\star Y}=\frac{g_{\star R }g_{\star }}{g_{\star R}^2+g_{\star }^2}$, where $g_{\star R}$ is the gauge coupling for composite $SU(2)_R$. Since in our model $SU(2)_{\rm L} \times SU(2)_{\rm R}$ gauge fields are irrelevant for signal study, we can choose $g_{\star R} \gg g_{\star }$. In this case, the lower bound on $g_{\star Y}$ becomes the lower bound on $g_{\star }$.  Therefore, our model requires $g_{\star }\geq 3\sqrt{\frac{3}{5}}$. For this collider analysis, we choose $g_{\star}=2.5$ as a benchmark point.

The masses for spin-1 composites are in general
constrained by EW precision tests.  
%
%
Since we choose the relevant states, i.e. $SU(2)_{\rm L} \times SU(2)_{\rm R}$ composite gauge bosons, to be heavy $\geq 4 $TeV,  the EW precision tests 
%
%
are not major constraints for the mass $m_\star$ of $\rho_{B-L}$ in our model. 
Moreover, the amount of compositeness of $Q^3_L$, i.e. $\sin \phi_{Q^3_{L}}$, is also not much constrained by EW precision tests for the same reason. We choose $\sin^2 \phi_{Q^3_{L}}=0.21$ in this study just for consistency with our previous paper on LHC signals \cite{Agashe:2016ttz}.

In addition, we have to consider bounds from direct searches at the LHC for these spin-1 states in the various decay channels with two SM particles.
%
%
However, for $\rho_{B-L}$, di-lepton and di-jet branching ratios are negligible due to the associated couplings being suppressed by  $\frac{g^2_Y}{g^2_{\star }}$ compared to the couplings to $N$ and $\tilde \ell$. The coupling for tops is not suppressed and there is a color multiplicity factor, but $(B-L)$ charge for top is $\frac{1}{6}$, which is factor of 3 smaller than $-\frac{1}{2}$ for $N$ and $\tilde \ell$ and
the heavy leptons are vector-like and also come in three generations . Total branching ratio to di-top is thus much smaller than dominant channel $N$ (as discussed in next section): we have checked that the bound from di-top searches is then very weak. The di-boson search 
provides usually one of the stringent bounds for warped/composite gauge bosons in the EW sector. However, in our model, the light composite gauge boson is $\rho_{B-L}$, which does not (directly) couple to Higgs. Therefore, we are safe 
%
%
from di-boson constraints also. In conclusion, there is no direct bound on $m_\star$ in our model, we simply choose $m_\star=2$TeV for rest of study. 

Next, we discuss the parameter in the lepton sector. The elementary-composite mixing, i.e., $\left| V_{\ell N} \right|^2$, is constrained by various experiments and the results are summarized in \cite{delAguila:2008pw} . We choose  $\left| V_{\ell N} \right|^2=0.001$ (as also done previously by us in  \cite{Agashe:2016ttz}) for all three generations to be consistent with these experimental bounds.
Finally, the constraint on the mass of $N$ and $\tilde \ell$ correlated with the choice of of $\left| V_{\ell N} \right|^2$. With the choice we make $\left| V_{\ell N} \right|^2=0.001$, however, 
%
%
the lower bound on $m_{N}$ is $O(100)$ GeV \cite{Das:2015toa} . 
%
%
Nevertheless, we require $\rho_{B-L}$ to decay into a pair of $N$ or $\tilde \ell$, which means $m_{N}< \frac{1}{2} m_\star$. Besides, since the $(B-L)$ gauge boson and $N$ all come from the same 
(i.e.composite)
sector, there should not be a big hierarchy between $m_{N}$ and $m_\star$. Taking all these considerations into account,  we choose $m_{N} = 750$ GeV in this study.

 
\subsubsection{Decay widths}
\label{subsec:X_production_decay}

In this section we show analytic expressions for decay widths for relevant 
%
%
particles. All the decay widths presented here
%
%
assume $m_\star > 2m_{N}\gg \textrm{mass of SM particles}$, thus masses of SM particles being reasonably neglected.  

\noindent{\textbf{Composite $\rho_{B-L}$}}

%
%
%
Decay width for each $\rho_{B-L}$ decay channel is shown below, which is computed using couplings in Eq.~(\ref{eq:Xproduction}) and Eq.~(\ref{eq:Xdecay}):
%
%
\bea \label{eq:Xwdith}
&&\Gamma(\rho_{B-L} \to N_i \bar N_i / \tilde \ell_i \bar{ \tilde \ell}_i ) = g^2_{\star }\left(1+2\frac{m^2_{N}}{m^2_{\star  }}\right) \sqrt{1-4\frac{m^2_{N}}{m^2_{\star }}} \frac{m_{\star  }}{48\pi }\nonumber \\
&&\Gamma(\rho_{B-L} \to t_R \bar t_R)= g^2_{\star }  \frac{m_{\star }}{288 \pi  }\nonumber\\
&&\Gamma(\rho_{B-L} \to t_L \bar t_L/b_L \bar b_L)= g^2_{\star } \sin^4 \phi_{Q^3_L}  \frac{m_{\star }}{288 \pi }\\
&&\Gamma(\rho_{B-L} \to \psi \bar \psi)= N_c Q^2_Y \frac{g_Y^4}{g^2_{\star }} \frac{m_{\star }}{24 \pi }  \nonumber
\eea
where $\psi$  denotes {\em light} (other than top/bottom) SM fermions, and $N_c$ shows the degree of freedom of corresponding fermion $\psi$: 3 for quarks and 1 for leptons. Also, 
%
the 
subscript $i$ for $N$ and $\tilde \ell$ is generation index ($i=e\, , \mu\, ,\tau$): note that these particles are vector-like. 
%

%
%

\noindent\textbf{Composite $SU(2)_R$ doublet $(N , \tilde \ell)$}

%
%
Their decays proceeded via the couplings in Eq.~(\ref{eq:Ndecay}), giving 
\bea\label{eq:Nwidth}
&&\Gamma(N \to W \ell) = y_{01}^2\frac{m_{N}}{32\pi} \nonumber\\
&&\Gamma(N \to H/Z \nu) =y_{01}^2 \frac{m_{N}}{64\pi} \nonumber \\
&&\Gamma(\tilde \ell \to W \nu)= y_{01}^2\frac{m_{N}}{32\pi} \\
&&\Gamma(\tilde \ell \to H/Z \ell)= y_{01}^2 \frac{m_{N}}{64\pi}. \nonumber
\eea
In principle, there will be three body decays via virtual $\rho_{W_L}$ or $\rho_{W_R}$. However, these 
%
%
are suppressed compared to 2 body decays given above due to 
%
%
$\rho_{W_L}$ and $\rho_{W_R}$ being heavy.

\subsection{Composite $SU(2)_L$ doublet}
\label{subsec:compositeSU2L}
In this section, we consider the case where the composite sector has  $SU(2)_L\times U(1)_Y$ global symmetry: we assume that
the composite hypercharge gauge boson is heavy, thus we keep only the composite 
$SU(2)_L$ gauge boson. 
Also, in the composite fermion sector, we add to the first model the composite partner of SM lepton doublet, denoted as $(\ell^h,\nu^h)$, with the singlet neutrino $N$ now not having a charged partner (unlike in the first model): for details, see
Sec.~\ref{sec:twosite}.
Our signal channel involves heavy spin-1 state (denoted by $\rho_{W_L}$) being produced via light quarks through mixing with elementary $W$ boson.
This is followed by $\rho_{W_L}$ decaying to $(\ell^h,\nu^h)$ pairs 
. Each $\ell^h$($\nu^h$) could further decay to $N$ and $W$($H/Z$) through Yukawa coupling, with each $N$ decaying to $\ell$ and $W$ or $\nu$ and 
$Z/H$. 
%
%

\subsubsection{Relevant Couplings}
\label{subsec:relevant_couplings_WL}

There are four types of couplings that we need to consider: (1) couplings between composite $\rho_{W_L}$ and SM fermions; (2) couplings between composite $\rho_{W_L}$ and composite $SU(2)_L$ lepton doublet $(\ell^h,\nu^h)$; 
(3) Yukawa couplings of $(\ell^h,\nu^h)$ to singlet $N$ and Higgs, and (4) Yukawa couplings of $N$ to SM $W\ell$ or $H \nu$ .

(1) The first type of coupling can be obtained by using Eq.~(\ref{eq:gauge_fermion_SU2L}) :
\bea
\delta \mathcal L_{(1)}=-\frac{g^2_W}{ \sqrt{2}g_{\star }} \rho^+_{W_L\mu}\bar\psi_L \gamma^\mu \psi'_L + \frac{g_{\star }}{ \sqrt{2}} \sin^2 \phi_{Q^3_L}\rho^+_{W_L\mu}\bar t_L \gamma^\mu b_L +\rm{ h.c} 
\label{eq:WLproduction}
\eea
where $g_{\star }$ is composite $\rho_{W_L}$ coupling. These couplings are responsible for the production of $\rho_{W_L}$ via light quarks inside proton and some of the $\rho_{W_L}$ decay channels. 
%
%

(2) The second type of coupling can be understood from Eq.~(\ref{eq:gauge_fermion_SU2L}) : 
\bea\label{eq:WLdecay}
\delta \mathcal L_{(2)}= \frac{ g_{\star}}{\sqrt{2}}  \rho^+_{W_L\mu}\bar \ell^h \gamma^\mu \nu^h +\rm {h.c}
\eea
This coupling leads to the decays of composite $\rho_{W_L}$ to $\ell^h$ and $\nu^h$.

(3) The third  type of coupling are obtained from Yukawa coupling in Eq.~(\ref{eq:gauge_fermion_SU2L}) and EWSB effect in Eq.~(\ref{eq:EWSB_SU2L}): 
\bea\label{eq:CompositeLdecay}
\delta \mathcal L_{(3)}&=& -\frac{g_{ W}}{\sqrt{2}} V_{L N}\frac{m_N}{m_L} W^+_{\mu}\bar N_L \gamma^\mu \ell^{h}_L- \frac{g_{ Z}}{2} V_{L N} \frac{m_N}{m_L}  Z_{\mu}\bar N_L\gamma^\mu \nu^{h}_L \nonumber \\
&-&\frac{g_{ W}}{\sqrt{2}} V_{L N} W^+_{\mu}\bar N_R \gamma^\mu \ell^{h}_R- \frac{g_{ Z}}{2} V_{L N}  Z_{\mu}\bar N_R\gamma^\mu \nu^{h}_R+ \frac{y_{11}}{\sqrt{2}} H \bar N_R \nu^{h}_L + \textrm {h.c.}
\eea
where $y_{11}$ is the Yukawa coupling defined in Eq.~(\ref{eq:LYukawa_SU2L}). These couplings lead to the decays of $\ell^{h}$ to $W$ and $N$ and $\nu^{h}$ to $H/Z$ and $N$.

(4) The fourth  type of coupling are obtained from Yukawa coupling in Eq.~(\ref{eq:gauge_fermion_SU2L}) and EWSB in Eq.~(\ref{eq:EWSB_SU2L}), which is essentially 
the same as Eq.~ (\ref{eq:Ndecay}): 
%
%
\bea\label{eq:Ndecay_WL}
\delta \mathcal L_{(4)}= \frac{g_{ W}}{\sqrt{2}} V_{\ell N} W^+_{\mu}\bar N_L \gamma^\mu \ell_L+  \frac{g_{ Z}}{2} V_{\ell N} Z_{\mu}\bar N_L\gamma^\mu \nu_L + \frac{y_{01}}{\sqrt{2}} H \bar N_R \nu_L + \textrm {h.c.}
\eea
These couplings lead to the decays of $N$ to $W$ and $\ell$ or $H$ and $\nu$. 

\subsubsection{Parameter Choice}
\label{subsec:Parameter_Choice_WL}

As mentioned in Sec.~\ref{sec:twosite}, in this model, we choose the global symmetry of composite sector to be $SU(2)_L \times  U(1)_Y$ for the purpose of this collider study only. 
However, the realistic model should have full $SU(2)_L \times SU(2)_R \times U(1)_X$ global symmetry in composite sector, in order to satisfy all EW precision bounds with a few TeV compositeness scale.
In other words, the model being studied here is to be viewed as 
its simplified version. 
%
%
This full model is almost the same as the model in our previous paper on LHC signals, thus a similar estimation of the bounds and parameter choices also apply here. Therefore, we choose the same value of $g_{\star }=3$ and $\sin^2{\phi_{Q^3_L}}=0.21$. In the present study, $m_\star$ is chosen to be $2.5$ TeV, sightly bigger value than the choice previously ($m_\star=2$TeV). As mentioned in \cite{Agashe:2016ttz} already, for these parameter choices, even with $m_\star$ being 2.5 TeV, some additional model building might still be needed to be fully safe from EW precision bounds.
%
%

As mentioned earlier, di-boson searches at the LHC usually set a stringent bound on composite Higgs models. However, in our model, the branching ratio of $\rho_{W_L}$ to di-boson is suppressed compared to that of standard  KK $W$. This is because the dominant decay channels of $\rho_{W_L}$ are $\rho_{W_L} \to \ell^h \nu^h$. As will be shown clearly in Sec.~\ref{subsec:KKL_production_decay}, the branching ratio to di-boson is $\sim \frac{1}{13}$. With our choice of $g_{\star}=3$,
%
%
we then checked that the current bound is $m_{\star}>1.5$TeV \cite{ATLAS:2016cwq} . Obviously, our choice of 
%
%
$m_{\star}$ is then safe from this bound.

Moreover, there are two lepton mixings in this model. $V_{LN}$ is the mixing between $\nu^h$ and $N$ and $V_{\ell N}$ is  the mixing between $N$ and SM $\nu$. $V_{\ell N}$ is the same mixing angle in our previous study of LHC signals (and in composite $(B-L)$ model
here). 
%
%
So, we choose $\left| V_{\ell N} \right|^2=0.001$ for all three generations in order to be safe with the relevant bounds and also for 
consistency with our other studies.
%
%
However, there is no direct bound on $V_{LN}$, i.e. the composite-composite mixing angle, which is the new feature in this model. 
%
%
Also, this mixing 
will be larger than the mixing between one composite and one elementary($V_{\ell N}$): this point was also mentioned in Sec.~\ref{sec:twosite}.
%
%
In this study, we choose $\left|V_{LN}\right|^2=0.01$.

Our signal 
%
%
assumes the following cascade decay: composite $\rho_{W_L}$ goes into $\ell^h$ and $\nu^h$ pair, which further decay to $N$ and $W/H/Z$. This requires the spectrum to satisfy $\frac{1}{2}m_{\star} > m_{L}> m_{N}$. Similarly to our previous cases (in \cite{Agashe:2016ttz} or composite $(B-L)$ model here), there is very weak bound (O(100)GeV)on $m_{L}$ and $m_{N}$ provided we make the choices of $V_{LN}$ and $V_{\ell N}$ discussed above. 
At the same time, too much gap between these states is unnatural. So, we pick $m_{ L} = 1000$ GeV  and $m_{N} = 500$ GeV.

\subsubsection{Decay widths}
\label{subsec:KKL_production_decay}

All the decay widths presented below 
%
%
are 
with the assumption $\frac{1}{2}m_{\star } > m_{L}> m_{N}\gg \textrm{mass of SM particles}$, thus masses of SM particles are reasonably neglected.  

%
%

\noindent\textbf{Composite $\rho_{W_L}$}

As mentioned above, $SU(2)_L$ doublet composite leptons are produced via decays of composite $\rho_{W_L}$ using the coupling in Eq.~(\ref{eq:WLdecay}). In addition, $\rho_{W_L}$ can decay into pair of SM fermions via the couplings in Eq.~(\ref{eq:WLproduction}).
Decay widths for composite $\rho_{W_L}$ are then given as
\bea \label{eq:WLwdith}
&&\Gamma(\rho_{W_L} \to \ell^{h}_i \nu^{h}_i ) = g^2_{\star }\left(1+2\frac{m^2_{L}}{m^2_{\star }}\right) \sqrt{1-4\frac{m^2_{L}}{m^2_{\star }} }\frac{m_{\star }}{24\pi } \nonumber \\
&&\Gamma(\rho_{W_L} \to WZ/WH)= g^2_{\star }  \frac{m_{\star }}{192\pi }\nonumber\\
&&\Gamma(\rho_{W_L} \to t_L b_L)= g^2_{\star }  \sin^4 \phi_{Q^3_L} \frac{m_{\star }}{16\pi }\\
&&\Gamma(\rho_{W_L} \to \psi \psi')= N_c  \frac{g_W^4}{g^2_{\star }} \frac{m_{\star }}{48 \pi }  \nonumber
\eea
%

\noindent\textbf{Composite $SU(2)_L$ doublet $(\ell ^h , \nu^h)$}

Similarly, using Eq.~\ref{eq:CompositeLdecay},
decay widths for composite $(\ell^{h},\nu^h)$ are given as

\bea \label{eq:L^hwdith}
&&\Gamma(\ell^h \to W N)= y_{11}^2\left( 1-\left(\frac{m_{N}}{m_{L}}\right)^4\right) \frac{m_{ L }}{32\pi}\nonumber \\
&&\Gamma(\nu^h \to H/Z N) =y_{11}^2\left( 1-\left(\frac{m_{N}}{m_{L}}\right)^4\right) \frac{m_{ L }}{64\pi} 
\eea%

Decay widths for $N$ are the same as Eq.~(\ref{eq:Nwidth}).


\section{Discovery Potential}
\label{sec:analysis_results}

In this section, we present our results for the phenomenological studies of the LHC signals for the model described in Sec.~\ref{sec:twosite}. In Sec.~\ref{subsec:X_channel}, we study the pair production of the singlet neutrino ($N$) via the on-shell decay of composite gauge boson $\rho_{B-L}$.
%
%
%
Once produced, each $N$ can decay into $\ell+W$ or $\nu+H/Z$, followed by SM $W$ decaying into $\ell\nu$ or $jj$.
Thus, the cascade decays of pair of $N$'s can result in several possible final states.
As our benchmark, in this paper, we consider the following decay channel.\footnote{To avoid notational clutter, we drop particle's charge.}
%
%
\bea
p p \rightarrow \rho_{B-L} \rightarrow N N, \;\;\; N \rightarrow \ell \;  W , \;\; (W \rightarrow j j), \;\; (W \rightarrow \ell \nu).
\eea
Namely, both $N$ decay first into a lepton and $W$ boson. One of such produced $W$ decays leptonically and the other decays hadronically. As a result, the final state consists of three leptons, two from direct decay of $N$ and one from leptonic decay of $W$, two jets from hadronically decaying $W$ and a neutrino in the form of missing energy or MET (thus labelled $3 \ell 2j +$MET). Since the production of $N$ is achieved by the $s$-channel $\rho_{B-L}$ composite gauge boson, we call this channel the ``$\rho_{B-L}$-channel''.

\noindent Sec.~\ref{subsec:WL-channel} is devoted to describing the study for the pair production of $N$ via decays of composite $SU(2)_L$ lepton doublet, $(\nu^h , \ell^h)$, which is produced by the on-shell decay of charged composite $SU(2)_L$ gauge boson $\rho_{W_L}$. 
Similarly to the above study, several final states are possible, again, depending on how $N$ and $W$ decay. In the current study, we consider the following cascade decay channel. 

\bea
p p \rightarrow \rho_{W_L} \rightarrow \; \ell^h \; \nu^h, \;\;\; \ell^h \rightarrow N \; W, \;\;\; \nu^h \rightarrow N \; H/Z, \;\;\; N \rightarrow \ell \; W, \\ 
H/Z \rightarrow b \; \tilde{b}, \;\;\; \text{Two} \;\; W \rightarrow j \; j, \;\;\; \text{One} \;\; W \rightarrow \ell \; \nu. \nonumber
\eea
In detail, cascade decay of composite gauge boson $\rho_{W_L}$ produces composite lepton doublet, $(\nu^h, \; \ell^h)$, which in turn decay into two singlet neutrino $N$, one $H/Z$ and one $W$. Subsequent decay of two $N$, then, produce two more $W$'s and two SM leptons. $H/Z$ decays into $b \; \bar{b}$ and two of the three $W$'s decay hadronically, rendering four jets, and one $W$ decays leptonically, producing one lepton and missing energy ($4j 2b 3 \ell +$MET). Notice that we combine contributions from processes with $H$ and $Z$ intermediate states. This is because resolutions of LHC detectors may not be able to distinguish the two cases, and at the same time, we achieve a slight increase in the signal rate. Since the singlet neutrino $N$ is produced by the decay of $SU(2)_L$ composite doublet lepton, we call this channel the ``composite $SU(2)_L$ doublet-channel''.

\begin{figure}
\center
\includegraphics[width=0.4\linewidth]{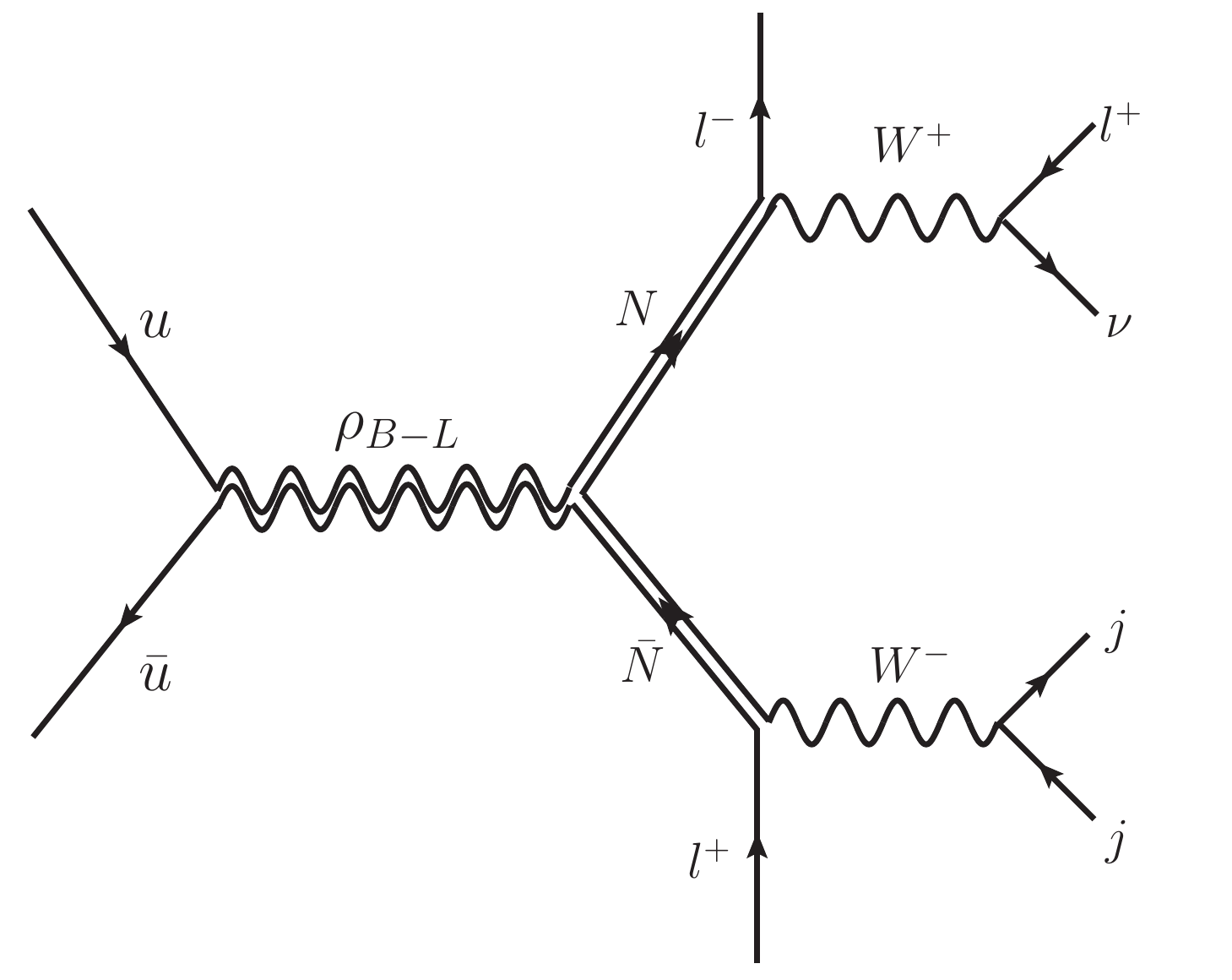}
\includegraphics[width=0.4\linewidth]{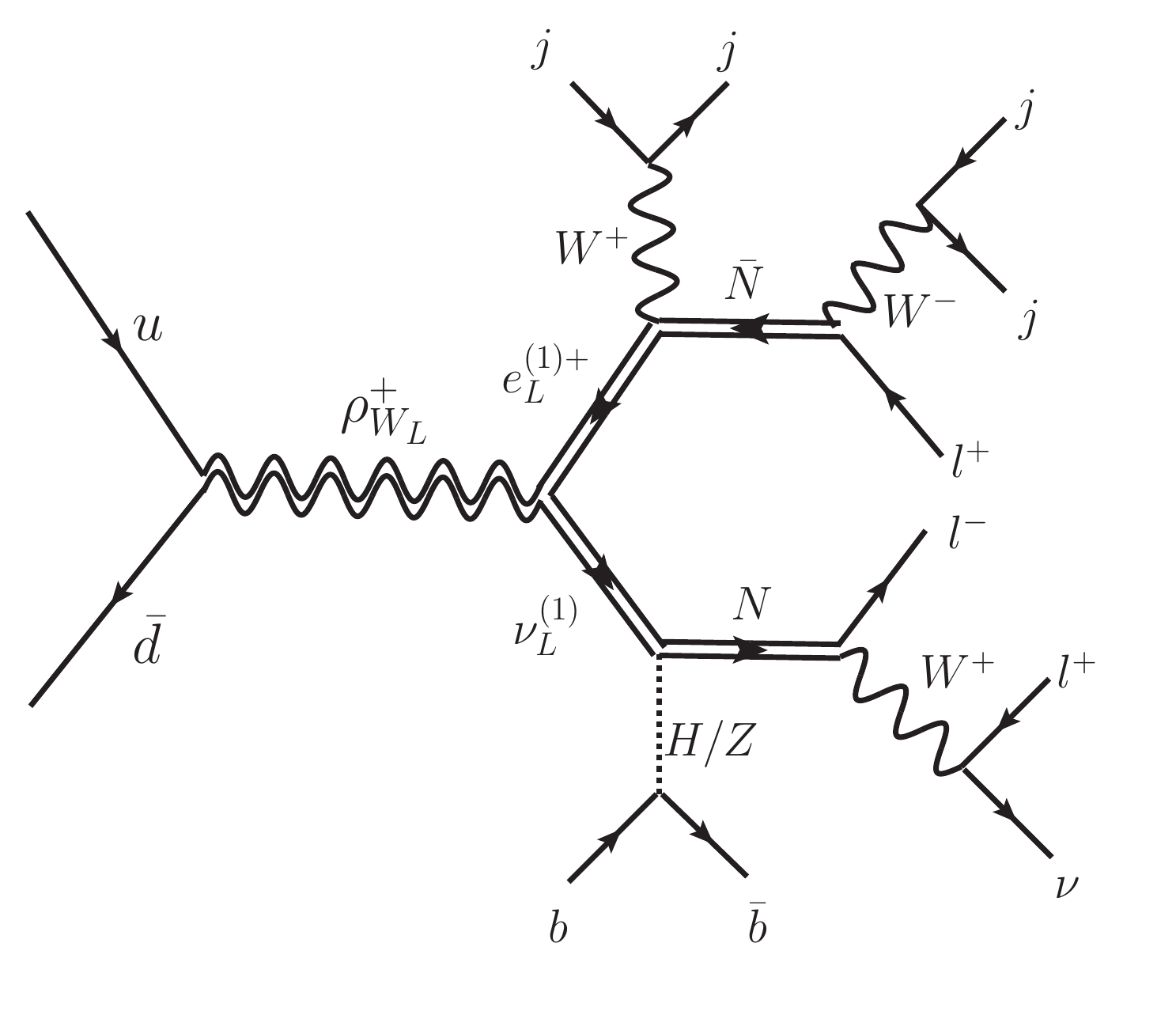}
\caption{The left panel shows Feynman diagram for the signal process of $\rho_{B-L}$-channel. The right panel shows Feynman diagram for the signal process of composite $SU(2)_L$ doublet-channel. Double (single) lines denote composite (SM) particles.}
\label{fig:signal}
\end{figure} 

\noindent The tree level Feynman diagrams for both signal processes are shown in Fig.~\ref{fig:signal}. The topology of our signal processes are characterized by several resonance peaks in various invariant mass distributions: invariant masses of S-channel gauge bosons, which we take to be $M_{\rho_{B-L}} = 2$ TeV  and  $M_{\rho_{W_L}} = 2.5$ TeV, invariant mass of singlet neutrino $N$, which we take $M_{N} = 750 \;(500)$ GeV for $\rho_{B-L}$(composite $SU(2)_L$ doublet)-channel, and invariant mass of composite $SU(2)_L$ doublet leptons, which we take $M_{\ell^h} =M_{\nu^h} = 1$ TeV. All of these invariant mass distributions, when successfully formed, will draw sharp distinctions between signal and SM backgrounds, allowing to achieve large significance. Notice, however, that both of our signal processes accompany \emph{single} neutrino in the form of missing energy. In order to reconstruct all (or most) of resonance peaks, therefore, the reconstruction of the longitudinal component of the neutrino momentum is a must. The presence of multiple leptons/jets in the signal processes then require that lepton/jet identification, i.e. 
the correct pairing of leptons and jets,
needs to be figured out along with neutrino momentum reconstruction. We show below that appropriate understanding of kinematics of the signal processes together with the help of symmetries, e.g. $SU(2)_L$, make this seemingly difficult task possible. We wish to emphasize the importance of reconstructing resonance peaks beyond simply achieving large $S/\sqrt{S+B}$. Accomplishing large enough significance is of course crucial in order to be able to reveal new physics out of overwhelming SM backgrounds. However, once discovered, immediate following task will be to understand the underlying physics that gives rise to such events. Successful reconstruction of all resonance peaks will answer the most important part of the questions: new particle content and spectrum.

\noindent In this study, we constrain ourselves to $\Delta R_{jj} \geq 0.4$ focusing only on regular (as opposed to boosted/fat) jets from $W$ decay. Inclusion of boosted $W$-jet contribution will yield increase in signal rate, resulting in larger significance $S/\sqrt{S+B}$. We leave study of boosted jet final states as well as study of several other decay channels mentioned above for future investigation.


\noindent Event simulations are performed by employing a sequence of simulation tools. We first created our two-site simplified model files using \textsc{FeynRules}~\cite{Alloul:2013bka} based on Heavy Vector Triplets models \cite{Pappadopulo:2014qza}. Then we used them as inputs model in a Monte Carlo event generator \textsc{MG5@aMC}~\cite{Alwall:2014hca} to generate parton level events. In this procedure, parton distribution functions parameterized by \textsc{NN23LO1}~\cite{Ball:2012cx} is used.  All the simulations are done at the leading order with a $\sqrt{s}=14$ TeV $pp$ collider. The generated parton level events are then streamlined to \textsc{Pythia 6.4}~\cite{Sjostrand:2006za} to take care of showering and hadronization/fragmentation. Since all our channels contain only regular jets, we directly pass on the output from \textsc{Pythia 6.4} to \textsc{Delphes 3}~\cite{deFavereau:2013fsa}. \textsc{Delphes 3}, interfaced with \textsc{FastJet}~\cite{Cacciari:2005hq, Cacciari:2011ma}, provides a way to incorporate the detector effects and jet formation. The jets are constructed with the anti-$k_t$ algorithm~\cite{Cacciari:2011ma} with a radius parameter $R=0.4$.


\subsection{$\rho_{B-L}$-channel}

\label{subsec:X_channel}

In this section, we present the results for $\rho_{B-L}$-channel: 
the relevant couplings for this channel are given in Sec.~\ref{subsec:compositeB-L}. 
In $\rho_{B-L}$-channel, as explained above, singlet $N$ is pair-produced via the decay of neutral composite $\rho_{B-L}$ gauge boson. This is possible because $\rho_{B-L}$ is the gauge boson of $U(1)_{B-L}$ and $N$ carries lepton number. The production of $\rho_{B-L}$ becomes possible thanks to the mixing between $\rho_{B-L}$ and elementary hypercharge gauge boson $B$, which then couples to quarks inside the proton. $N$ decays dominantly to $\ell W$ using Yuakwa coupling. As outlined above, we consider the case where one $W$ decays to $jj$ and the other to $\ell \nu$. Consequently, the final state consists of $2j3\ell$+MET. 

\noindent There are multiple invariant mass variables that will turn out to be very efficient and crucial in background suppression. These include $M_{jj}$, $M_{\ell_\nu \nu}$, $M_{\ell\ell_\nu \nu}$, $M_{\ell jj}$, $M_{\ell \ell \ell}$ and $M_{\rm All}$, where $M_{\rm All}$ is the invariant mass constructed from all visible objects, not including MET,
and $\ell_\nu$ is the lepton that groups with neutrino to reconstruct leptonically decaying $W$. Due to the fact that there are three leptons in the process, particle groupings, e.g. $\{\ell j j\}$ v.s. $\{\ell \ell_\nu \nu\}$ which reconstruct $N$ separately, is not obvious a priori. We will describe our grouping prescription below and show that we can reconstruct all particle bumps, resolving ambiguity in particle pairings. When properly  reconstructed, signal distribution of these variables will be peaked at $M_{jj} = M_{\ell_\nu \nu} = M_W$, $M_{\ell\ell_\nu \nu} = M_{\ell jj} = M_N$, and $M_{\rm All} = M_{\rho_{B-L}}$. The distribution of $M_{\ell \ell \ell}$ does not have direct connection to resonance peak, nevertheless, it will provide a very strong cut.

\noindent There are several SM backgrounds we need to consider and we describe them one by one now. As it will become relevant soon, we simulate processes to leading order in QCD and QED couplings. Namely, the set of diagrams considered in the numerical simulations is those with minimum order in QCD and QED couplings.\\

\noindent {\bf{(1) $\bf{jj \ell \ell W}$:}} The relevant process is $pp \rightarrow j j \ell \ell W, \; W \rightarrow \ell \nu$. This is the dominant background with largest cross section. Since two leptons (not from $W$ decay) in this process arise mostly from near-on-shell $Z$ decay, $M_{\ell \ell}$ distribution is sharply peaked at $M_Z$. On the other hand, in signal, they are from direct decays of $N$ and hence smoother $M_{\ell \ell}$ distribution. For this reason, the condition $M_{\ell \ell} \neq m_Z$ will achieve significant background suppression. However, we find that practically, $M_{\ell \ell \ell}$ distribution provides slightly sharper distinction between signal and backgrounds. Most of the other invariant mass variables described above will be important in reducing this background. \\

\noindent {\bf{(2) $\bf{t \bar{t} \ell \ell}$:}} The relevant process is $pp \rightarrow t \bar{t} \ell \ell$, with each top decaying to $b+W$. One of $W$'s decays hadronically giving $jj$, the other leptonically to $\ell \nu$. Since we use inclusive event selection criteria, i.e. $2j 3 \ell$ + X, this process, even if it contains extra two $b$'s, is indeed one of backgrounds. Background reduction can be achieved in a similar manner as $\bf{jj \ell \ell W}$.\\

\noindent {\bf{(3) $\bf{t \bar{t} W}$:}} The relevant process is $pp \rightarrow t \bar{t} W$, with each top decaying to $b+W$. All three $W$'s decay leptonically generating $\ell \ell \ell$+MET. In order to pass the event selection criteria, both $b$'s must be un-tagged as $b$-jet and detected as regular jets. This helps reduce this background. \\

\noindent {\bf{(4) $\bf{W W \ell \ell}$:}} The relevant process is $pp \rightarrow W W \ell \ell$, with one $W$ decaying to $jj$, the other to $\ell \nu$. However, naively, one can think that this process is already included in $\bf{jj \ell \ell W}$: just let one of $W$ decays to $jj$ and notice that the final state is precisely that of $\bf{jj \ell \ell W}$. Nonetheless, we include this background separately and yet we do not have double counting issue. This is because the leading QED coupling order for this process is QED=6 in \textsc{MG5@aMC}, and this corresponds to next-to-leading order for $\bf{jj \ell \ell W}$ (leading ordre is QED=4) and hence not included there.\footnote{As a further sanity check, we explicitly checked, using parton-level events, that $\bf{jj \ell \ell W}$ does not contain any event with \emph{two} intermediate $W$'s.}

%
%
\noindent Defining $N_\ell$ and $N_j$ as the number of isolated leptons and regular jets, respectively, 
we select events using the following selection criteria:
\bea
& N_\ell & > 2 \;\;\; \text{with} \; \vert \eta_\ell \vert < 2.5 \nonumber \\
& N_j & > 1 \;\;\; \text{with} \; \vert \eta_j \vert < 3 \label{eq:selection_criteria_X_channel}
\eea
In addition, we impose a set of basic cuts like $p_{Tj} > 20$ GeV and $p_{T\ell_1} > 200$ GeV at parton level event simulation for signal and backgrounds, where $P_{T\ell_1}$ means $P_T$ of the hardest lepton. This rather hard cut ($p_{T\ell_1} > 200$ GeV) is designed to improve background statistics for data analysis. We reimpose such cuts, together with $\Delta R > 0.4$ for \emph{all} possible pair of objects chosen, on objects (hardest two jets, and three leptons) in events that pass selection criteria of Eq.~(\ref{eq:selection_criteria_X_channel}). We use $p_T$ to evaluate hardness of the reconstructed objects. 

\noindent Next, we discuss the way we reconstruct all invariant mass of each particles.

\begin{itemize}
\item[I.] {\bf{Reconstruction of $p_{\nu z}$:}} For a given choice of lepton, a candidate for $\ell_\nu$, the z-component of the neutrino's momentum can be obtained by requiring
\bea
M_W^2 = (p_{\ell_\nu} + p_\nu)^2
\label{eq:pvz_reconstruction}
\eea
where $M_W$ is the mass of the SM $W$ boson. Neutrino four-momentum $p^{\mu}_{\nu}$ is constructed using the missing transverse momentum $\slashed{E}_T$ to get $p_x$ and $p_y$ and using masslessness of neutrino to compute $E_\nu$ from $p_x, \; p_y$ and $p_{\nu z}$. The equation $(p_{\ell_\nu} + p_\nu)^2 = M_W^2$ is, therefore, a quadratic equation for $p_{\nu z}$. For each choice of $\ell_\nu$, we vary $M_W$, starting from the central value $M_W = 80$ GeV, by $M_W \pm \Delta$ in step size of 2 GeV until the quadratic equation finds real solution(s). We choose maximum step to be $\vert \Delta \vert \leq 80$ so that the net mass $(M_W-\Delta)$ is still positive semi-definite. We repeat this for all three leptons and keep all possible solutions, if exist, for each lepton.
If no solution exists after all three leptons, we drop the corresponding event. Notice that this procedure may end up giving more than one $\ell_\nu$ 
, and furthermore, for each $\ell_\nu$, there may exist more than one solution for $p_{\nu z}$. In step II., we will specify the criteria by which we pick up one unique solution. \\

\item[II.] {\bf{Reconstruction of $M_N$:}} Once longitudinal component of neutrino's momentum (or equivalently full $p^\nu_\mu$) is reconstructed by step I., for a given choice of $\ell_\nu$, we then determine $\ell_j$ and $\ell_W$, the lepton combined with jet pair to reconstruct $N$ ($\ell_j$) and the lepton grouped with $W$ (i.e. $\ell_\nu \nu$) to form $N$ ($\ell_W$), respectively, by minimizing
\bea
\vert M_{jj\ell_j} - M_{\ell_\nu \ell_W \nu} \vert.
\label{eq:l_j_vs_l_W}
\eea
%
This is motivated by the fact that both $\{jj \ell_j\}$ and $\{\ell_W \ell_\nu \nu \}$ comes from the decay of on-shell $N$. If correctly chosen, $\{jj \ell_j\}$ and $\{\ell_W \ell_\nu \nu \}$ should have invariant mass peaked at the same value ($M_N$), thus giving a small difference of two invariant masses. Whereas wrong pairing would tend to give much larger difference.
We repeat this procedure for all three possible choices of $\ell_\nu$. Final decision is made for the combination $\{\ell_\nu, \ell_W, \ell_j\}$ that renders minimum value for Eq.~(\ref{eq:l_j_vs_l_W}).
\end{itemize}
%
%
%

\begin{figure}
    \centering
    \includegraphics[width = 7.5 cm]{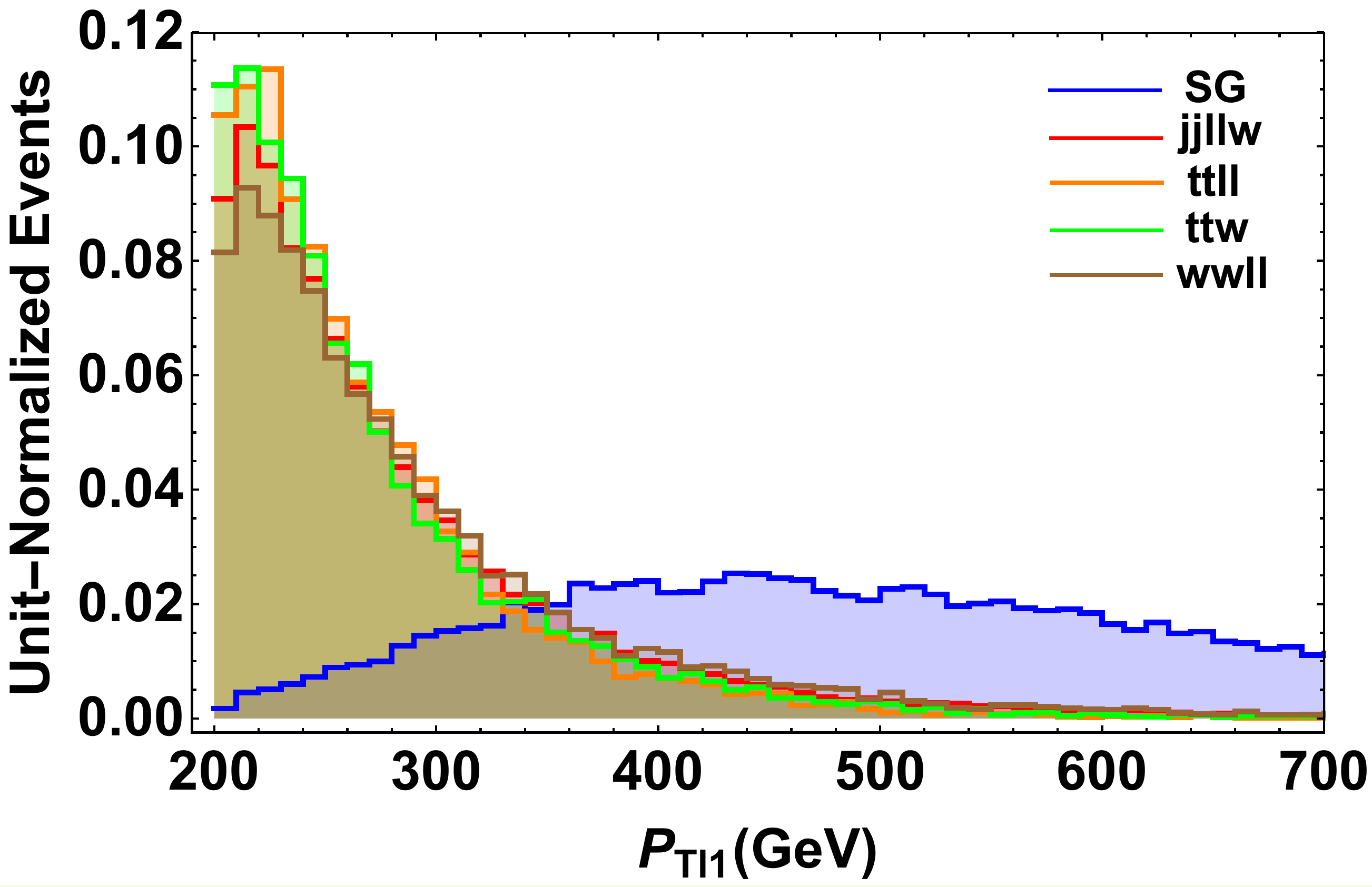}
    \includegraphics[width = 7.5 cm]{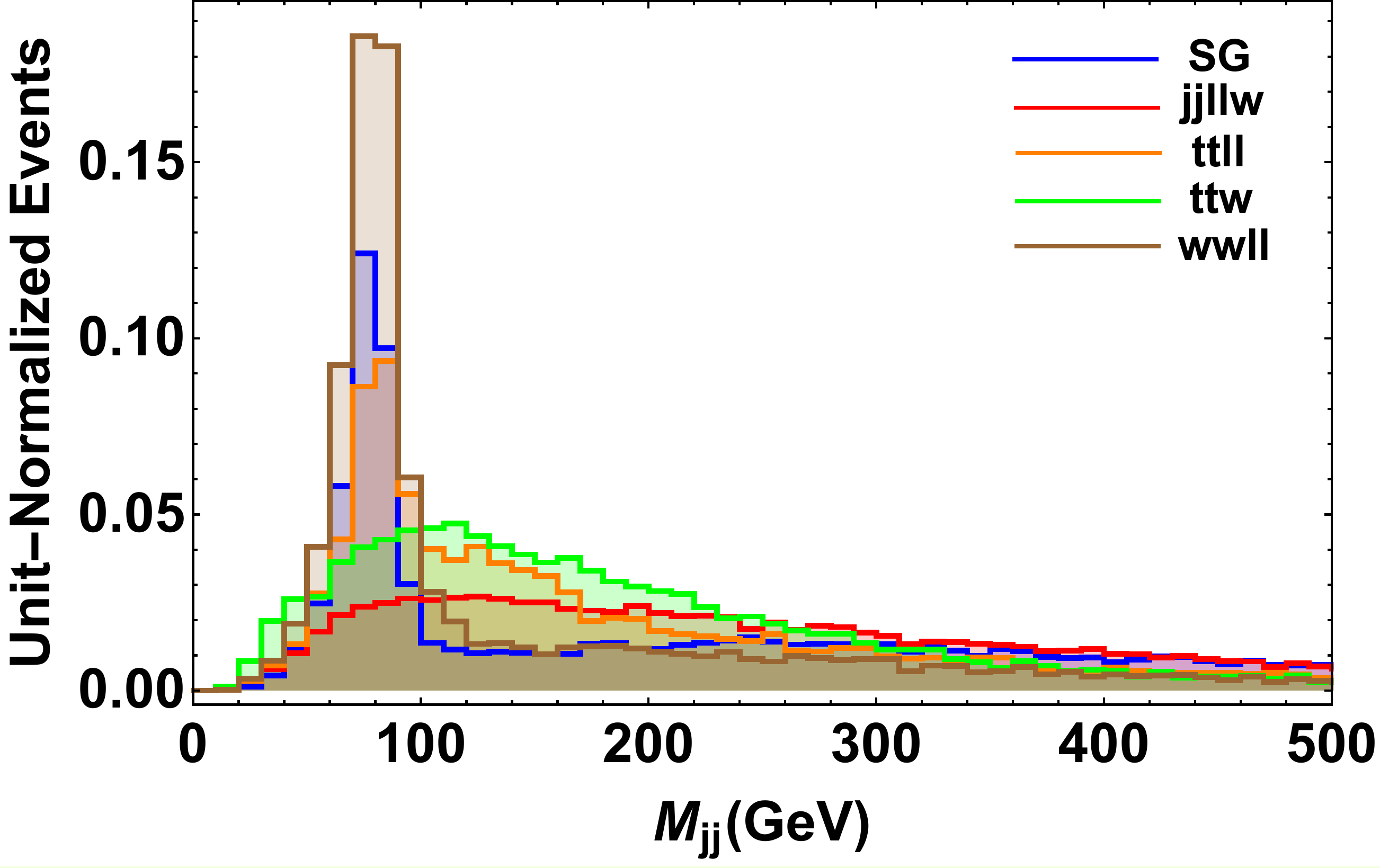}
    \includegraphics[width = 7.5 cm]{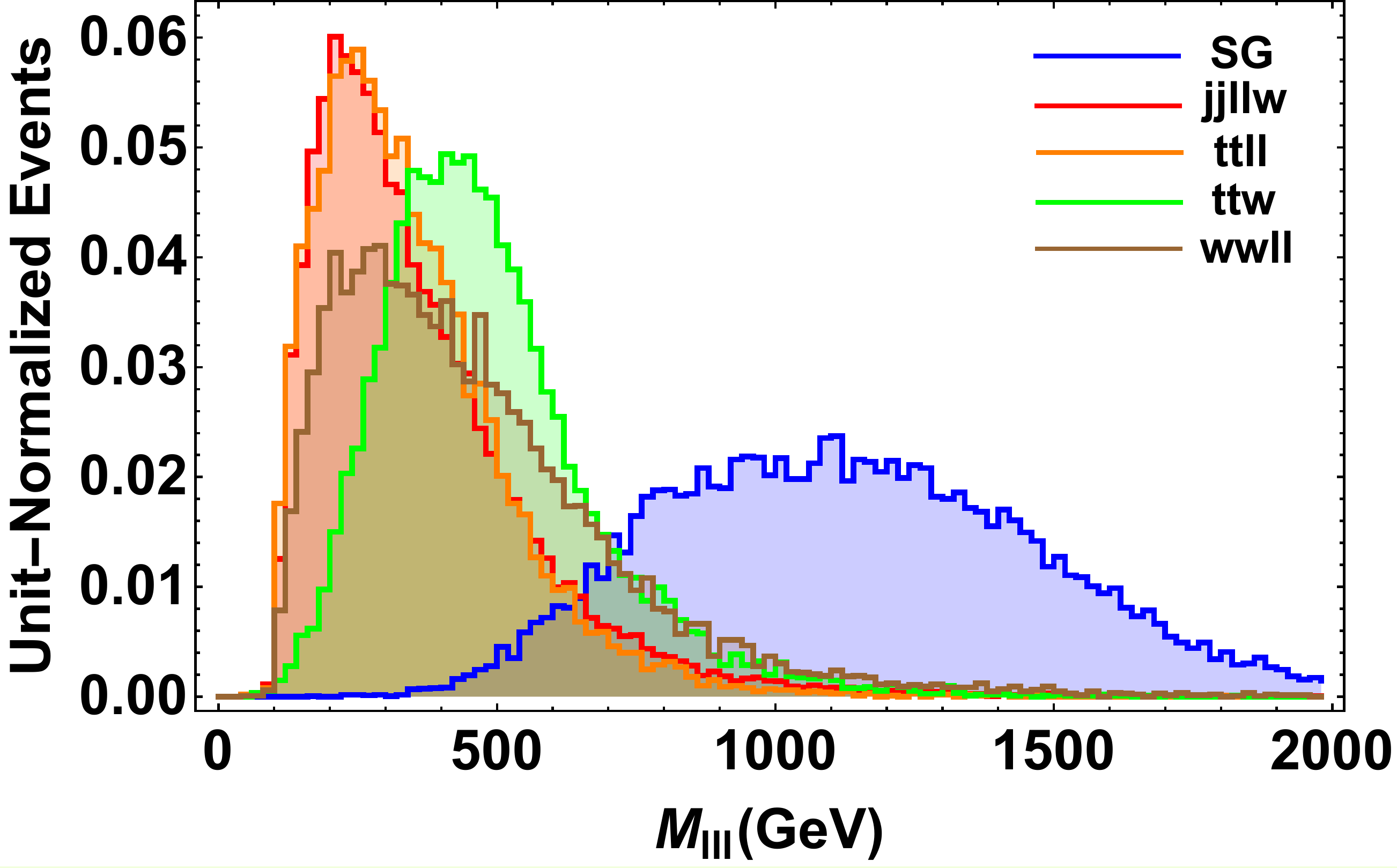}
    \includegraphics[width = 7.5 cm]{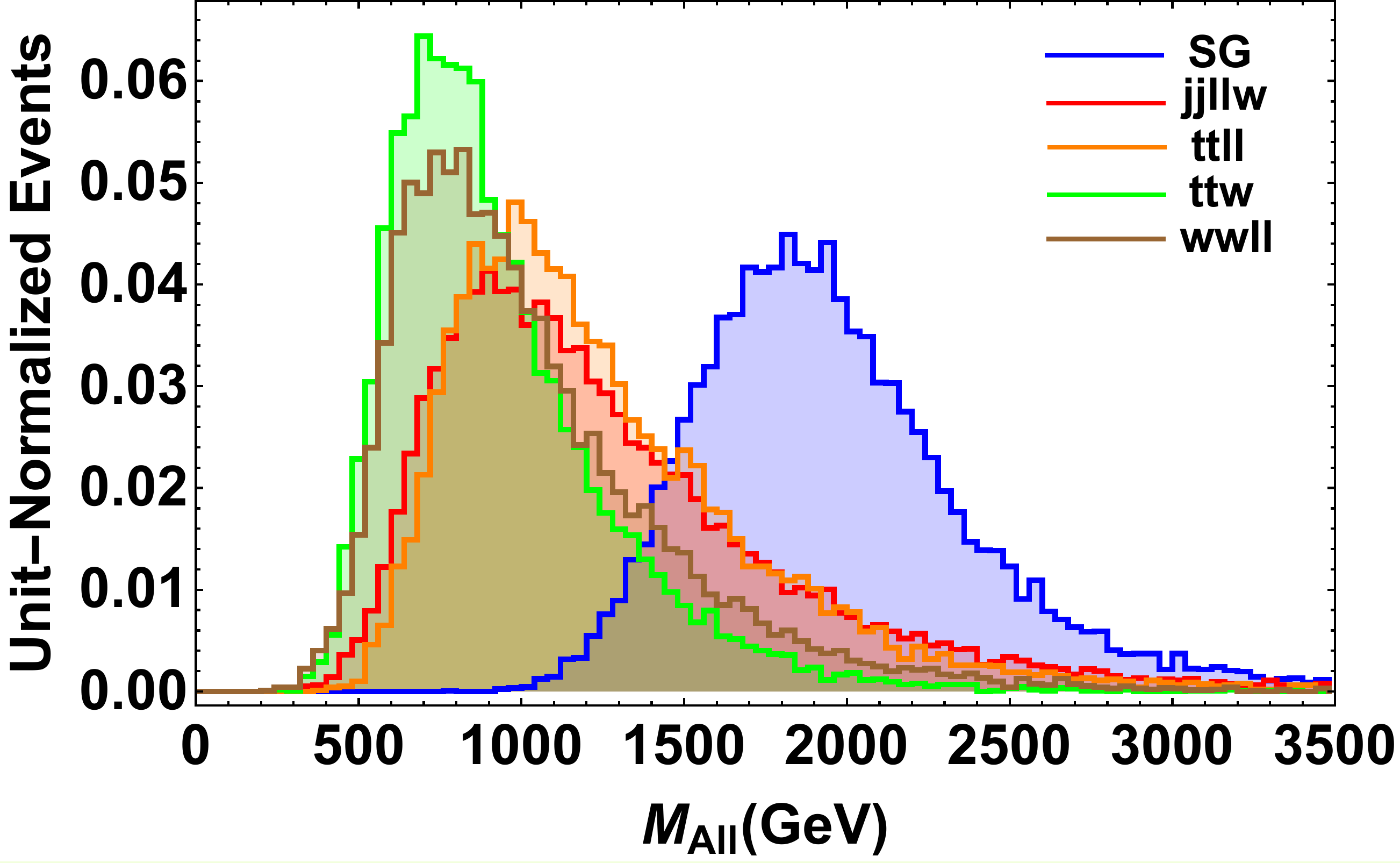}
    \includegraphics[width = 7.5 cm]{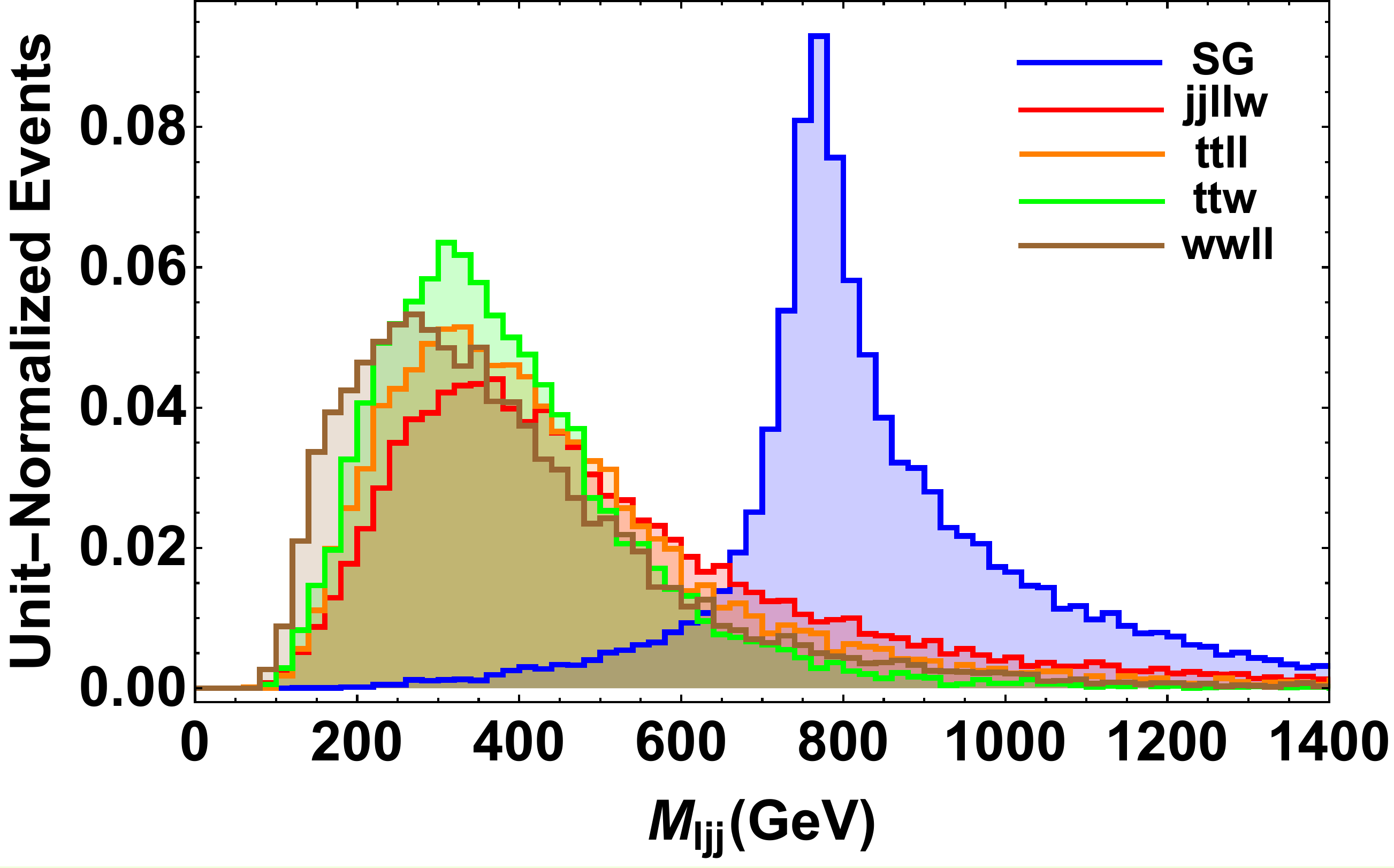}
    \includegraphics[width = 7.5 cm]{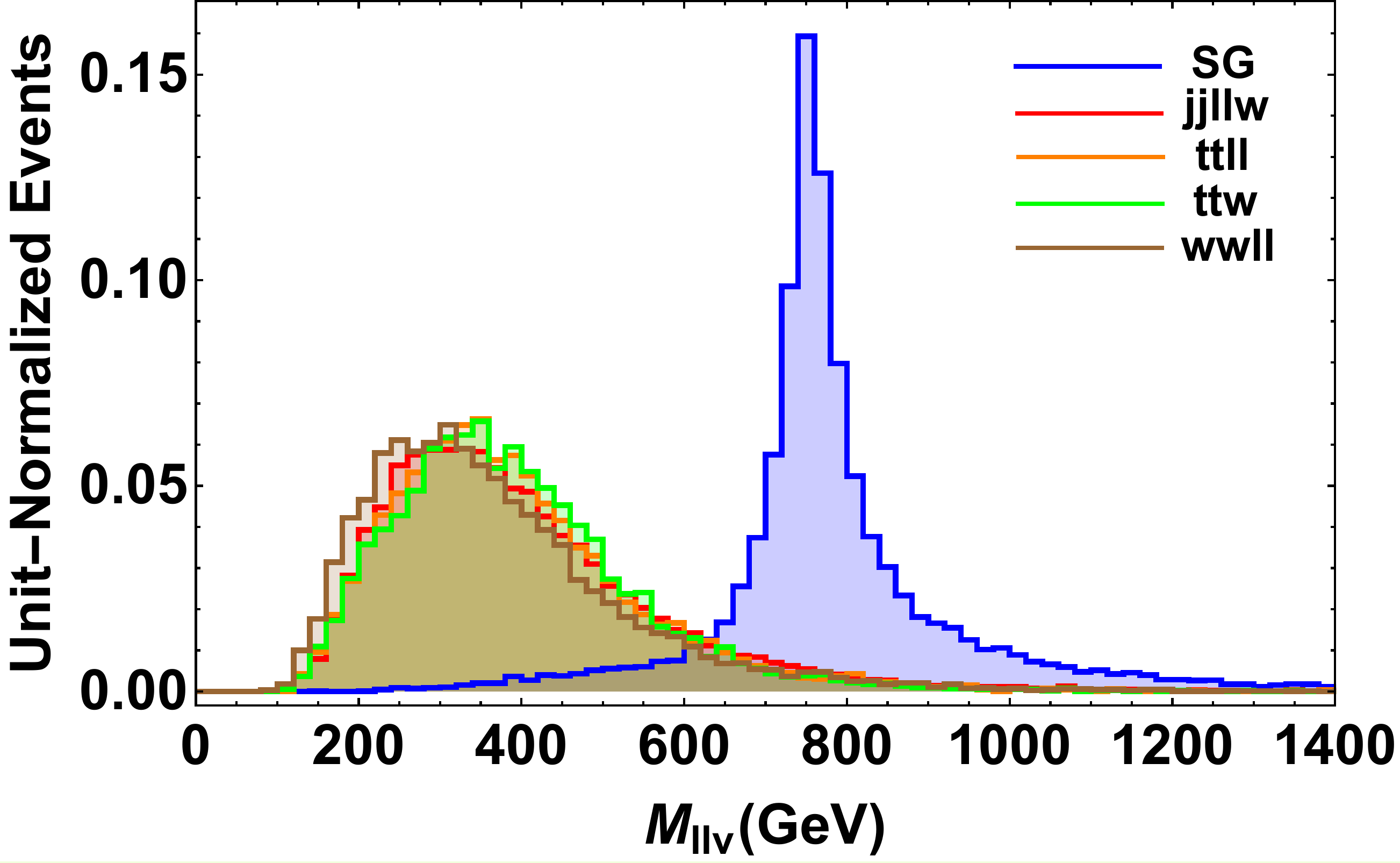}
    \caption{$\rho_{B-L}$-Channel: $p_{T\ell_1}$ (top row, left), $M_{jj}$ (top row, right),
$M_{\ell \ell \ell}$ (middle row, left), $M_{\rm All}$ (middle row, right), $M_{\ell j j}$ (bottom row, left), $M_{\ell \ell \nu}$ (bottom row, right) for signal (solid blue) and backgrounds (solid, $jj\ell \ell W$-red, $t\bar{t} \ell \ell$-orange, $t\bar{t} W$-green, $\ell \ell W W$-brown)
}
\label{fig:X_channel_plots}
\end{figure}
\noindent In Fig.~\ref{fig:X_channel_plots}, we show distributions of various kinematic variables for signal and background events that pass selection criteria and basic cuts. Invariant mass variables are reconstructed following the prescription described above. The invariant mass distributions of signal events evidently show that our reconstruction prescription for invariant mass distributions is very successful. In particular, the distributions of $M_{jj}$ (top row, right), $M_{\rm All}$ (middle row, right), $M_{\ell jj}$ (bottom row, left), $M_{\ell_\nu \ell \nu}$ (bottom row, right) are peaked at the position expected from the input values. Also shown is $M_{\ell \ell \ell}$ which reveals apparent separation between signal and background, supplying a very efficient cut for background reduction.

\noindent We employed a set of cuts to achieve significant $S/\sqrt{S+B}$ for our analysis. We provide the cut flows for signal and the major SM backgrounds in Table~\ref{tab:X_channel}. We find that the $\rho_{B-L}$-channel may provide a sensitivity to discover $N$ and $\rho_{B-L}$ by $\sim 4.8 \sigma$ with an integrated luminosity of $\mathcal{L} = 3000 \; {\rm fb}^{-1}$.


\begin{table}[t]
\begin{adjustwidth}{-1cm}{}  
\centering
\begin{tabular}{|c|c|c|c|c|c|}
\hline 
Cuts & {\bf{Signal}} & $\bf{jj \ell \ell W}$ &$\bf{t\bar{t}\ell \ell}$&$\bf{t\bar{t}W}$ & {$\bf{ WW \ell \ell}$}  \\
\hline \hline
No cuts & $4.88\times 10^{-2}$  & $1.39\times 10^2$&7.26& 2.50& $7.61 \times 10^{-1}$\\
$N_{\ell}\geq 3$, $N_j\geq 2$ with basic cuts &$ 2.88 \times 10^{-2}$ &4.26&$2.14 \times 10^{-1} $&$5.32 \times 10^{-2} $ &$3.95 \times 10^{-2} $  \\
$M_{\ell\ell\ell} \in [600,\,\infty]$ GeV &$2.76\times 10^{-2} $&$4.49 \times 10^{-1} $&$1.62 \times 10^{-2} $& $1.16 \times 10^{-2} $ & $9.01 \times 10^{-3} $ \\
$M_{\ell j j} \in [700,\,1200]$ GeV &$2.14 \times 10^{-2}$&$1.01 \times 10^{-1} $&$3.21 \times 10^{-3} $&$1.08 \times 10^{-3} $ & $1.03 \times 10^{-3} $ \\
$M_{\ell \ell \nu} \in [700,\,1000]$ GeV & $1.79 \times 10^{-2}$&$4.51 \times 10^{-2} $&$1.39 \times 10^{-3} $ & $3.48 \times 10^{-4} $ & $3.97 \times 10^{-4} $  \\	
$p_{T\ell_1} \in [300,\,\infty]$ GeV & $1.68 \times 10^{-2}$&$3.12 \times 10^{-2} $&$5.99 \times 10^{-4} $ & $2.04 \times 10^{-4} $ &  $2.99 \times 10^{-4} $ \\
$M_{j j} \in [0,\,500]$ GeV & $1.46 \times 10^{-2}$&$1.40\times 10^{-2} $&$3.42 \times 10^{-4} $ &  $1.39 \times 10^{-4} $&$1.96 \times 10^{-4} $ \\
$M_{\rm{All}} \in [0,\,2500]$ GeV & $1.40 \times 10^{-2}$&$1.13 \times 10^{-2} $& $2.78 \times 10^{-4} $&$1.39 \times 10^{-4} $&  $1.70 \times 10^{-4} $\\
\hline
$S/B$ & 1.18 & -- & --& -- & --  \\
$S/\sqrt{S+B}$ ($\mathcal{L}=300$ fb$^{-1}$) & 1.51 & -- & --& -- & --  \\
$S/\sqrt{S+B}$ ($\mathcal{L}=3000$ fb$^{-1}$) & 4.77 & -- & --& -- & -- \\
\hline
\end{tabular}
\caption{Cut flows for signal and major background events in terms of their cross sections. The cross sections are in fb. The cross sections in the first row are such that for signal and for backgrounds, a set of minimum cuts ($p_{Tj}  > 20$ GeV, $p_{T\ell} > 10$ GeV, $\Delta R > 0.4$ etc.) at event generation level are applied to avoid IR-divergence. In the second row, the same basic cuts ($P_{Tj}>20 $ GeV, $P_{T\ell_1}> 200 $ GeV, $| \eta_j | < 3$, $| \eta_l |  < 2.5$, $\Delta R\geq 0.4$ for \emph{all} pairs of objects) are reimposed on both signal and backgrounds events. \label{tab:X_channel} }
\end{adjustwidth}
\end{table}





\subsection{Composite $SU(2)_L$ doublet-channel}
\label{subsec:WL-channel}

As we briefly described at the beginning of the section, in composite $SU(2)_L$ doublet-channel, the singlet neutrino $N$ is produced via the on-shell decay of composite $SU(2)_L$ lepton doublet, which, in turn, is produced by the on-shell decay of composite gauge boson $\rho_{W_L}$. 
The relevant couplings for this channel are given in Sec.~\ref{subsec:compositeSU2L}.
%
%
As already mentioned in the introduction, we want to emphasize that this production channel for singlet $N$ is largely model-independent, making its study important and well-motivated. It is model-independent in a sense that (i) the production is via couplings that always exist and sizable and (ii) it is largely independent of the representation of the singlet $N$ under global symmetries of the composite sector. In order to see this, first notice that composite $SU(2)_L$ doublet lepton $(\nu^h, \ell^h)$ is present in any composite Higgs models (or its 5D dual Randall-Sundrum model) whenever leptons exhibit partial-compositeness. The same is true for composite $\rho_{W_L}$, even when we do not extend our electroweak symmetry to LR symmetric version. Then, the coupling between composite lepton doublet and composite $\rho_{W_L}$ exists by global symmetry of the composite sector (or bulk gauge symmetry in 5D picture). Moreover, the coupling of composite lepton doublet with singlet $N$ and Higgs must be present simply because this is the coupling that generates the mass of the SM neutrino mass by the seesaw mechanism. Finally, the production of the composite $\rho_{W_L}$ can be achieved simply by the composite-elementary mixing between composite $\rho_{W_L}$ and its elementary partner, which then couples to quarks inside the proton. Therefore, we see that, provided masses of composite particles are light enough and are such that their on-shell decays are kinematically allowed (hence resonance-enhancement), we can ensure enough rate for signal process at the LHC, which is, again, independent of specifics of the model.

\noindent Once produced, singlet $N$ decays (using Yuakwa coupling) to a lepton and $W$ boson. As can be seen from the Feynman diagram shown in Fig.~\ref{fig:signal}, the signal process contains three $W$'s, two leptons and one either $H$ or $Z$. We choose to study the case when two of the three $W$'s decay hadronically producing four jets and one leptonically producing lepton+MET. Since there are three $W$'s, there will be three ways that this occurs. $H/Z$ decays to $b \bar{b}$. The final state, therefore, consists of $4j 2b 3\ell +$MET.

\noindent There are several invariant mass variables which are, as we will explain momentarily, all fully reconstructible and crucial in revealing signal process out of large SM backgrounds. Those invariant mass variables include, $M_{\rho_{W_L}}$, $M_{\ell^h}$, $M_{N}$, $M_{H}$, and $M_{W}$, where $M_{\rho_{W_L}}$ is the reconstructed invariant mass of composite $\rho_{W_L}$, and similarly for others as is evident from their names. Due to $SU(2)_L$ symmetry, $\ell^h$ and its partner $\nu^h$ are degenerate and we use $M_{\ell^h}$ to denote the mass for $SU(2)_L$ composite doublet lepton. Because of combinatorics in $W$ decays, i.e. two $W \rightarrow j j$ and one $W \rightarrow \ell \bar{\nu}$, the identification of a set of particles coming from decay of $N$ (and similarly for $\ell^h/\nu^h$) is \emph{not} fixed. For example, in case when $W$ from direct decay of $\ell^h$ decays leptonically, both $N$ will decay eventually into $\ell jj$. On the other hand, if leptonically decaying $W$ is from $N$, then one $N$ decays into $\ell jj$, but the other into $\ell \ell +$MET. For this reason, the sets of particles that reconstruct invariant masses of composite particles will vary event by event and this explains why we denote invariant mass variables by the \emph{name} of corresponding particle, e.g. $M_{N}$, instead of its daughter SM particles, say, $M_{j j\ell}$ like in our earlier paper \cite{Agashe:2016ttz} . If successfully reconstructed, the signal distributions of these variables will be peaked at $M_{\rho_{W_L}} = 2.5$ TeV, $M_{\ell^h} = 1$ TeV, $M_{N} = 500$ GeV, $M_{H} \approx 125$ GeV, and $M_{W} \approx 80$ GeV, respectively.

\noindent There are several SM backgrounds we need to consider and we now describe them one by one in order of significance. \\

\noindent {\bf{(1) $\bf{t\bar{t}jj\ell\ell}$:}} The relevant process is $pp \rightarrow t \bar{t} jj \ell \ell$, followed by subsequent decay of $t \to b W$, and similarly for $\bar{t}$. One of $W$ from top decays hadronically and the other leptonically. In principle, one may consider even more inclusive background, $pp \rightarrow 4j 2b \ell \ell W, \; W \rightarrow \ell \nu$, where we need $W$ to obtain the third lepton from its decay without violating lepton number conservation. However, we found that generating such an inclusive process in a Monte Carlo event generator like \textsc{MG5@aMC} is rather impractical mainly due to proliferation of Feynman diagrams. As the next best plan, we decided to consider the above described process, which we think is inclusive enough to capture relevant SM backgrounds for our study. The kind of processes that we are missing by not considering the most inclusive one will be those with soft pure QCD jets (including b-jets). In order to strengthen our argument, therefore, we impose hard $p_T$ cut on the hardest jet, i.e. $p_{Tj_1} > 100$ GeV, where $j_1$ is the jet with the largest $p_T$. This is the background with largest cross section. As already mentioned, we simulate processes to leading order in QCD and QED couplings. For later purpose, we mention that the QED coupling order relevant for this background is QED=6. Background reduction will be achieved by means of a combination of various invariant mass cuts. Since two leptons (those not from $W$ decay) in this background will mostly come from near-on-shell decay of $Z$ boson, the distribution of the di-lepton invariant mass, $M_{\ell\ell}$, will be sharply peaked at the mass of the $Z$ boson, $M_Z$. However, since two leptons in the signal process arise from direct decay of $N$, they do not reconstruct $M_Z$. Consequently, the condition $M_{\ell\ell} \neq M_Z$ provides a very efficient cut. Other useful cuts are $M_{\rho_{W_L}}$, $M_{\ell\ell\ell}$ and $M_{N}$ cuts. $P_{T\ell_1}$ will also be very useful. \\

\noindent {\bf{(2) $\bf{t\bar{t} t\bar{t}}$:}} The relevant process is $pp \rightarrow t \bar{t} t\bar{t}$, where each  top decays to $t  \rightarrow 
b \; W$. Out of four $W$'s produced in top decays, one decays hadronically and the remaining three decays leptonically, resulting in three leptons and MET. In order to pass the selection criteria, however, two of the four $b$'s must be tagged as $b$-jets and the other two must be un-tagged as regular two jets, leading to a large reduction of the background. In addition, $M_{\rho_{W_L}}$ and $M_{N}$ cuts will be particularly efficient for reduction of this background. \\

\noindent {\bf{(3) $\bf{t\bar{t}\ell\ell W}$:}} The relevant process is $pp \rightarrow t\bar{t} \ell \ell W$, with subsequent decay of tops to $b+W$. Two $W$'s decay hadronically rendering four jets and the third one leptonically. Similarly to $\bf{t\bar{t}jj\ell\ell}$, the lepton pair comes mostly from decay of on-shell $Z$ (and off-shell photon). Therefore, $M_{\ell\ell} \neq M_Z$ will remove most of this background. Several other invariant mass cuts will be useful. One may readily notice that if one of the hadronically decaying $W$'s is the one not from top, this process has the precisely the same final state (before top decays) as $\bf{t\bar{t}jj\ell\ell}$ and one would worry about double counting issue. However, we claim there is no such issue. The resolution is that we are working at leading order in QCD and QED couplings, and the leading QED coupling order that gives rise to this process is QED=8, which is next-to-leading order for $\bf{t\bar{t}jj\ell\ell}$ background (leading order was QED=6 for this one), hence not captured there. \\

\noindent {\bf{(4) others:}} There are several other processes that can contribute to SM backgrounds. One of them is $pp \rightarrow t \bar{t} W W W$, with three $W$'s decay leptonically and the other two hadronically. Another is $pp \rightarrow H/Z \; \ell \ell W W W$, with two $W$'s decay hadronically and the third one leptonically. However, these processes (i) require high QED coupling order (QED=10) leading to parametric suppression compared to above three cases and (ii) multiplicity of the final state is equal or even greater than above major processes (i.e. either comparable or additional phase space suppression). As a result, cross section of these processes are much smaller, leading to at most $O(1)$ events before selection criteria and hence no effects in the final results. For this reason, we do not consider these backgrounds explicitly.   \\

\noindent Defining $N_\ell$, $N_b$ and $N_j$ as the number of isolated leptons, b-tagged jets and non-b-tagged jets, respectively, 
we select events using the following selection criteria:
\bea
& N_\ell & > 2 \;\;\; \text{with} \; \vert \eta_\ell \vert < 2.5 \nonumber \\
& N_b & > 1 \;\;\; \text{with} \; \vert \eta_b \vert < 3 \label{eq:selection_criteria_WL_channel}\\
& N_j & > 3 \;\;\; \text{with} \; \vert \eta_j \vert < 3. \nonumber
\eea
In addition, we impose a set of basic cuts $p_{Tj} / p_{Tb} > 20$ GeV, $p_{T\ell} > 10$ GeV, $\Delta R > 0.4$, and so on at parton level event simulation, partly to avoid possible IR-divergence issues for background simulations. We reimpose such cuts on objects (hardest four jets, two $b$-jets, and three leptons) in events that pass selection criteria of Eq.~(\ref{eq:selection_criteria_WL_channel}). We use $p_T$ to evaluate hardness of the reconstructed objects. We also explicitly impose $\Delta R > 0.4$ for \emph{all} possible pair of objects chosen out of above selected objects.
%
%

\noindent Now, we describe how we reconstruct invariant masses of each particle. 

\begin{itemize}
\item[I.] {\bf{Reconstruction of $p_{\nu z}$:}} The first step towards the reconstruction of all resonance peaks is the reconstruction of longitudinal component of neutrino momentum, $p_{\nu z}$. There are three leptons in the process and each of these can potentially pair with neutrino to form $W$. For each choice of lepton, call it $\ell_\nu$, we first solve the quadratic equation $(p_{\ell_\nu} + p_\nu)^2 = M_W^2$, where $p_\nu^\mu$ is formed using the missing transverse momentum $\slashed{E}_T$ to get $p_x$ and $p_y$ and using masslessness to compute $E_\nu$ from $p_x, \; p_y$ and $p_{\nu z}$. The equation $(p_{\ell_\nu} + p_\nu)^2 = M_W^2$ is, therefore, a quadratic equation for $p_{\nu z}$. For each choice of $\ell_\nu$, we vary $M_W$, starting from the central value $M_W = 80$ GeV, by $M_W \pm \Delta$ in step size of 1 GeV until the quadratic equation finds real solution(s). We choose maximum step to be $\vert \Delta \vert \leq 80$ so that the net mass $(M_W-\Delta)$ is still positive semi-definite. We repeat this for all three leptons and choose the lepton(s) with minimum $\Delta$, i.e. we choose the lepton(s) which reconstruct the $p_{\nu z}$ such that the computed $W$ mass, i.e. $(p_{\ell_\nu} + p_\nu)^2$, is closest to the central value of 80 GeV. If no solution exists after all three leptons, we drop the corresponding event. Notice that this procedure may end up giving more than one $\ell_\nu$ with the same deviation factor $\Delta$, and furthermore, for each $\ell_\nu$, there may exist more than one solution for $p_{\nu z}$. In step III., we will specify the criteria by which we pick up one unique solution. \\

\item[II.] {\bf{Jet pairing:}} In signal process, four jets arise from decay of two $W$'s. So, we determine jet pairing (i.e.~figuring out jet pair from each $W$ boson) as follows. We consider all possible jet pairings. Then for each jet pairing $jj_1, jj_2$, e.g. $jj_1 \equiv (j_1,j_2)$ and $jj_2 \equiv (j_3,j_4)$,  we compute the following quantity (``$L^2$-distance'' function): 
\bea
d_{jj} = (M_{jj_1} - M_W)^2 + (M_{jj_2} - M_W)^2.
\eea
We choose the pairing with minimum $d_{jj}$. \\

\item[III.] {\bf{Reconstruction of $M_N$:}} Next, we want to determine sets of particles coming from $N$ decay. Let's call each set $N_1$ and $N_2$. Depending on how $W$ from $N$ decays (to $jj$ v.s. to $\ell \nu$), the final set can be either $\ell j j$ or $\ell \ell_\nu \nu$, where $\ell_\nu$ is the lepton that pairs with neutrino to form a $W$.\footnote{Recall that, at the stage, there still can be several $\ell_\nu$ choices, together with the possibility of two $p_{\nu z}$ solutions. Unique solution will be determined by the determination of the set $N_1$ and $N_2$.} Thus, there are two possibilities we need to consider: (i) both $N$ decay to $\ell j j$ and (ii) one to $\ell jj$ and the other to $\ell \ell_\nu \nu$. In step I. we build a set that collects lepton(s), $\ell_\nu$, and corresponding $p_{\nu z}$'s. For each such $\ell_\nu$, we can now simply figure out the remaining two leptons. In case (i), these will be paired with jet-pairs to form $N$. Using result of step II. there are only two combinations of $\{\ell j j\}$. For each such constructed set $N_1$ and $N_2$, we compute the distance function:
\bea
d_N = (M_{N_1} - M_{N_2})^2.
\label{eq:d_N}
\eea
In case (ii), on the other hand, one lepton goes with jet pair (two choices, $jj_1$ or $jj_2$), and the other lepton groups with $\ell_\nu$. For latter, we consider all choices of $\ell_\nu$ and corresponding $p_{\nu z}$ that pass the criteria of step I. Similarly to the case (i), for each set $N_1$ and $N_2$, we compute the distance Eq.~(\ref{eq:d_N}). For the set containing $\ell_\nu$, we use fully reconstructed $p_\nu^{\mu}$ to calculate invariant mass. At the end, we choose the set $N_1$ and $N_2$ with minimum $d_N$.\footnote{At this stage, we have made a unique decision for $\ell_\nu$ and corresponding $p_{\nu z}$, unless numerical coincidence happens by complete accidence. Considering the amount of significant figures in the data, this will be very unlikely and indeed, we have not found one case.} \\

\item[IV.] {\bf{Reconstruction of $M_{\ell^h}$:}} Finally, we determine the set of particles from the decay of $\ell^h$ and $\nu^h$. Thanks to $SU(2)_L$ symmetry, we know that $M_{\ell^h}=M_{\nu^h}$ and using this we consider all the combinations and choose the one that minimizes
\bea
d_{\ell^h} = (M_{\ell^h} - M_{\nu^h})^2.
\eea
To be more specific, in case (i) of step III. both $N_1$ and $N_2$ are set of the form $\{\ell j j \}$. One of this will combine with $\{\ell_\nu \nu\}$ to render $\ell^h$ and the other $\{b\bar{b}\}$ to make up $\nu^h$. In case (ii) of step III., on the other hand, one of $N_1$/$N_2$ will groups with $\{jj\}$ to give $\ell^h$ and other with $\{b\bar{b}\}$ to construct $\nu^h$. We simply consider all these combinations and compute invariant masses for each choice. At the end, we simply choose the one with minimum $d_{\ell^h}$. 
\end{itemize}
%
%

\begin{figure}
    \centering
    \includegraphics[width = 7.5 cm]{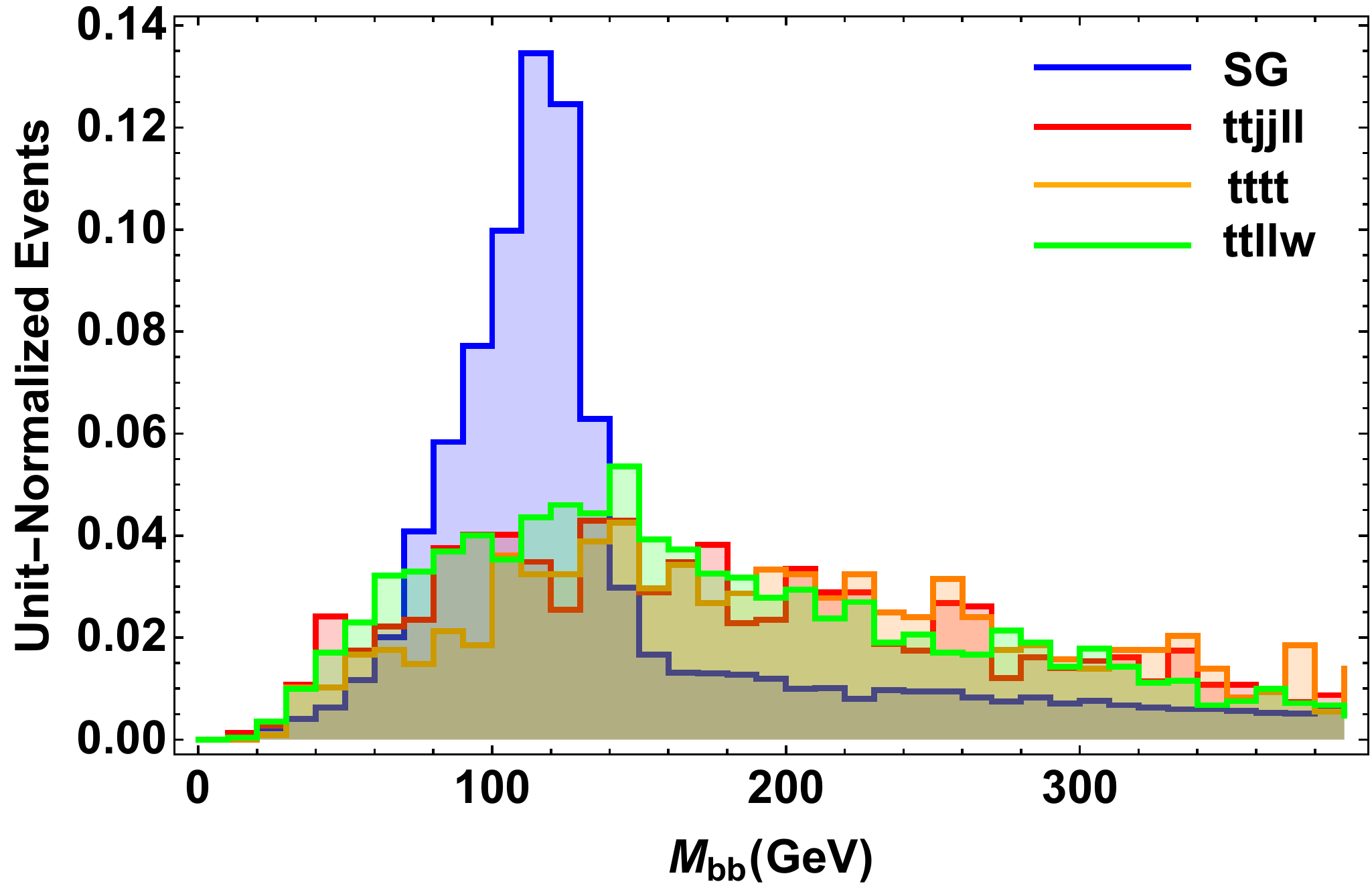}
    \includegraphics[width = 7.5 cm]{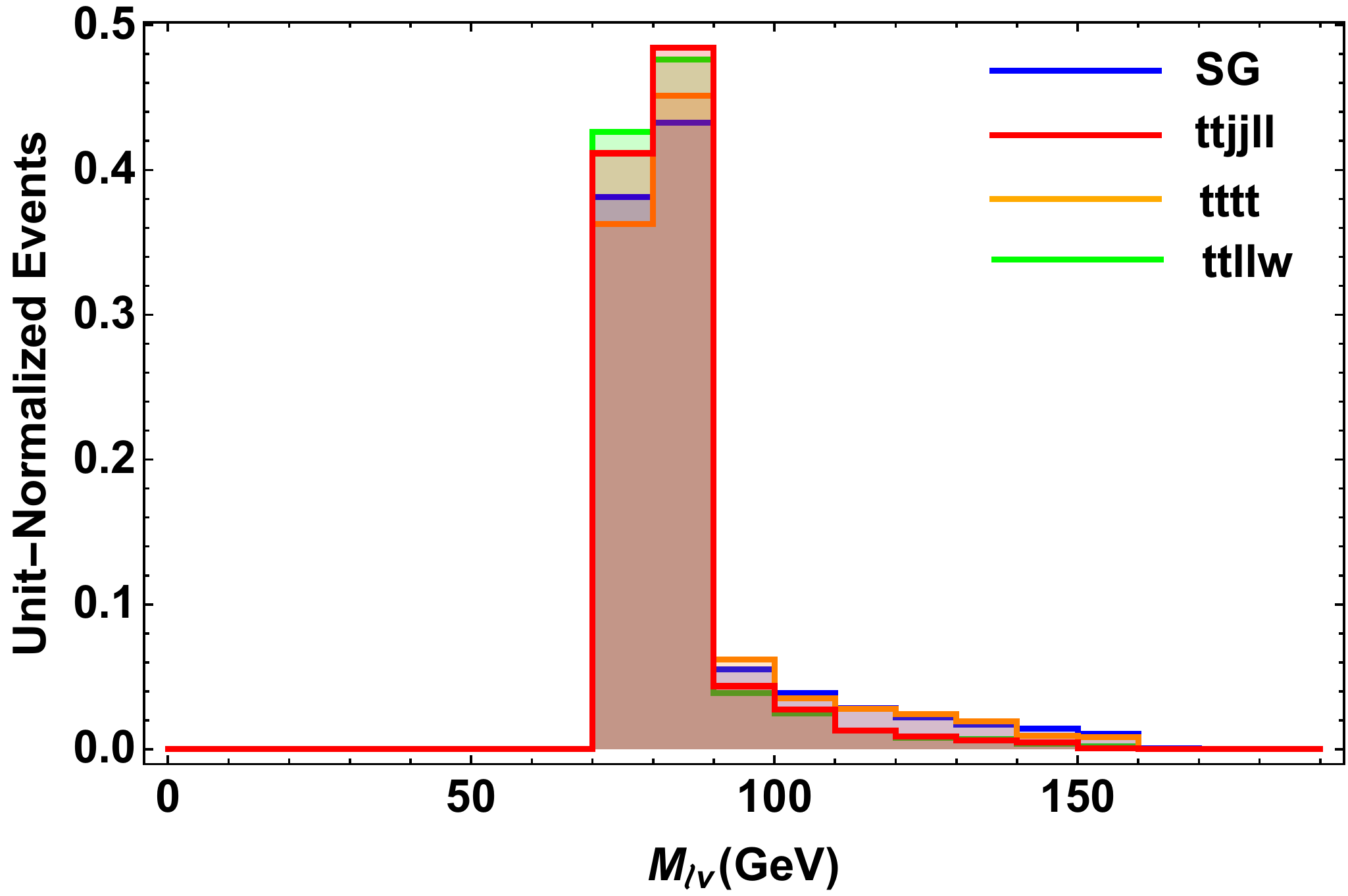}
    \includegraphics[width = 7.5 cm]{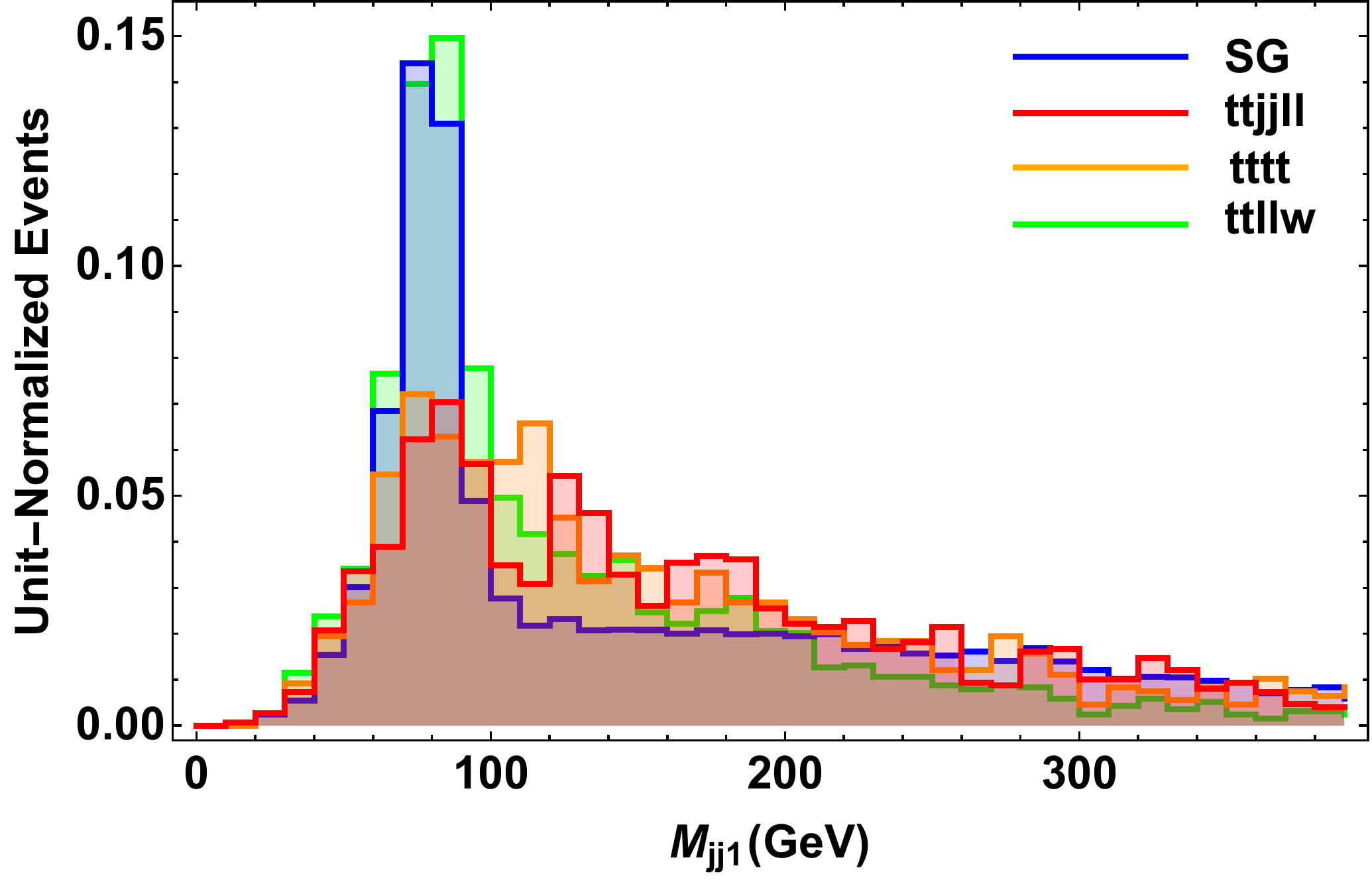}
    \includegraphics[width = 7.5 cm]{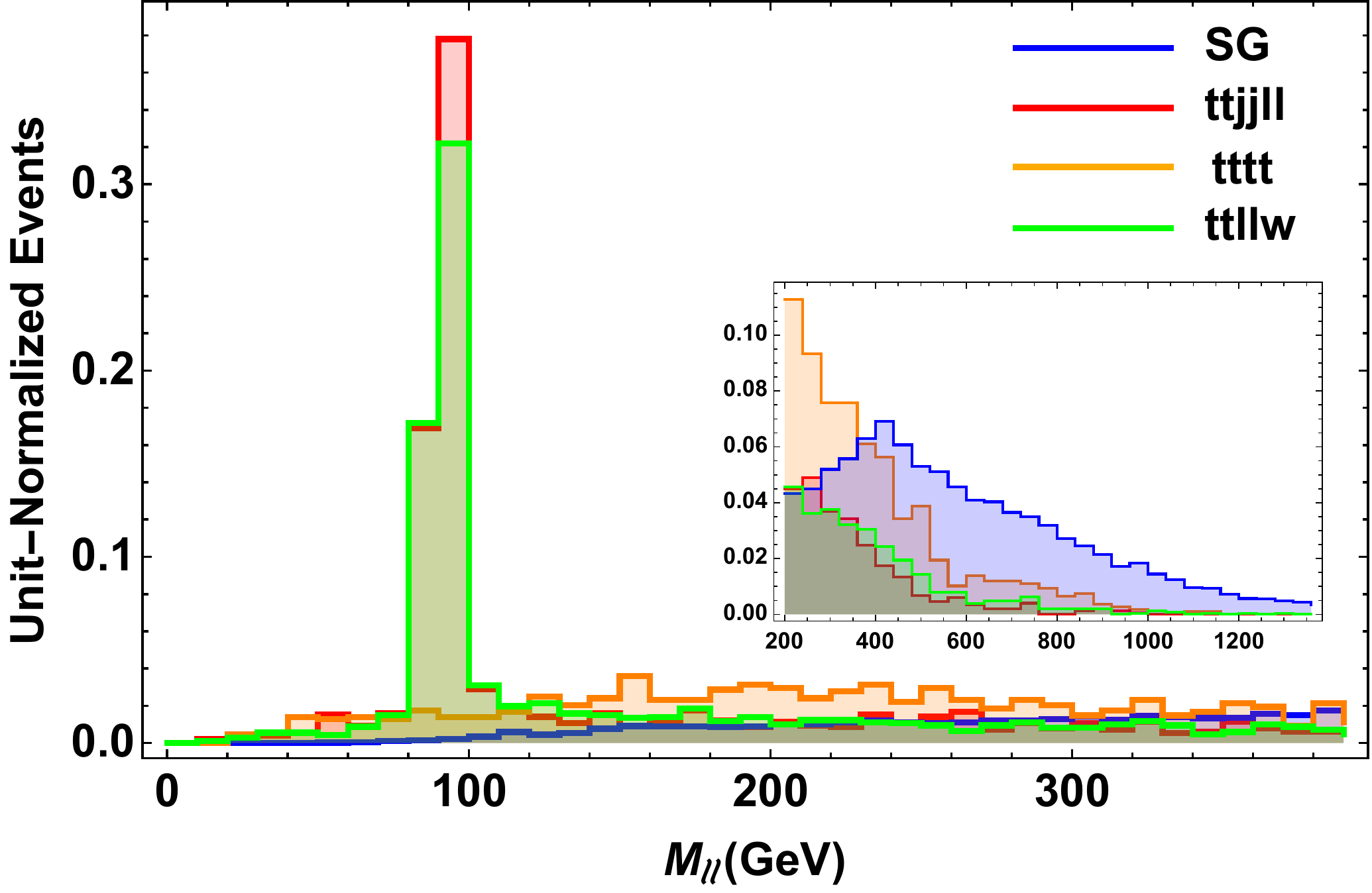}
    \includegraphics[width = 7.5 cm]{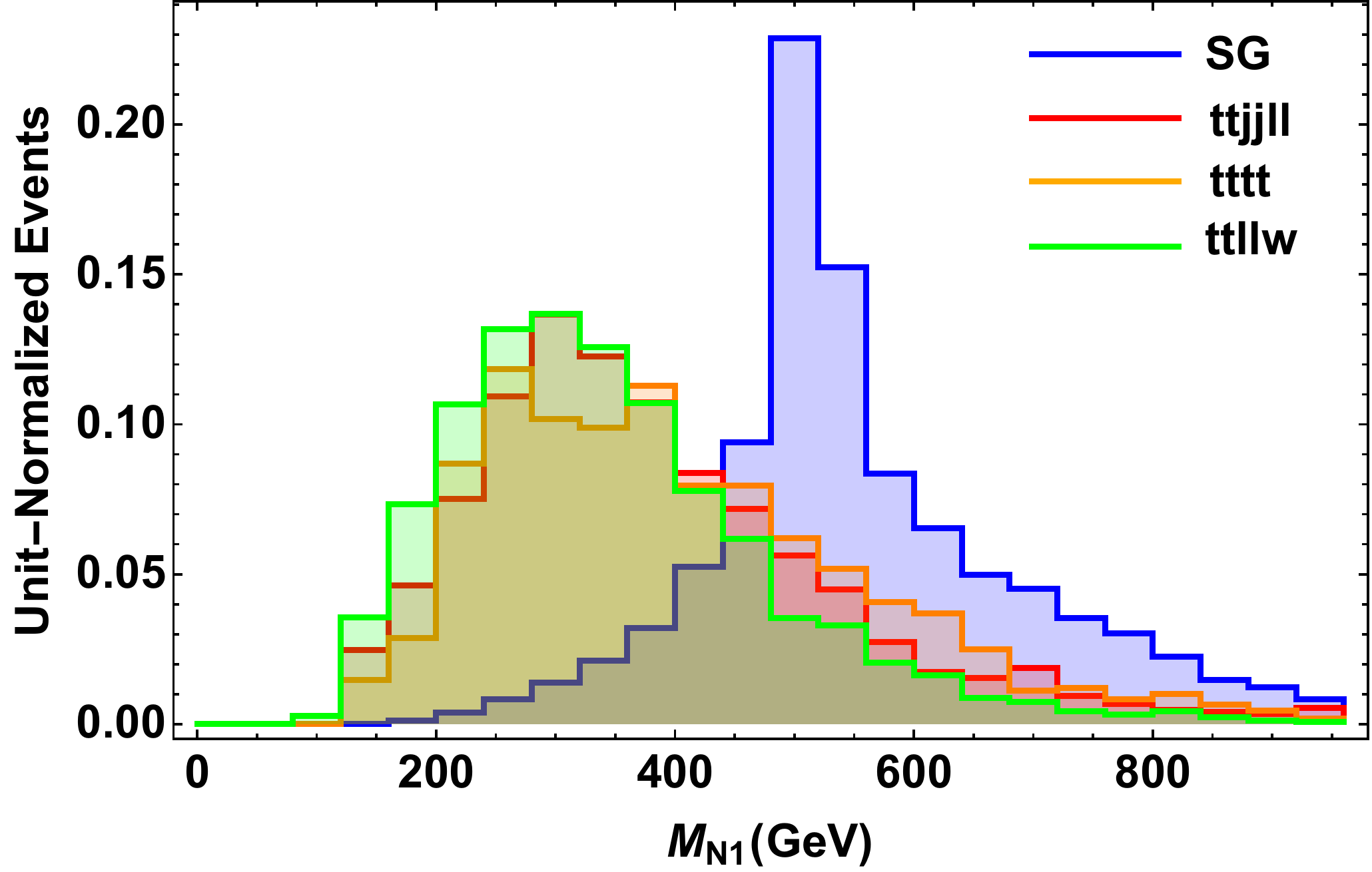}
    \includegraphics[width = 7.5 cm]{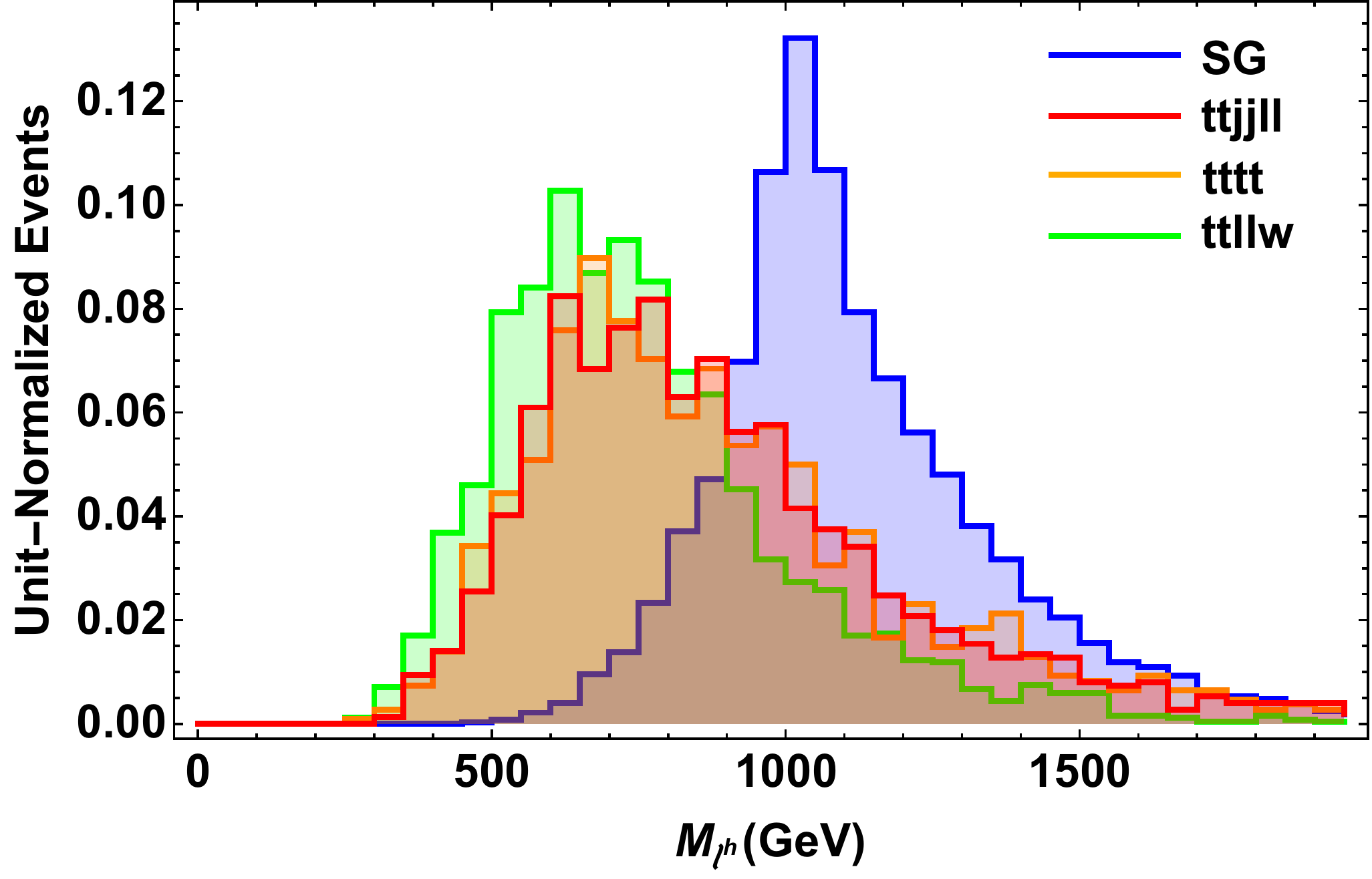}
    \includegraphics[width = 5 cm]{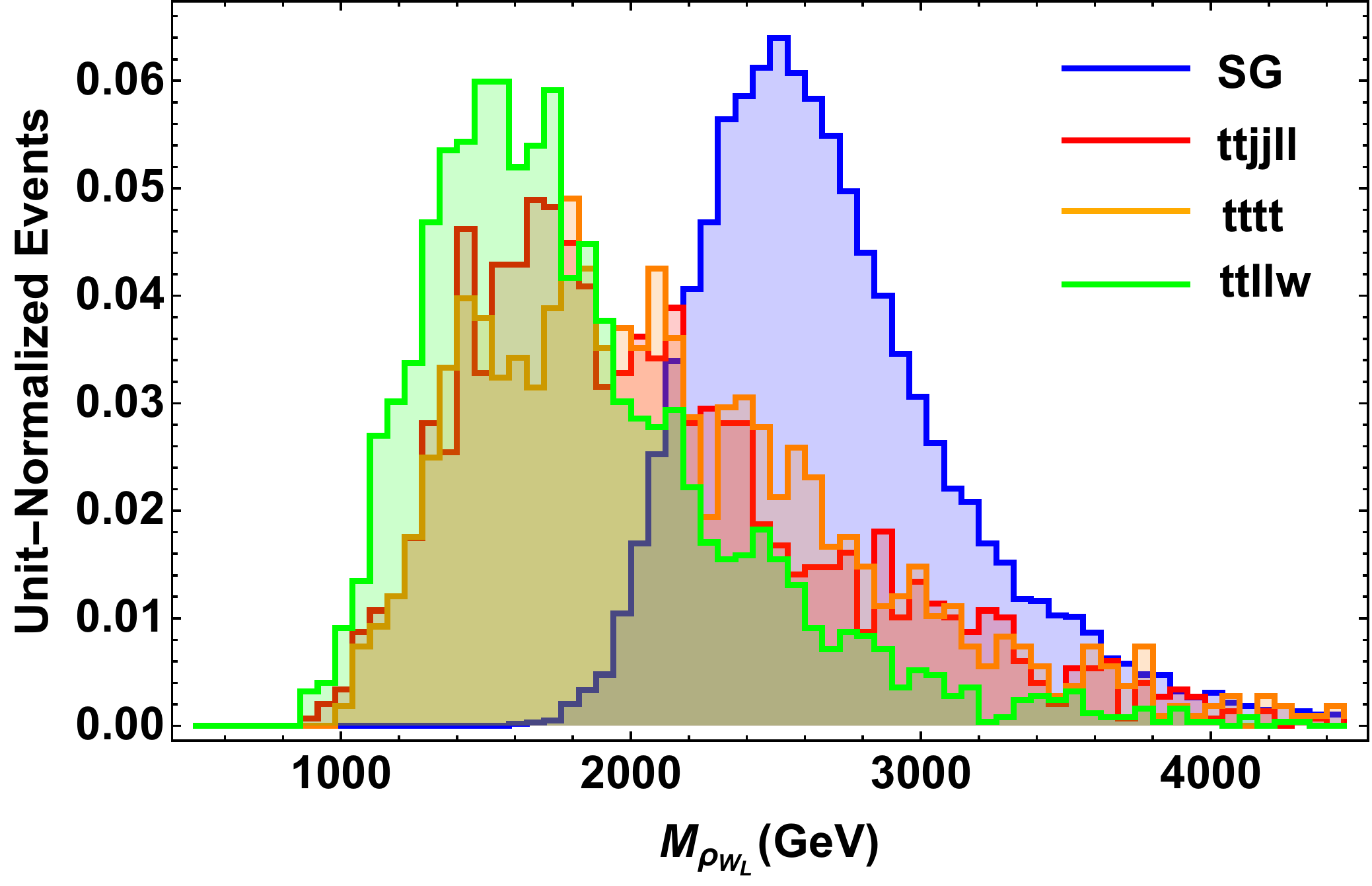}
    \includegraphics[width = 5 cm]{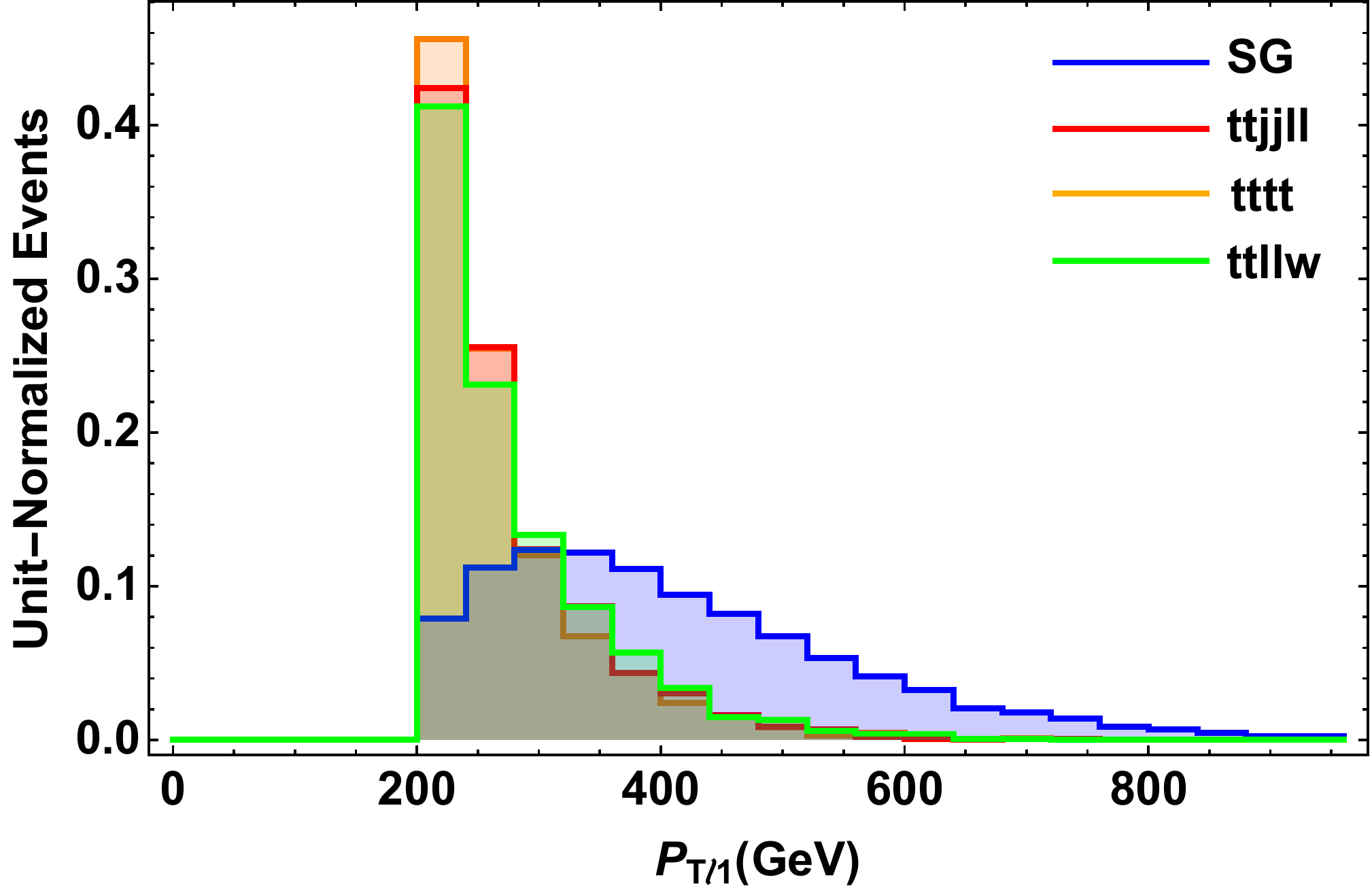}
    \includegraphics[width = 5 cm]{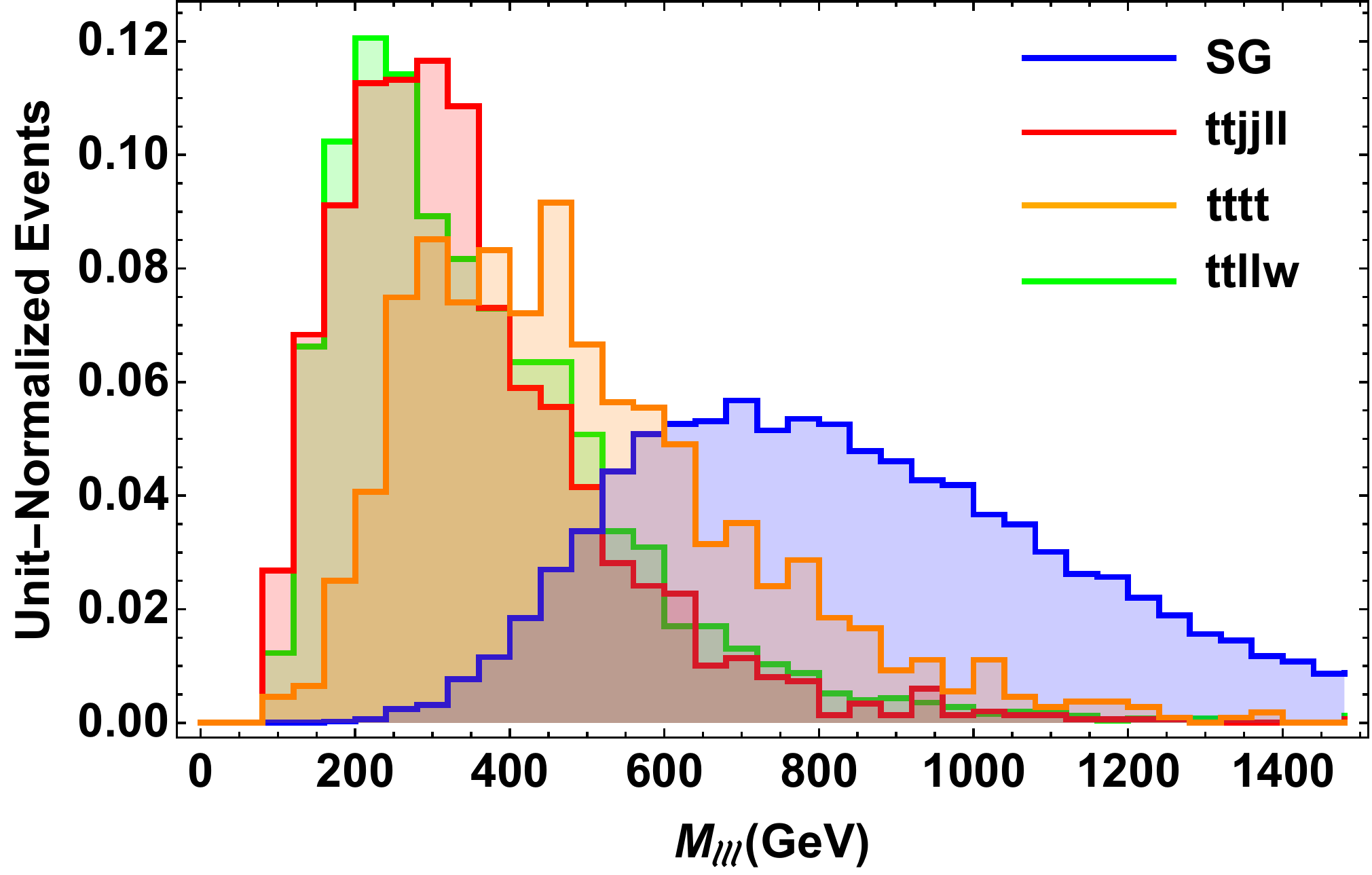}
    \caption{Composite $SU(2)_L$ doublet-Channel: Distributions of variables:
$M_{bb}$ (top row, left), $M_{\ell_\nu \nu}$ (top row, right), $M_{jj_1}$ (second row, left), $M_{\ell \ell}$ (second row, right), $M_{N_1}$ (third row, left), $M_{\ell^h}$ (third row, right), $M_{\rho_{W_L}}$ (bottom row, left), $P_{T \ell_1}$ (bottom row, middle) and $M_{\ell\ell\ell}$ (bottom row, right) for signal (solid blue) and backgrounds (solid, $t\bar{t}jj\ell\ell$-red, $t\bar{t}t\bar{t}$-orange, ${t\bar{t} \ell \ell W}$-green)
}
\label{fig:WL-channel_plots}
\end{figure}
\noindent Fig.~\ref{fig:WL-channel_plots} shows distributions of various variables for signal and backgrounds constructed using the above described procedure for events that pass selection criteria and basic cuts. Remarkably, as seen clearly from signal distributions, our reconstruction prescription for invariant mass distributions is very successful. For example, the distributions of $M_{\ell_\nu \nu}$ (top row, right), $M_{jj_1}$ (second row, left), $M_{N_1}$ (third row, left), $M_{\ell^h}$ (third row, right) and $M_{\rho_{W_L}}$ (bottom row, left) reveal well-developed resonance peak with the position matched well with the input values. Considering the number of intermediate states in the process (see Fig.~\ref{fig:signal}) and high multiplicity of leptons and jets, such outcome is rather surprising. We also present distributions for $M_{bb}$, which is peaked at around $M_H$ for signal, while smoother for backgrounds, $M_{\ell \ell}$ for non-$\ell_\nu$ two leptons, which is sharply peaked at $M_Z$ for {\bf{$\bf{t\bar{t}jj\ell\ell}$}} and {\bf{$\bf{t\bar{t}\ell\ell W}$}}, $M_{\ell\ell\ell}$, which provides yet another very strong cut, and $p_T$ of the hardest lepton. All of these features are as expected. In particular, we see that $M_{\ell \ell}$, $M_{N_1}$, $M_{\ell^h}$, $M_{\rho_{W_L}}$, $M_{\ell\ell\ell}$ and $p_{T\ell_1}$ draw sharp distinctions between signal and backgrounds, providing strong cuts to attain significant backgrounds reduction.

\noindent We performed analysis by applying a series of these kinematic cuts. We provide the cut flows for signal and the major SM backgrounds in Table~\ref{tab:WL_channel}. We find that the composite $SU(2)_L$ doublet-channel may provide a sensitivity to uncover warped seesaw nature in largely model-independent way by $\sim 3.6 \sigma$ with an integrated luminosity of $\mathcal{L} = 3000 \; {\rm fb}^{-1}$. Notice, however, that in this analysis, we have selected events with $W$'s decaying into regular jet pair. Given the heaviness of $\ell^h$ and $N$ from which it is produced, we expect that including boosted $W$ events with fat jet final state will provide a significant increase in the signal rate, and hence larger significance.


\begin{table}[t]
\begin{adjustwidth}{-0.5cm}{}
\centering
\begin{tabular}{|c|c|c|c|c|}
\hline 
Cuts & {\bf{Signal}} & $\bf{t\bar{t}jj \ell \ell}$ & $\bf{t\bar{t}t\bar{t}}$ & {$\bf{t\bar{t}\ell\ell W}$}  \\
\hline \hline
No cuts & $7.87\times10^{-2}$  & 6.86 & $3.45\times10^{-1}$ & $1.65\times10^{-2}$ \\
$N_{\ell}>2$, $N_j>3$, $N_b>1$ with basic cuts & $7.62\times10^{-3}$ & $4.59\times10^{-1}$ & $1.38\times10^{-2}$ & $1.22\times10^{-3}$ \\
$p_{T\ell_1}\geq200$ GeV, $p_{Tj_1}\geq100$ GeV & $6.94\times10^{-3}$ & $6.61\times10^{-2}$ & $1.49\times10^{-3}$ & $1.66\times10^{-4}$ \\
$M_{\ell \ell} \in [110,\,\infty]$ GeV & $6.87\times10^{-3}$ & $2.46\times10^{-2}$ & $1.33\times10^{-3}$ & $7.19\times10^{-5}$ \\
$M_{\rho_{W_L}} \in [2000,\,\infty]$ GeV & $6.73\times10^{-3}$ & $1.14\times10^{-2}$ & $7.11\times10^{-4}$ & $2.00\times10^{-5}$ \\
$M_{\ell\ell\ell} \in [450,\,\infty]$ GeV & $6.42\times10^{-3}$ & $3.33\times10^{-3}$ & $4.72\times10^{-4}$ & $9.04\times10^{-6}$  \\	
$M_{N_1} \in [400,\,\infty]$ GeV & $5.97\times10^{-3}$ & $2.17\times10^{-3}$ & $3.13\times10^{-4}$ & $5.08\times10^{-6}$  \\		
\hline
$S/B$ & 2.40 & -- & -- & -- \\
$S/\sqrt{S+B}$ ($\mathcal{L}=300$ fb$^{-1}$) & 1.12 & -- & -- & -- \\
$S/\sqrt{S+B}$ ($\mathcal{L}=3000$ fb$^{-1}$) & 3.56 & -- & -- & -- \\
\hline
\end{tabular}
\caption{Cut flows for signal and major background events in terms their cross sections. The cross sections are in fb. The numbers in the first row (``No cuts'') are cross sections obtained with basic cuts at the generation level to avoid divergence (for both signal and backgrounds). In the second row, the same basic cuts are reimposed, in addition to $\Delta R > 0.4$ for all selected object pairs, to both signal and background events along with multiplicity requirements for b-jet, non-b-jet and leptons.  
\label{tab:WL_channel} }
\end{adjustwidth}
\end{table}


\section{Conclusions and Outlook}
\label{conclude}


In \cite{Agashe:2015izu}, we argued that (i) warped seesaw is a {\em natural} implementation of the essence of the original seesaw paradigm {\em and} that (ii) composite {\em TeV}-mass singlet neutrinos play a crucial role in SM neutrino mass generation. Then, in a previous paper on LHC signals \cite{Agashe:2016ttz}, we studied production of these singlet neutrinos using mechanisms analogous to 4D LR models, but with important and interesting differences. In this paper, we considered production of singlet neutrinos from decays of particles {\em beyond} 4D LR models. 

In the earlier work \cite{Agashe:2016ttz}, we considered models where the {\em composite} sector has an extended EW {\em global} symmetry, $SU(2)_L \times SU(2)_R \times U(1)_X$. Such a left-right symmetric structure is motivated in the context of composite Higgs or 5D warped framework by consistency with EW precision tests. In addition, we assumed degeneracy of masses of composite gauge bosons. The couplings of the light quarks to composite $W^{ \pm }_R$ and $Z^{ \prime }$ required for their production, and hence that of the singlet neutrino via their decay, then were established in part by EWSB induced mixing between composite $W^{ \pm }_L$ and $W^{ \pm }_R$ (and similarly for neutral gauge bosons); mass degeneracy was crucial for enhancing the size of this mixing. This effect has to be combined with the elementary-composite gauge boson mixing. In this case, composite $W^\pm_R$ couples to {\em left}-handed quarks. This is to be contrasted with the usual 4D LR models, where $W^{\pm}_R$ \emph{directly} couples to {\em right}-handed quarks. For more detail, see \cite{Agashe:2016ttz}.

In this paper, we show that taking the same model (with $X = \frac{1}{2}( B - L )$) and yet exploring different regions of parameter space can result in different production channels for singlet neutrino with remarkable {\em qualitative} differences. In the first part of this paper, as a more general consideration, we choose composite $W_R^3$ and $(B - L)$ to be {\em non}-degenerate. At first sight, such non-degeneracy seems to suppress $Z^{ \prime }$ signal. Remarkably, however, there is actually an emergence of another new neutral channel signal. The point is that composite $W_R^3$ and $(B - L)$ are now {\em separately} mass eigenstates, i.e. their ``hypercharge'' and $Z^{ \prime }$ combinations appearing in the degenerate case are nowhere close to being mass eigenstates. Both these spin-1 composites, $W^3_R$ and $(B - L)$, mix with elementary hypercharge. This effect suffices to couple them to light quarks, with{\em out} need of EWSB in addition. This is to be contrasted with EWSB mixing {\em also} being required for coupling composite $W^{ \pm }_R$/$Z^{ \prime }$ to light quarks in this model. In fact, in this sense, the status of composite $W^3_R$ and $(B-L)$ in the non-degenerate case is rather similar to composite $W_L$'s. Moreover, the singlet neutrino couples to both $W_R^3$ and $(B-L)$. So, provided the lighter of the two is light enough, it can be produced at the LHC with a significant rate, and thus, producing the singlet neutrino through its decay. 
We show that $4.8 \sigma$ signal can be achieved with 3000 fb$^{ -1 }$ luminosity for 2 TeV composite $(B - L)$ gauge boson and 750 GeV singlet neutrino.

It is worth emphasizing a distinction from 4D LR here: there is {\em no} analogue in that model of $W_R^3$ and $(B - L)$ as separate mass eigenstates because only their heavy combination, $Z^{ \prime }$, corresponding to the broken gauge symmetry, is in play, orthogonal one being SM hypercharge. Even if we focus only on the lightest state in the composite/5D case above, note that its couplings are different from that of $Z^{ \prime }$ (whether 4D LR model or the degenerate case in 5D), allowing disambiguation between the two. We studied this interesting channel here, taking as an illustration the case of a composite $(B - L)$ being lighter than $W^3_R$.

Then, in the second part of this paper, we demonstrated the possibility of obtaining a signal from singlet neutrino which is {\em independent} of gauge couplings of singlet neutrino, e.g. $W_R$ and $(B-L)$ are too heavy to be relevant or neutrino is a singlet of $SU(2)_R \times U(1)_X$ also. Namely, decays of composite $SU(2)_L$ {\em doublet} leptons can produce the composite singlet neutrino, accompanied by Higgs boson/longitudinal $W/Z$. Note that the associated coupling originates purely in the composite sector, thus is sizable. It is also robust feature of this model in the sense that the {\em same} coupling is also involved in the Dirac mass term part of the neutrino mass seesaw. At the same time, the composite doublet leptons are ``guaranteed'' to have significant couplings to light quarks and thus substantial production rate at the LHC, via elementary and composite {\em left}-handed $W$. On the other hand, for the singlet neutrino signal either from decays of $W_R^3/X$ as studied in the first part of this paper, or from decays of $Z^{ \prime }/ W_R^{ \pm }$ studied in the previous paper, we need certain representations of the SM singlet neutrino under the extended EW symmetry. In addition, degeneracy of spin-1 states was invoked in the earlier work for the purpose of obtaining a significant coupling of light quarks to $Z^{ \prime }/ W_R^{ \pm }$ for their production. 
We show that $\sim 4 \sigma$ signal can be achieved with 3000 fb$^{-1}$ luminosity for the following spectrum: 2.5 TeV composite $\rho_{W_L}$, $1$ TeV composite $SU(2)_L$ doublet lepton and $500$ GeV singlet neutrino.

Even though we studied singlet neutrino production via particles not featured in the usual LR models, either $W_R^3$ or $(B - L)$ or composite lepton doublet, the resulting signals from these have appreciable similarities to the standard ones from usual LR $W_R^{ \pm }/Z^{ \prime }$. Specifically, we get multi-leptons, in association with SM gauge bosons generically. However, there do exist important differences as well. In particular, we do get extra Higgs/$Z$ in pair-production of singlet neutrinos from decay of $SU(2)_L$ doublet composite leptons as compared to $Z^{ \prime }$. 
The different detailed topologies involved also imply that the kinematic distributions of common part of the final state, say, leptons, can also be rather distinct.

In fact, in part of the parameter space, the situation might be even better as follows. For example, in the case of composite $(B - L)$ being lighter than {\em all} others, {\em merely} comparing the {\em other} decay channels of this spin-1 state {\em and} the production/decay of other heavy {\em resonances} can readily reveal nature of the underlying model. In other words, we are concerned here with more qualitative differences between the signals for the various models, which can afford model characterization even before any detailed analyses are available (for example, related to extra bosons in final state mentioned above). This is because there is no charged channel at all to be seen at the LHC in this example. Furthermore, it is interesting that there are {\em no} diboson decays for the composite $(B - L)$. Both these features are in sharp contrast to 4D LR models, where $Z^{ \prime }$ (neutral channel) obviously has a charged counterpart, i.e. $W^{ \pm }_R$. In fact, $W^{ \pm }_R$ is typically easier to discover because it is lighter. Moreover, both $Z^{ \prime }$ and $W^{ \pm }_R$ decay to dibosons with similar rate as singlet neutrinos. Once again, such dramatic differences in the signal afford {\em discrimination} between the two models essentially {\em right at the time of discovery}.

Relatedly, even within the context of the composite/5D model the singlet neutrino decay channel can have significant implications for discovery of EW spin-1 composites. Regardless of the seesaw model, top/EW gauge boson/Higgs are the usual discovery channel for such composite EW gauge bosons. However, diboson is not available for composite $(B-L)$ as already mentioned above. Composite $(B-L)$ does decay to top quarks, but that might be {\em diluted} by the new, singlet neutrino channel. Amazingly then, for this ``new'' particle, the decay into singlet neutrino might instead be the way to go even for {\em discovery}. Whereas, in the earlier case of degenerate composites, $W^{ \pm }_R$ or $Z^{ \prime }$, diboson or top channels were available for discovery, and the singlet neutrino channel was more for {\em testing} seesaw itself.

Finally, a word about possible future work related to this framework. One direction would be to explore LHC signals in other regions of parameter space, for example, the (more challenging) case of heavier composite singlet/doublet neutrinos/charged leptons such that spin-1 composites can{\em not} decay into pair of them. Also, the idea of a 100 TeV hadron collider is being discussed a lot: it will be interesting to determine the reach of this proposed collider as far as the singlet neutrino is concerned.

It is worth noting that there are {\em beyond} high-energy collider phenomenological aspects to TeV-mass singlets. For example, we plan to study leptogenesis in this set-up, which could occur at $O(\textrm{TeV})$ temperatures, cf.~at super-high scales for original seesaw. In addition, we have constraints from lepton flavor-violation arising from {\em virtual} effects of $\lesssim$ TeV-mass singlet neutrino/composite charged leptons. We assumed some sort of flavor symmetries here in order to be consistent with those precision tests. It might be worthwhile exploring these in more detail. The bottomline is that our work has opened up new avenues for studies of the {\em natural} seesaw idea for neutrino mass within the framework of a warped extra dimension/composite Higgs, whether at the LHC or other non-(high-energy) collider areas.


\section*{Acknowledgements}

%
%

%

We would like to thank Chien-Yi Chen, Roberto Contino, Bhupal Dev,  Shrihari Gopalakrishna, Doojin Kim and Rabindra Mohapatra for discussions and David Curtin and Jack Collins for help with simulations. 
This work was supported in part by NSF Grant No.~PHY-1620074 and the Maryland Center for Fundamental Physics. 

\appendix


\section{One Elementary-Two Composite Gauge Boson Mixing}
\label{non-deg_mix}

As mentioned in the main text, in the two-site basis, we have elementary hypercharge gauge boson (denoted by $B_{ \rm elem }$) mixing with 
composite $X$ (labelled $\tilde{ \rho }_X$) and composite $W_R^3$ (called $\tilde{ \rho }_{ W_R^3}$).
Their gauge couplings (to other particles in the corresponding sector {\em only}) 
are $g_{ \rm elem }$, $g_{ \star }$ and $G_{ \star }$ respectively.
The diagonal mass terms for $\tilde{ \rho }_X$ and $\tilde{ \rho }_{ W_R^3}$ are denoted by $m_{ \star }$ and $M_{ \star }$.
In addition, there are mass mixing terms between elementary and composites, along with the appropriate diagonal mass term
for elementary gauge boson.
Note that, even {\em before} EWSB, these are then the gauge/weak eigenstates, i.e. not mass eigenstates. 

The mass matrix has the form: 
\bea
M^2 & = & 
\left( 
\begin{array}{ccc}
\frac{ g_{ \rm elem }^2 }{ g_{ \star }^2 } m_{ \star }^2 + \frac{ g_{ \rm elem }^2 }{ G_{ \star }^2 } M_{ \star }^2 & - \frac{ g_{ \rm elem } }{ g_{ \star } } m_{ \star }^2 &  - \frac{ g_{ \rm elem } }{ G_{ \star } } M_{ \star }^2 \\
- \frac{ g_{ \rm elem } }{ g_{ \star } } m_{ \star }^2 & m_{ \star }^2 & 0 \\
 - \frac{ g_{ \rm elem } }{ G_{ \star } } M_{ \star }^2 & 0 & M_{ \star }^2
\end{array}
\right)
\eea
We can check that the above matrix has vanishing determinant, which is as expected based on it being made up of the ``usual'' two elementary-composite $2 \times 2$ blocks.  This results in one field with zero eigenvalue, corresponding to the SM hypercharge gauge boson.

This matrix is diagonalized by the rotation\footnote{in general, this would be unitary, for example, for mass matrix of charged gauge bosons
so that we keep that notation.}
\bea
U^{ \dagger } M^2 U & = & M^2_{ \rm diag }
\eea
where we have a product of three effectively $2 \times 2$ rotations:
\bea
U & = & U_{ 12 } U_{ 13 } U_{ 23 } 
\eea
with
\bea
U_{ 12 } & = & %
\left( 
\begin{array}{ccc}
c & -s & 0 \\
s & c & 0 \\
0 & 0 & 1 
\end{array}
\right)
\\
U_{ 13 } & = & 
\left( 
\begin{array}{ccc}
C & 0 & - S \\
0 & 1 & 0  \\
S & 0 &  C
\end{array}
\right)
\\
U_{ 23 } & = & 
\left( 
\begin{array}{ccc}
1 & 0 & 0 \\
0 & c_{ \star } & - s_{ \star }  \\
0 & s_{ \star } & c_{ \star }
\end{array}
\right)
\eea
Here, the 
sines of the first two mixing angles are given by 
\bea
s & = & \frac{ g_{ \rm elem } }{ \sqrt{ g_{ \rm elem }^2 + g_{ \star }^2 } } 
\\
S & = & \frac{ g_{ \rm elem } c }{ \sqrt{ g_{ \rm elem }^2 c^2  + G_{ \star }^2 } } 
\eea
and $c$ and $C$ are the corresponding cosines:
\bea
c & = & \sqrt{ 1 - s^2 } \left( = \frac{ g_{ \star } }{ \sqrt{ g_{ \rm elem }^2 + g_{ \star }^2 } } \right)
\\
C & = & \sqrt{ 1 - S^2 }  \left( = \frac{ G_{ \star } }{ \sqrt{ g_{ \rm elem }^2 c^2 + G_{ \star }^2 } } \right)
\eea
Finally,  $c_{ \star }$ and $s_{ \star }$ are cosine/sine of last/2-3 rotation angle ($\theta_{ \star })$:
\bea
\tan 2 \theta_{ \star } & = & \frac{ - 2 M_{ \star }^2 \frac{ g_{ \rm elem } }{ G_{ \star } }  \;  s \left( C + c \; S \frac{ g_{ \rm elem } }{ G_{ \star } } \right) }{ M_{ \star }^2 \left( 1 + c^2 \frac{ g_{ \rm elem }^2  }{ G_{ \star }^2 } - s^2 \frac{ g_{ \rm elem }^2 }{ G_{ \star }^2} \right) 
- m_{ \star }^2 \left( 1 + \frac{ g_{ \rm elem }^2 }{ g_{ \star }^2 } \right)  }
\eea
The relation between the mass (denoted by $B^{ (0) }$, $\rho_X$
and $\rho_{ W_R^3 }$\footnote{These models can be identified with zero-mode and KK models of 5D model. This choice of notation for the heavy mass eigenstates will be explained below.}) and weak/gauge eigenstate bases is given by 
\bea
\left( 
\begin{array}{c}
B^{ (0) } \\
\rho_{X} \\
\rho_{ W_R^3 }
\end{array}
\right)
& = & U_{ 23 }^{ \dagger } U_{ 13 }^{ \dagger } U_{ 12 }^{ \dagger } 
\left( 
\begin{array}{c}
B_{ \rm elem }\\
\tilde{\rho}_{X} \\
\tilde{\rho}_{ W^3_R }
\end{array}
\right)
\eea
so that we have the massless and two heavy eigenstates:
\bea
B^{ (0) } & = & c \; C \;  B_{ \rm elem } + C \; s \;  \tilde{ \rho }_X + S \;  \tilde{ \rho }_{ W_R^3 } \\
\rho_{X} & = & - \left( s \;  c_{ \star }  + s_{ \star } \;  S \;  c \right) B_{ \rm elem } + \left( c \;  c_{ \star } - s_{ \star } \;  S \;  s \right) 
\tilde{ \rho }_X 
+ s_{ \star } C \tilde{ \rho }_{ W_R^3 } \\
\rho_{ W_R^3 } & = & \left( s_{ \star } \;  s - c_{ \star } \;  S \;  c \right) B_{ \rm elem } - \left( s_{ \star } \;  c + c_{ \star } \;  S \;  s \right) \tilde{ \rho }_X + c_{ \star } \;  C \; \tilde{ \rho }_{ W_R^3 }
\label{mass-weak}
\eea
%

%
%
%
%

The elementary fermions have a charge $Q_Y$ under $B^{ \rm elem }$  only
so that they 
couple to $B^{ (0) }$ with strength $Q_Y g_{ \rm elem } c C$.
Thus, if $g_Y$ denotes the SM hypercharge gauge coupling, we can identify
\bea
g_Y& = & c C g_{ \rm elem } \nonumber \\
& \approx & g_{ \rm elem }, \; \hbox{assuming} \; g_{ \rm elem } \ll g_{ \star}, G_{ \star }
\eea
%
Whereas, composite fermions couple only to $\tilde{ \rho }_X$ and $\tilde{ \rho }_{ W_R^3 }$ with charges denoted 
by $Q_X$ and $Q_{W_R^3}$
respectively. One can check that their coupling to $B^{ (0) }$ is given by $g_Y \left( Q_{ W^3_R } + Q_ X \right)$, where
we have
%
%
$Q_Y=Q_{W^3_R}+Q_X$
%
In particular, in the first model that we study, we identify $X =\frac{1}{2} ( B - L )$, with the singlet neutrino $N^{ (1) }$ having $Q_{W^3_R}=\frac{1}{2}$
 and
$Q_X = -\frac{1}{2}$ (thus $Q_Y = 0$).

Assuming 
\bea
M_{ \star } & \gg & m_{ \star } \\
g_{ \rm elem } & \ll & g_{ \star }, \; G_{ \star } \; \hbox{i.e.}, \; s, \; S  \ll 1 
\eea
we get
\bea
s_{ \star } & \approx & - s S  \left( \approx \frac{ g_{ \rm elem } }{ g_{ \star } } \frac{ g_{ \rm elem } }{ G_{ \star } } \right) 
\eea
i.e. in this case, composite $W-X$ mixing angle is given roughly by {\em product} of $B_{ \rm elem }-\tilde{ \rho }_X$ and $B_{ \rm elem } -\tilde{ \rho }_{ W_R^3 }$ mixing angles
({\em second} order in elementary-composite mixing, thus negligible for purpose of LHC signals).
Actually, 
\bea
M_{ \star } ^2 - m_{ \star }^2 & \sim & O \left( M_{ \star }^2 \right), \; \rm for \; example \\
M_{ \star } & \sim & 2 m_{ \star }
\eea
suffices to give the above negligibly small size of $W-X$ mixing angle, i.e.
\bea
s_{ \star } & \sim & s S  \left( \sim \frac{ g_{ \rm elem }}{ g_{ \star } } \frac{ g_{ \rm elem } }{ G_{ \star } } \right) 
\eea
Similarly, in the opposite limit, composite $(B-L)$ is heavier than $W_R^3$, we find that the composite $X-W$ mixing angle is doubly-suppressed as above.

Thus, {\em generically},  the two composite masses are $ O(1)$ different, 
we see that the two heavy mass eigenstates ($\rho_{ W_R^3 }$ and $\rho_X$ in Eq.~(\ref{mass-weak})) are to a good approximation the same as the
weak basis, i.e., composite $W$ and $(B-L)$ ,which, a posteriori, explains the labelling of the 2 massive eigenstates in Eq.~(\ref{mass-weak}). The leading deviation from this identification stemming from mixing with {\em elementary}
hypercharge, with composite-composite mixing being even smaller.

Therefore,
the couplings of $\rho_X$ (whether it is the lighter or heavier massive eigenstate) to $N^{ (1) }$ and light quarks, which to a very good approximation only couple to $B_{ \rm elem }$,
%
%
are given by
$\approx -\frac{1}{2} g_{ \star }$ and $- Q_Y \frac{g^2_Y}{g_{ \star }}$ respectively.
Similarly, couplings of $\rho_{ W_R^3 }$ to $N^{ (1) }$ and light quarks are given by
$\approx \frac{1}{2} G_{ \star }$ and $-Q_Y \frac{g^2_Y}{G_{ \star }}$ respectively.
Of  course, for the study of the LHC signals, it suffices to keep only the lighter of these two states,
since that production will dominate.

Whereas in the special case of 
{\em degeneracy}, we can show 
%
%
that the {\em product} of couplings of light quark and singlet neutrino to {\em each} of the above 
{\em heavy} mass eigenstates vanishes. This is 
as expected, since one of them corresponds to the ``heavy'' (or KK of the 5D model) hypercharge which decouples from singlet neutrino, whereas
the other one is composite (or KK of 5D model ) $Z^{ \prime }$ which decouples from light quarks instead.



\begin{thebibliography}{99}

\bibitem{Agashe:2015izu} 
  K.~Agashe, S.~Hong and L.~Vecchi,
  Phys.\ Rev.\ D {\bf 94}, no. 1, 013001 (2016)
  doi:10.1103/PhysRevD.94.013001
  [arXiv:1512.06742 [hep-ph]].


\bibitem{inverse}
  R.~N.~Mohapatra,
  Phys.\ Rev.\ Lett.\  {\bf 56}, 561 (1986);
%
  R.~N.~Mohapatra and J.~W.~F.~Valle,
  Phys.\ Rev.\  D {\bf 34}, 1642 (1986).
  %
  For further model-building/studies, see, for example, 
  P.~S.~B.~Dev and R.~N.~Mohapatra,
  Phys.\ Rev.\ D {\bf 81}, 013001 (2010)
  doi:10.1103/PhysRevD.81.013001
  [arXiv:0910.3924 [hep-ph]];
  %
  A.~Abada and M.~Lucente,
  Nucl.\ Phys.\ B {\bf 885}, 651 (2014)
  doi:10.1016/j.nuclphysb.2014.06.003
  [arXiv:1401.1507 [hep-ph]].



\bibitem{original}
P.~Minkowski,
{\em Phys. Lett.} {\bf B67} (1977) 421;
T.~Yanagida in {\em Workshop on Unified Theories, KEK Report
79-18}, p.~95,
 1979.
M.~Gell-Mann, P.~Ramond and R.~Slansky, {\em Supergravity},
p.~315.
\newblock Amsterdam: North Holland, 1979;
S.~L. Glashow, {\em 1979 Cargese Summer Institute on Quarks and
Leptons},
 p.~687.
\newblock New York: Plenum, 1980;
R.~N. Mohapatra and G.~Senjanovic,
 {\em Phys. Rev. Lett.} {\bf 44}, 912
 (1980).


\bibitem{Huber:2003sf}
  S.~J.~Huber, Q.~Shafi,
  Phys.\ Lett.\  {\bf B583}, 293-303 (2004).
  [hep-ph/0309252]; 
%
  C.~Csaki, C.~Grojean, J.~Hubisz, Y.~Shirman and J.~Terning,
  Phys.\ Rev.\ D {\bf 70}, 015012 (2004)
  [hep-ph/0310355].
%
  G.~Perez and L.~Randall,
  JHEP {\bf 0901}, 077 (2009)
  [arXiv:0805.4652 [hep-ph]];
%
  C.~Csaki, C.~Delaunay, C.~Grojean and Y.~Grossman,
  JHEP {\bf 0810}, 055 (2008)
  [arXiv:0806.0356 [hep-ph]];
  %
  M.~Carena, A.~D.~Medina, N.~R.~Shah and C.~E.~M.~Wagner,
  Phys.\ Rev.\ D {\bf 79}, 096010 (2009)
  [arXiv:0901.0609 [hep-ph]].


\bibitem{Agashe:2016ttz} 
  K.~Agashe, P.~Du and S.~Hong,
  arXiv:1612.04810 [hep-ph].




\bibitem{Agashe:2003zs} 
  K.~Agashe, A.~Delgado, M.~J.~May and R.~Sundrum,
  JHEP {\bf 0308}, 050 (2003)
  doi:10.1088/1126-6708/2003/08/050
  [hep-ph/0308036].
  
\bibitem{Mohapatra:2016twe} 
  R.~N.~Mohapatra,
  Nucl.\ Phys.\ B {\bf 908}, 423 (2016).
  doi:10.1016/j.nuclphysb.2016.03.006


\bibitem{Agashe:2006at} 
  K.~Agashe, R.~Contino, L.~Da Rold and A.~Pomarol,
  Phys.\ Lett.\ B {\bf 641}, 62 (2006)
  doi:10.1016/j.physletb.2006.08.005
  [hep-ph/0605341].


\bibitem{Das:2015toa} 
For production of singlet neutrinos via their mixing with doublet ones, see, for example,
A.~Das and N.~Okada,
  Phys.\ Rev.\ D {\bf 93}, no. 3, 033003 (2016)
  doi:10.1103/PhysRevD.93.033003
  [arXiv:1510.04790 [hep-ph]]; 
  %
  A.~Das, P.~S.~Bhupal Dev and N.~Okada,
  Phys.\ Lett.\ B {\bf 735}, 364 (2014)
  doi:10.1016/j.physletb.2014.06.058
  [arXiv:1405.0177 [hep-ph]];
  A.~Das and N.~Okada,
  Phys.\ Rev.\ D {\bf 88}, 113001 (2013)
  doi:10.1103/PhysRevD.88.113001
  [arXiv:1207.3734 [hep-ph]];
  %
  T.~Saito {\it et al.},
  Phys.\ Rev.\ D {\bf 82}, 093004 (2010)
doi:10.1103/PhysRevD.82.093004
  [arXiv:1008.2257 [hep-ph]];
%
  N.~Haba, S.~Matsumoto and K.~Yoshioka,
  Phys.\ Lett.\ B {\bf 677}, 291 (2009)
  doi:10.1016/j.physletb.2009.05.042
  [arXiv:0901.4596 [hep-ph]].
  For NLO corrections, see, for example,
  A.~Das, P.~Konar and S.~Majhi,
  JHEP {\bf 1606}, 019 (2016)
  doi:10.1007/JHEP06(2016)019
  [arXiv:1604.00608 [hep-ph]].




\bibitem{Contino:2006nn} 
  R.~Contino, T.~Kramer, M.~Son and R.~Sundrum,
  JHEP {\bf 0705}, 074 (2007)
  doi:10.1088/1126-6708/2007/05/074
  [hep-ph/0612180]; 
%
  D.~Pappadopulo, A.~Thamm, R.~Torre and A.~Wulzer,
  JHEP {\bf 1409}, 060 (2014)
  doi:10.1007/JHEP09(2014)060
  [arXiv:1402.4431 [hep-ph]];
  M.~Low, A.~Tesi and L.~T.~Wang,
  Phys.\ Rev.\ D {\bf 92}, no. 8, 085019 (2015)
  doi:10.1103/PhysRevD.92.085019
  [arXiv:1507.07557 [hep-ph]].

 

\bibitem{Carena:2002dz} 
For warped models, see, for example, 
  M.~Carena, E.~Ponton, T.~M.~P.~Tait and C.~E.~M.~Wagner,
  Phys.\ Rev.\ D {\bf 67}, 096006 (2003)
  doi:10.1103/PhysRevD.67.096006
  [hep-ph/0212307]; 
  %
  H.~Davoudiasl, J.~L.~Hewett and T.~G.~Rizzo,
  Phys.\ Rev.\ D {\bf 68}, 045002 (2003)
  doi:10.1103/PhysRevD.68.045002
  [hep-ph/0212279].


\bibitem{Agashe:2016rle} 
  K.~Agashe, P.~Du, S.~Hong and R.~Sundrum,
  arXiv:1608.00526 [hep-ph].




\bibitem{Carena:2004zn} 
For warped models, see, for example, 
  M.~Carena, A.~Delgado, E.~Ponton, T.~M.~P.~Tait and C.~E.~M.~Wagner,
  Phys.\ Rev.\ D {\bf 71}, 015010 (2005)
  doi:10.1103/PhysRevD.71.015010
  [hep-ph/0410344].



\bibitem{Agashe:2005vg} 
 See, for example,
  K.~Agashe, R.~Contino and R.~Sundrum,
  Phys.\ Rev.\ Lett.\  {\bf 95}, 171804 (2005)
  doi:10.1103/PhysRevLett.95.171804
  [hep-ph/0502222].
  K.~Agashe, A.~Delgado and R.~Sundrum,
  Annals Phys.\  {\bf 304}, 145 (2003)
  doi:10.1016/S0003-4916(03)00013-7
  [hep-ph/0212028].


  
\bibitem{delAguila:2008pw} 
  F.~del Aguila, J.~de Blas and M.~Perez-Victoria,
  Phys.\ Rev.\ D {\bf 78}, 013010 (2008)
  doi:10.1103/PhysRevD.78.013010
  [arXiv:0803.4008 [hep-ph]].
  A.~Atre, T.~Han, S.~Pascoli and B.~Zhang,
  JHEP {\bf 0905}, 030 (2009)
  doi:10.1088/1126-6708/2009/05/030
  [arXiv:0901.3589 [hep-ph]].


\bibitem{ATLAS:2016cwq} 
  The ATLAS collaboration [ATLAS Collaboration],
  ATLAS-CONF-2016-062.
  M.~Aaboud {\it et al.} [ATLAS Collaboration],
  Phys.\ Lett.\ B {\bf 765}, 32 (2017)
  doi:10.1016/j.physletb.2016.11.045
  [arXiv:1607.05621 [hep-ex]].



  
\bibitem{Alloul:2013bka} 
  A.~Alloul, N.~D.~Christensen, C.~Degrande, C.~Duhr and B.~Fuks,
  Comput.\ Phys.\ Commun.\  {\bf 185}, 2250 (2014)
  doi:10.1016/j.cpc.2014.04.012
  [arXiv:1310.1921 [hep-ph]].
  
\bibitem{Pappadopulo:2014qza} 
  D.~Pappadopulo, A.~Thamm, R.~Torre and A.~Wulzer,
  JHEP {\bf 1409}, 060 (2014)
  doi:10.1007/JHEP09(2014)060
  [arXiv:1402.4431 [hep-ph]].


\bibitem{Alwall:2014hca} 
  J.~Alwall {\it et al.},
  JHEP {\bf 1407}, 079 (2014)
  doi:10.1007/JHEP07(2014)079
  [arXiv:1405.0301 [hep-ph]].
  
\bibitem{Ball:2012cx} 
  R.~D.~Ball {\it et al.},
  Nucl.\ Phys.\ B {\bf 867}, 244 (2013)
  doi:10.1016/j.nuclphysb.2012.10.003
  [arXiv:1207.1303 [hep-ph]].
  
\bibitem{Sjostrand:2006za} 
  T.~Sjostrand, S.~Mrenna and P.~Z.~Skands,
  JHEP {\bf 0605}, 026 (2006)
  doi:10.1088/1126-6708/2006/05/026
  [hep-ph/0603175].
    
  
  
\bibitem{deFavereau:2013fsa} 
  J.~de Favereau {\it et al.} [DELPHES 3 Collaboration],
  JHEP {\bf 1402}, 057 (2014)
  doi:10.1007/JHEP02(2014)057
  [arXiv:1307.6346 [hep-ex]].
  
\bibitem{Cacciari:2005hq} 
  M.~Cacciari and G.~P.~Salam,
  Phys.\ Lett.\ B {\bf 641}, 57 (2006)
  doi:10.1016/j.physletb.2006.08.037
  [hep-ph/0512210].
  
\bibitem{Cacciari:2011ma} 
  M.~Cacciari, G.~P.~Salam and G.~Soyez,
  Eur.\ Phys.\ J.\ C {\bf 72}, 1896 (2012)
  doi:10.1140/epjc/s10052-012-1896-2
  [arXiv:1111.6097 [hep-ph]].





\end{thebibliography}
\end{document}